\documentclass[prd,twocolumn,nofootinbib,showpacs,10pt]{revtex4-1}
\usepackage{textcomp}
\usepackage{amsmath}
\usepackage{amsfonts}
\usepackage[bbgreekl]{mathbbol}
\usepackage{calc}
\usepackage{mathrsfs}
\usepackage{amssymb}
\usepackage{amsmath}
\usepackage{array}
\usepackage{color}
\usepackage{bbm}
\usepackage{graphicx}
\usepackage{hyperref}

\renewcommand{\P}{\mathcal{P}}
\newcommand{\res}{\mathcal{R}}
\renewcommand{\S}{{\rm S}}
\newcommand{\R}{{\rm R}}

\newcommand{\e}{\epsilon}

\newcommand{\man}{\mathcal{M}}

\newcommand{\varLie}{\mathsterling}
\newcommand{\vardelta}{\bbdelta}
\newcommand{\Lie}{\mathcal{L}}

\newcommand{\GW}[1]{\breve{#1}}

\renewcommand{\O}{\mathcal{O}}

\newcommand{\beq}{\begin{equation}}
\newcommand{\eeq}{\end{equation}}

\begin{document}
\title{Gauge and motion in perturbation theory} 
\author{Adam Pound} 
\affiliation{Mathematical Sciences, University of Southampton, Southampton,
United Kingdom, SO17 1BJ}
\date{\today}

\begin{abstract}
Through second order in perturbative general relativity, a small compact object in an external vacuum spacetime obeys a generalized equivalence principle: although it is accelerated with respect to the external background geometry, it is in free fall with respect to a certain \emph{effective} vacuum geometry. However, this single principle takes very different mathematical forms, with very different behaviors, depending on how one treats perturbed motion. Furthermore, any description of perturbed motion can be altered by a gauge transformation. In this paper, I clarify the relationship between two treatments of perturbed motion and the gauge freedom in each. I first show explicitly how one common treatment, called the Gralla-Wald approximation, can be derived from a second, called the self-consistent approximation. I next present a general treatment of smooth gauge transformations in both approximations, in which I emphasise that the approximations' governing equations can be formulated in an invariant manner. All of these analyses are carried through second perturbative order, but the methods are general enough to go to any order. Furthermore, the tools I develop, and many of the results, should have broad applicability to any description of perturbed motion, including osculating-geodesic and two-timescale descriptions.
\end{abstract}

\maketitle

\section{Introduction}
In general relativity, a point test mass travels on a timelike geodesic of the spacetime geometry it resides in, satisfying
\beq\label{geodesic-bg}
\frac{D^2z^\mu}{d\tau^2} = 0,
\eeq
where $z^\mu(\tau)$ is the particle's coordinate position, $\tau$ is proper time as measured in the spacetime's metric $g_{\mu\nu}$, $\frac{D}{d\tau}:=u^\mu \nabla_{\mu}$, $u^\mu:=\frac{dz^\mu}{d\tau}$, and $\nabla_\mu$ is the covariant derivative compatible with $g_{\mu\nu}$.

Now suppose we take a step beyond the test-particle approximation and treat the mass as a small, extended, gravitating object surrounded by a vacuum region. The mass creates a nonlinear perturbation $h_{\mu\nu}$ atop the vacuum background metric $g_{\mu\nu}$. This perturbation accelerates the object, and it no longer moves on a geodesic of $g_{\mu\nu}$. Furthermore, it also does not move on a geodesic of the full spacetime's metric ${\sf g}_{\mu\nu}=g_{\mu\nu}+h_{\mu\nu}$. However, if the  object is sufficiently similar to the structureless point mass---uncharged, slowly spinning, nearly spherical, and compact---then it \emph{does} travel on a geodesic of an \emph{effective} metric $\tilde g_{\mu\nu}$. In this case, Eq.~\eqref{geodesic-bg} is replaced by~\cite{Pound:12a}
\beq\label{geodesic-eff}
\frac{\tilde D^2z^\mu}{d\tilde \tau^2} = \O(\e^3),
\eeq
where $\e$ is a small quantity proportional to the object's diameter $d$ and mass $m$, and $\tilde\tau$, $\frac{\tilde D}{d\tilde\tau}:=\tilde u^\mu \tilde\nabla_{\mu}$, and $\tilde u^\mu:=\frac{dz^\mu}{d\tilde \tau}$ are now defined with respect to $\tilde g_{\mu\nu}$. What is the effective metric $\tilde g_{\mu\nu}$? It is a certain smooth vacuum solution $g_{\mu\nu}+h^\R_{\mu\nu}$, where the \emph{regular field} $h^\R_{\mu\nu}=h_{\mu\nu}-h^\S_{\mu\nu}$ is obtained from $h_{\mu\nu}$ by subtracting a certain \emph{singular field} $h^\S_{\mu\nu}$, which encodes the local information about the object's mass and multipole structure and diverges on $z^\mu$.

These results are derived in gravitational self-force theory, in which one examines the limit $\e\to0$ and constructs an asymptotic expansion of the full metric ${\sf g}_{\mu\nu}$ around the background $g_{\mu\nu}$. The titular self-force in this theory is the force that accelerates the object with respect to the background metric; I refer the reader to Refs.~\cite{Mino-Sasaki-Tanaka:97, Quinn-Wald:97, Detweiler-Whiting:03} for the seminal work in this field and Refs.~\cite{Gralla-Wald:08, Pound:10a, Pound:12a,Pound:12b,Gralla:12,Harte:12}, together with the reviews~\cite{Poisson-Pound-Vega:11,Pound:15a}, for rigorous and comprehensive treatments. Because Eq.~\eqref{geodesic-eff} tells us that all objects, regardless of internal composition, fall freely in a vacuum gravitational field, it can be loosely thought of as a generalized equivalence principle. 

However, this simple principle masks two important subtleties: how the form of the equation of motion and its solutions, and even perturbation theory as a whole, depend strongly on one's basic treatment of perturbed motion; and how perturbed motion depends inherently on one's choice of gauge. 

In self-force theory, various descriptions of perturbed motion are in use. In the Gralla-Wald formalism developed in Refs.~\cite{Gralla-Wald:08,Gralla:12}, one expands both $h_{\mu\nu}$ and $z^\mu$ in powers of $\e$. The equation of motion~\eqref{geodesic-eff} then takes the form of a sequence of equations, given by~\eqref{z0}, \eqref{z1_generic}, and \eqref{z2_generic} below; these are evolution equations for the zeroth-order worldline $z^\mu_0$ and for deviation vectors that describe the object's movement away from $z^\mu_0$. Because the deviations eventually grow large (due to dissipation of energy through gravitational radiation, for example), this formalism is restricted to spacetime domains of finite, $\e$-independent size. An alternative description of perturbed motion is provided by the self-consistent formalism developed in Refs.~\cite{Pound:10a, Pound:12a, Pound:12b}, which expands $h_{\mu\nu}$ but not $z^\mu$ in the limit $\e\to0$, such that $z^\mu$ retains its dependence on $\e$. The equation of motion then takes the form~\eqref{geodesic-eff} given above, or in terms of background-metric quantities, the form~\eqref{2nd-geo} given below. By avoiding the expansion of $z^\mu$, this formalism constructs an approximation valid on spacetime domains of asymptotically large size, such as $\sim1/\e$. Other treatments, such as the osculating-geodesics approximation~\cite{Mino:05,Pound-Poisson:08a,Warburton-etal:12,Pound:15a} and the two-timescale expansion of the Einstein equation~\cite{Hinderer-Flanagan:08}, are intermediate between Gralla-Wald and self-consistent approximations, utilizing expansions of the worldline to build a self-consistent approximation. In this paper I will not discuss these intermediate treatments, as they are easily understood on the basis of the extreme two; I refer the reader to Ref.~\cite{Pound:15a} for a description of how that understanding arises in the case of the osculating-geodesics approximation.

Now, supposing we have adopted one or the other treatment, consider the effect of a gauge transformation. The test-particle equation~\eqref{geodesic-bg} is manifestly invariant under a diffeomorphism $\varphi$: $z^\mu$ and $g_{\mu\nu}$ transform to $z'^\mu=\varphi(z^\mu)$ and $g'_{\mu\nu}(z')=\varphi_*g_{\mu\nu}(z)$ (where $\varphi_*$ is the push forward), and the equation of motion is the same in terms of these new quantities as it was in terms of the old ones. Of course, the field equation is also invariant: if $g_{\mu\nu}$ is a solution to the Einstein equation, then so is $g'_{\mu\nu}$. Can we say the same about the perturbed equation of motion~\eqref{geodesic-eff} and the field equation for the effective metric $\tilde g_{\mu\nu}$ that appears in that equation? Several issues arise when answering this question. A gauge transformation is induced by a near-identity diffeomorphism $\varphi$, and so the transformations $z'^\mu=\varphi(z^\mu)$ and ${\sf g}'_{\mu\nu}=\varphi_*{\sf g}_{\mu\nu}$ should still apply when expanded in the limit $\e\to0$. In coordinates, a gauge transformation reads $x^\mu\to x^\mu-\e\xi^\mu+\O(\e^2)$, and hence one expects $z^\mu\to z^\mu-\e\xi^\mu+\O(\e^2)$; however, the form of $z^\mu$, and therefore the effect of $\varphi$ on it, depends entirely on which representation of perturbed motion one uses, and this plays out in nonobvious ways in the metric. Furthermore, a transformation law for ${\sf g}_{\mu\nu}$ does not imply a transformation law for the effective metric $\tilde g_{\mu\nu}$. How does this particular piece of the total metric transform? That is, if $h_{\mu\nu}$ is split into $h^\S_{\mu\nu}$ and $h^\R_{\mu\nu}$ in a particular gauge, how does that split behave under a gauge transformation?

In this paper, I present a unified framework to tackle these questions. I first clarify the relationship between the self-consistent and Gralla-Wald formalisms. In particular, I show explicitly how the latter  can be derived from the former. I then describe how, in both formalisms, with appropriate transformation laws for  $h^\S_{\mu\nu}$ and $h^\R_{\mu\nu}$, the governing field equations and equation of motion can be made manifestly gauge-invariant, just as they are for a test particle in a background. 

This work extends previous results in several ways. Primarily, it carries explicit calculations to second order in perturbation theory, the order at which significant complexity arises.  More broadly, it provides very general descriptions of the self-consistent formalism, which has previously been formulated almost exclusively in the Lorenz gauge. More narrowly, in clarifying the relationship between the two formalisms, it makes concrete many ideas that had previously only been suggested~\cite{Pound:10a} or roughly sketched~\cite{Pound:14c}, and in the case of the Gralla-Wald formalism, it obtains a covariant and reparametrization-invariant second-order equation of motion, which had previously been obtained only in locally inertial coordinates~\cite{Gralla:12} or in parametrization-dependent form~\cite{Pound:15a}. On gauge transformations, it goes beyond earlier work~\cite{Barack-Ori:01,Gralla-Wald:08,Pound:10b,Gralla:11,Gralla:12,Pound-Merlin-Barack:14,Pound:14c,Shah-Pound:15,Pound:15a} by providing the first clear and complete explication of the gauge freedom in the self-consistent formalism, for which only sketches~\cite{Pound:10b} or incomplete descriptions~\cite{Pound:15a} are available in the literature; because this formalism is a nontrivial alteration of standard perturbation theory, determining the action of gauge transformations within it requires care. The analysis also yields, for the first time, a second-order transformation law for the self-force, generalizing the standard Barack-Ori result~\cite{Barack-Ori:01} from first order. 

One advantage of clearly formulating gauge freedom in the self-consistent formalism is that it illuminates what is and is not invariant in long-term evolutions. The central practical utility of self-force theory is in modeling binary systems of compact objects, particularly extreme-mass-ratio inspirals (EMRIs)~\cite{Barack:09,Amaro-Seoane-etal:14}. EMRIs evolve on a long timescale of size $\sim1/\e$, inversely proportional to the rate of energy emission in gravitational waves, and accurate models must track the waveform on that timescale. However, in formalisms designed for long-term evolution, the waveform, which one expects to be invariant, is a functional of a gauge-dependent worldline. In this paper I perform a preliminary analysis of this issue. Based on informal error estimates, I argue that while the self-force (and more generally, any small perturbations of motion) are pure gauge on domains of size $\sim \e^0$, they contain invariant content in any well-behaved gauge on the large domains of size $\sim 1/\e$, and that the gauge-dependent aspect of the waveform represents a small, negligible correction to its gauge-invariant content.

I organize these analyses as follows. Section~\ref{formalisms} summarizes the self-consistent and Gralla-Wald formalisms. Section~\ref{worldline-expansion} shows how the latter can be derived from the former; the methods I use can also be applied in any other formalism in which the accelerated motion is (implicitly or explicitly) expanded around a nearby worldline, particularly osculating-geodesics and two-timescale approximations. Section~\ref{gauge} describes the gauge freedom in each formalism, deriving transformation laws for the various representations of the worldline, the physical perturbations $h_{\mu\nu}$, the singular and regular fields $h^\S_{\mu\nu}$ and $h^\R_{\mu\nu}$, and the self-force. For simplicity, I restrict my attention to smooth transformations. The paper concludes in Sec.~\ref{conclusion} with a discussion of practical issues, particularly the issue of long-term dynamics.

Before proceeding to that material, I note one subtlety I gloss over: the specific choice of regular field $h^\R_{\mu\nu}$. The split of $h_{\mu\nu}$ into $h^\S_{\mu\nu}$ and $h^\R_{\mu\nu}$ is far from unique, and any number of choices can be made which preserve the core properties that (i) $z^\mu$ is a geodesic of $\tilde g_{\mu\nu}=g_{\mu\nu}+h^\R_{\mu\nu}$, and (ii), $\tilde g_{\mu\nu}$ is a smooth vacuum metric~\cite{Pound-Miller:14,Pound:15a}. Furthermore, one can also choose practical splits into $h^\S_{\mu\nu}$ and $h^\R_{\mu\nu}$ for which one or the other of those core properties is not satisfied. For example, with the split defined in Ref.~\cite{Gralla:12}, property (i) is not satisfied, and with the one defined in Ref.~\cite{Harte:12}, property (ii) is not. Here I assume the effective metric is chosen such that both (i) and (ii) are satisfied, but I place no further constraints upon it. The obvious choice is thus the one presented in Refs.~\cite{Pound:12a,Pound:12b,Pound-Miller:14} (a generalization of the Detweiler-Whiting regular field~\cite{Detweiler-Whiting:03} from first order), but other choices could also be made.

I work in geometrized units with $G=c=1$ and the metric signature $(-,+,+,+)$. Indices are raised and lowered with the background metric $g_{\mu\nu}$, and a semicolon denotes the covariant derivative compatible with that metric. Functionals $P$ of a function $f$ (or tensor field $f$) are written as either $P(x;f)$ or $P[f](x)$, as convenient, where $x^\mu$ is the coordinate label of a point. Quantities with a breve over them, such as $\GW{z}_n^\mu$ and $\GW{h}^n_{\mu\nu}$, refer to a Gralla-Wald expansion; corresponding quantities without breves refer to a self-consistent expansion. The expansion parameter $\e$ is treated as a purely formal one, to be set equal to 1 at the end of a calculation. Throughout the paper, I specialize to objects which, through order $\e^3$, are nonspinning and spherical.

\section{Self-consistent and Gralla-Wald approximations}\label{formalisms}
In this section, I briefly describe the two treatments of perturbed motion. This sets the stage for subsequent sections. It also  serves to generalize past descriptions  of the self-consistent approximation, making the formalism more amenable to gauge transformations.

\subsection{Self-consistent}
The self-consistent approximation treats the metric perturbation as a functional of the object's self-accelerated, center-of-mass worldline. It is encapsulated in an asymptotic series of the form\footnote{All series in this paper are to be thought of as asymptotic rather than convergent, meaning I actually assume that for each integer $N>0$, the perturbation can be written as $h_{\mu\nu}(x,\e)=\sum_{n=0}^N\e^n h^n_{\mu\nu}(x;z)+o(\e^N)$, where ``$o(f(\e))$'' means ``goes to zero faster than $f(\e)$ in the limit $\e\to0$''. Generically, $\ln\e$ terms also appear in the series, but these terms do not disrupt the series' well-orderedness~\cite{Pound:12b}. Here I absorb them into the coefficients $h^n_{\mu\nu}(x;z)$.}
\beq\label{SC_expansion}
h_{\mu\nu}(x,\e)=\sum_{n>0}\e^n h^n_{\mu\nu}(x;z),
\eeq
together with an equation of motion of the form
\beq\label{SC_force_expansion}
\frac{D^2z^\mu}{d\tau^2} = F^\mu(\tau,\e)= \sum_{n\geq0} \e^n F_n^\mu[h^1,\ldots,h^n],
\eeq
where $z^\mu$ are the coordinates on the worldline $\gamma$ representing the object's center of mass~\cite{Pound:10a,Pound:15a}. The perturbations $h^n_{\mu\nu}$ are functionals of $z^\mu$, and $z^\mu$ in turn depends on $h_{\mu\nu}$ through Eq.~\eqref{SC_force_expansion}. The worldline $\gamma$ is accelerated with respect to the background spacetime, and according to Eq.~\eqref{SC_force_expansion}, its acceleration is $\e$ dependent. Hence, through its dependence on $z^\mu$, each term $h^n_{\mu\nu}(x;z)$ likewise depends on $\e$. This $\e$ dependence distinguishes the approximation from an ordinary Taylor expansion.

The idea of a self-consistent approximation has been around since the inception of gravitational self-force theory~\cite{Mino-Sasaki-Tanaka:97,Quinn-Wald:97}, and it is by now fairly well developed~\cite{Pound:10a,Pound:10b,Pound:12a, Pound:12b,Pound-Miller:14, Pound:15a}. However, aside from a generalization in Ref.~\cite{Pound:12b}, the approximation has always been formulated in a specific gauge, the Lorenz gauge. In that formulation, inspired by post-Minkowski theory~\cite{Blanchet:14,Poisson-Will:14}, one imposes the Lorenz gauge condition to split the exact, fully nonlinear Einstein equation into a weakly nonlinear wave equation plus an elliptic constraint equation. The wave equation is solved perturbatively for the functionals $h^n_{\mu\nu}(x;z)$ while leaving $\gamma$ unspecified, reducing the constraint equation to evolution equations for the object's matter degrees of freedom, principal among them the equation of motion governing $\gamma$. This formulation makes each functional $h^n_{\mu\nu}(x;z)$ a solution to a wave equation, with the wave operator being exactly the same at each order. The more general formulation in Ref.~\cite{Pound:12b} allows one to choose a different wave operator, but even in this more general formulation, the wave operator is always the same at each order; there is no flexibility to adopt a gauge in which the operator takes different forms at different orders.

This limitation runs counter to the usual usage of gauge freedom, in which one can impose a different gauge condition at each order and thence work with different field equations at each order. So we require a more general formulation, one which allows more flexibility in one's choice of gauge. I present such a formulation here.

\subsubsection{General formulation}\label{SC_general_formulation}
To begin, first consider the exact Einstein equation $R_{\mu\nu}[{\sf g}] = 0$ in a vacuum region \emph{outside} the object. After substituting the expansion~\eqref{SC_expansion}, this equation becomes 
\beq\label{EFE-pregauge}
\sum_{n>0}\e^n\delta R_{\mu\nu}[h^n] = \sum_{n>1} \e^n S^n_{\mu\nu}[h^1,\ldots,h^{n-1}], 
\eeq
where $\delta R_{\mu\nu}[j]:= \frac{d}{d\lambda}R_{\mu\nu}[g+\lambda j]\big|_{\lambda=0}$ is the linearized Ricci tensor constructed from a perturbation $j_{\mu\nu}$, and the quantities on the right-hand side,
\beq\label{S}
S^n_{\mu\nu}[j^1,\ldots,j^{n-1}] := - \frac{1}{n!}\frac{d^n}{d\lambda^n}R_{\mu\nu}[g+\sum_p\lambda^p j^p]\big|_{\lambda=0},
\eeq
are made up of strictly nonlinear combinations of their arguments. Explicitly, $S^1_{\mu\nu} = 0$ and $S^2_{\mu\nu} = -\delta^2 R_{\mu\nu}[h^1,h^1]$, where $\delta^2 R_{\mu\nu}$ is the second-order Ricci tensor; the $\delta$ notation is defined in Appendix~\ref{gauge-identities}.

We wish to solve Eq.~\eqref{EFE-pregauge} subject to 
\begin{enumerate}
\item[(i)] some desired (e.g., retarded) boundary conditions
\item[(ii)] a ``matching condition'' which states that at small distances $r$ from $\gamma$, the solution takes the generic form of a metric outside a compact object, scaling as $h^n_{\mu\nu}\sim 1/r^n$.\footnote{Note that $\gamma$ lives in the smooth manifold of the background spacetime, not that of the perturbed spacetime, making this condition sensible even if the small object is a black hole. See Refs.~\cite{Gralla-Wald:08,Pound:10a,Pound:15a} for precise formulations in the language of matched asymptotic expansions. One works outside the object and imposes the matching condition, in lieu of involving the object's interior, to overcome two problems: First, an expansion in the limit of small mass and size fails to be sensible at distances $\sim\e$ from the object (and in the object's interior); there, the object's own gravity is strong, and the field $h_{\mu\nu}$ it contributes to ${\sf g}_{\mu\nu}$ is comparable to, and has stronger curvature than, the external metric $g_{\mu\nu}$. So one should assume an expansion of the form~\eqref{SC_expansion} only in a region sufficiently far (i.e., at distances $\gg\e$) from the object. Second, if we work in the object's interior, we must specify its internal structure and composition, and we must design different approximations for black holes than for material bodies.}
\item[(iii)] a center-of-mass condition that makes the metric effectively mass-centered on $\gamma$, as discussed in Refs.~\cite{Pound:10a,Pound:12a,Pound:12b,Pound:15a}.
\end{enumerate}
After obtaining a vacuum solution outside the object, we can analytically extend it down to all points $x\notin\gamma$, in a manner described in Refs.~\cite{Pound:10a,Pound:12b}; because of its $1/r^n$ behavior, the extended field will not be of uniform accuracy very near $\gamma$, and it will diverge precisely on $\gamma$, but it will agree with the physical solution everywhere else.

Since the functions $h^n_{\mu\nu}(x;z)$ depend on $z^\mu$, we cannot solve Eq.~\eqref{EFE-pregauge} by equating explicit coefficients of $\e$ without forcing the functional argument $z^\mu$ to be independent of $\e$; intuitively, this is the case because the Einstein equation is overdetermined, constrained by the linearized Bianchi identity $\nabla^\nu(\delta R_{\mu\nu}-\frac{1}{2}g_{\mu\nu}g^{\alpha\beta}\delta R_{\alpha\beta})=0$, which prevents us from freely specifying the center-of-mass worldline $\gamma$. So instead of solving Eq.~\eqref{EFE-pregauge} directly, we construct the asymptotic series~\eqref{SC_expansion} by dividing the Einstein equation into two parts: field equations that (speaking roughly) determine the functionals $h^n_{\mu\nu}(x;z)$, and constraint equations that determine the functionals' argument $z^\mu$.

To accomplish this, we first write Eq.~\eqref{EFE-pregauge} with a gauge condition already imposed. Choose a set $\{E^n_{\mu\nu}\}_n$ of second-order, linear differential operators that can be obtained from the linearized Ricci tensor via a linear gauge transformation; that is, for each tensor $f_{\mu\nu}$ there must exist a vector field $\xi^\mu$ satisfying $E^n_{\mu\nu}[f+\Lie_{\xi}g]=\delta R_{\mu\nu}[f]$, where $\Lie$ is a Lie derivative. Define the functionals $h^n_{\mu\nu}(x;z)$ to be solutions to
\beq\label{EFE-wgauge}
E^n_{\mu\nu}[h^n] = S^n[h^1,\ldots,h^{n-1}] \quad (x\notin\gamma)
\eeq
\emph{for arbitrary} $\gamma$, subject to the desired boundary and matching conditions. For some choices of $E^n_{\mu\nu}$, there may be no solution to Eq.~\eqref{EFE-wgauge} for arbitrary $\gamma$. However, if we choose each $E^n_{\mu\nu}$ to be symmetric hyperbolic, the solutions should be guaranteed to exist. So henceforth, assume we have made such a choice. 

The sequence of equations~\eqref{EFE-wgauge} form our field equations. They are equivalent to the Einstein equation~\eqref{EFE-pregauge} in a particular gauge, and hence they take care of a large portion of that equation. What is left is a gauge condition, or constraint equation, which reads
\beq\label{EFE-gauge}
\sum_{n\geq1} \e^n C^n_{\mu\nu}[h^n] = 0 \quad (x\notin\gamma),
\eeq
where
\beq\label{C}
C^n_{\mu\nu}[h^n]:=\delta R_{\mu\nu}[h^n]-E^n_{\mu\nu}[h^n].
\eeq

Equations~\eqref{EFE-wgauge} and \eqref{EFE-gauge} are the desired division of the Einstein equation. By solving Eqs.~\eqref{EFE-wgauge}, one obtains the series~\eqref{SC_expansion} as a sum of functionals of two types of free functions: the worldline $\gamma$, which characterizes the object's mean motion, and a set of multipole moments, which characterize the object itself and can be defined as tensors on $\gamma$. By enforcing the condition~\eqref{EFE-gauge}, one determines the equation of motion~\eqref{SC_force_expansion} as well as evolution equations for the multipole moments.\footnote{For moments of quadrupole order and higher, the Einstein equation alone does not fully determine the moments' evolution. The evolution of these moments is freely specified, either directly or via a specification of the object's composition.} These facts about the solutions follow from the algorithmic solution method presented most thoroughly in Refs.~\cite{Pound:12b,Pound:15a}, which I will not recapitulate but will draw conclusions from here and below.  

Because each $h^n_{\mu\nu}$ depends on $\e$ (through $\gamma$), Eq.~\eqref{EFE-gauge} cannot be solved by setting $C^n_{\mu\nu}[h^n] = 0$. Instead, it is to be solved by substituting the expansion~\eqref{SC_force_expansion} and only then solving order by order in $\e$ \emph{while holding $z^\mu$ (and $\frac{dz^\mu}{d\tau}$) fixed}. This leads to a sequence of equations for $F^\alpha_n(\tau;z)$, in which $z^\mu$ is still held fixed. By holding $z^\mu$ and $\frac{dz^\mu}{d\tau}$ fixed during this procedure, rather than expanding their $\e$ dependence, I preserve the particular accelerated worldline that satisfies the chosen center-of-mass condition. Solving the sequence of constraint equations for $F^\alpha_n(\tau;z)$ yields a better and better approximation to the equation of motion of that particular worldline, \emph{without ever expanding the worldline itself}. 

In practice, of course, we have access to only a finite number of terms $F_n^\mu$ in the expansion of the force. If we truncate the expansion~\eqref{SC_force_expansion} at order $n$, then we work with an approximation $z^\mu_n$ to $z^\mu$; this approximant satisfies
\beq\label{zn-eqn}
\frac{D^2z^\mu_n}{d\tau^2} = \sum_{p=0}^n\e^p F^\mu_p(\tau;z_n),
\eeq
and the $n$th-order self-consistent approximation to the field is a functional of this worldline, given by $h_{\mu\nu}(x,\e)=\sum_{p=0}^n\e^p h^p_{\mu\nu}(x;z_n)$. However, to reduce the burden of notation, I will typically write expressions in terms of $z^\mu$ instead of its approximant $z^\mu_n$.

Thus far I have not referred to the decomposition $h_{\mu\nu}=h^{\S}_{\mu\nu}+h^{\R}_{\mu\nu}$ with which the paper began. This is a decomposition of the extended field, and its singular piece $h^{\S}_{\mu\nu}$ diverges on $\gamma$. A convenient, precise way of choosing the split is described in Refs.~\cite{Pound:10a,Pound:12a, Pound:12b, Pound:15a}. With a straightforward extension of the analyses in Refs.~\cite{Pound:12b,Pound:15a}, one can show that no matter what the forces $F^\mu_p(\tau;z_n)$ in Eq.~\eqref{zn-eqn} turn out to be, one can always choose the decomposition of $h_{\mu\nu}$ such that the regular field possesses the two core properties that (i)  Eq.~\eqref{zn-eqn} is equivalent to the geodesic equation in $\tilde g_{\mu\nu}=g_{\mu\nu}+h^\R_{\mu\nu}$, and (ii) $\tilde g_{\mu\nu}$ is a smooth vacuum metric on $\gamma$. 

\subsubsection{Field and motion through second order}
Let me make this more concrete at first and second order. At these orders, the regular fields $h^{\R n}_{\mu\nu}$ satisfy the vacuum version of Eq.~\eqref{EFE-wgauge}:
\begin{subequations}\label{hRSC}
\begin{align}
E^1_{\mu\nu}[h^{\R1}] &= 0,\label{hR1SC}\\
E^2_{\mu\nu}[h^{\R2}] &= -\delta^2 R_{\mu\nu}[h^{\R1},h^{\R1}];\label{hR2SC}
\end{align}
\end{subequations}
note that these equations, unlike Eq.~\eqref{EFE-wgauge}, are not restricted to points off $\gamma$. The singular fields satisfy the nonvacuum equations
\begin{subequations}\label{hSSC}
\begin{align}
E^1_{\mu\nu}[h^{\S1}] &= 8\pi \bar T^1_{\mu\nu}[z],\label{hS1SC}\\
E^2_{\mu\nu}[h^{\rm s 2}] &= 8\pi \bar T^2_{\mu\nu}[z,\delta m],\label{hs2SC}\\
E^2_{\mu\nu}[h^{\rm Snl}] &= -\delta^2 R_{\mu\nu}[h^{\S1},h^{\S1}]\nonumber\\
							&\quad -2\delta^2 R_{\mu\nu}[h^{\S1},h^{\R1}]\quad(x\notin\gamma),\label{hSnlSC}
\end{align}
\end{subequations}
where I have written $h^{\S2}_{\mu\nu}$ as the sum of two parts, $h^{\rm s2}_{\mu\nu}+h^{\rm Snl}_{\mu\nu}$, the first of which is sourced by a ``stress-energy'' and the latter of which is sourced by nonlinearities. 

The stress-energy source terms I have introduced in Eq.~\eqref{hSSC} are derived quantities defined from the extended fields $h^n_{\mu\nu}$ according to
\beq
\bar T^n_{\mu\nu}:=\frac{1}{8\pi}E^n_{\mu\nu}[h^n]-S^n_{\mu\nu}[h^1,\ldots,h^{n-1}].\label{TSC}
\eeq
One can loosely, though not precisely~\cite{Pound:12b,Pound:15a}, think of these quantities as the sources of everything in $h^n_{\mu\nu}$ that is \emph{not} generated by nonlinearities. They are supported only on $\gamma$, and they are uniquely specified by the multipole moments that characterize the object. For a nonspinning object, at the first two orders they (or their trace-reversals $T^n_{\mu\nu}:=\bar T^n_{\mu\nu}-\tfrac{1}{2}g_{\mu\nu}g^{\alpha\beta}\bar T^n_{\alpha\beta}$) are given by
\begin{subequations}\label{TSC-1,2}
\begin{align}
T^1_{\mu\nu} &= \int_\gamma mu_\mu u_\nu \delta(x,z)d\tau,\label{T1}\\
T^2_{\mu\nu} &= \int_\gamma \frac{1}{4}\overline{\delta m}_{\mu\nu} \delta(x,z)d\tau,\label{T2}
\end{align}
\end{subequations}
where $\delta(x,z):=\frac{\delta^4(x-z)}{\sqrt{-g}}$ is a covariant delta function, $m$ is the object's leading-order mass, $\delta m_{\mu\nu}$ is a correction to the object's monopole moment, and the bar denotes trace-reversal. In addition to the equation of motion for $\gamma$, the constraint equation~\eqref{EFE-gauge} dictates the specific forms of $T^n_{\mu\nu}$,  determining a (gauge-dependent) expression for $\delta m_{\mu\nu}$ and determining that $m$ is constant. Note that $T^1_{\mu\nu}$ is simply the stress-energy of a point particle of mass $m$ moving on $\gamma$; in this sense, the small object can be thought of as a point mass.

Unlike Eq.~\eqref{EFE-wgauge}, the equations~\eqref{hRSC} and \eqref{hSSC} include $\gamma$ in their domain, with the exception of Eq.~\eqref{hSnlSC}. Equation~\eqref{hSnlSC} is restricted to points off $\gamma$ because the nonlinear source term in it is generically ill defined as a distribution on any region including $\gamma$. In the same way, the total first-order field $h^1_{\mu\nu}=h^{\S1}_{\mu\nu}+h^{\R1}_{\mu\nu}$ satisfies a field equation with a distributional source,
\begin{align}
E^1_{\mu\nu}[h^1] &= 8\pi \bar T^1_{\mu\nu}[z] ,\label{h1SC}
\end{align} 
but the total second-order physical field $h^2_{\mu\nu}=h^{\rm s2}_{\mu\nu}+h^{\rm Snl}+h^{\R2}_{\mu\nu}$ does not, instead satisfying an equation with a pointwise source,
\beq
E^2_{\mu\nu}[h^2] = -\delta^2 R_{\mu\nu}[h^1,h^1]  \quad(x\notin\gamma).\label{h2SC}
\eeq
However, by moving all curvature terms to the left-hand side, we \emph{can} write the field equation in the distributional form
\beq
E^2_{\mu\nu}[h^2]+\delta^2 R_{\mu\nu}[h^1,h^1]=8\pi\bar T^2_{\mu\nu}.\label{h2SC-distributional}
\eeq
The left-hand side is well defined as a distribution because the non-distributional singularities in $\delta R_{\mu\nu}[h^2]$ cancel those in $\delta^2R_{\mu\nu}[h^1,h^1]$.
We can also combine the first- and second-order equations to form\footnote{More generally, we may write the complete Einstein equation as the distributional equation
\beq
\sum _n\e^n \left(E^n_{\mu\nu}[h^n]-S^n_{\mu\nu}\right) = 8\pi\sum_n\e^n\bar T^n_{\mu\nu},
\eeq
or simply
\beq
R_{\mu\nu}\left[g+\sum _n\e^n h^n\right] = 8\pi\sum_n\e^n\bar T^n_{\mu\nu},
\eeq
where all tensors are understood to live on the background.
}
\begin{align}
\delta R_{\mu\nu}[\e h^1+&\e^2 h^2]+\e^2\delta^2R_{\mu\nu}[h^1,h^1] \nonumber\\
 &= 8\pi\e\bar T^1_{\mu\nu}+8\pi\e^2\bar T^2_{\mu\nu}+\O(\e^3).\label{distributional-EFE-SC-dR}
\end{align}

All of the above equations involve functionals of $\gamma$. The final ingredient in the second-order-accurate approximation is the equation of motion~\eqref{zn-eqn}. It can be written in terms of the regular field as Eq.~\eqref{2nd-geo}. As shown in Sec.~\ref{geodesic_expansion_in_h}, this is equivalent to the geodesic equation~\eqref{geodesic-eff} in the smooth vacuum metric $\tilde g_{\mu\nu}$.

\subsubsection{Gauge choices and practical implementations}
The structure of the approximation becomes more transparent in a particular gauge. In the Lorenz-gauge formulation, we have $E^n_{\mu\nu}[h^n]=-\frac{1}{2}(\Box h^n_{\mu\nu}+2R_\mu{}^\alpha{}_\nu{}^\beta h^n_{\alpha\beta})$ and $C^n_{\mu\nu}=\nabla_{(\mu}\nabla^\alpha\bar h^n_{\nu)\alpha}$ for all $n$, and Eq.~\eqref{EFE-gauge} can be replaced with $\sum_n \e^n \nabla^\alpha\bar h^n_{\nu\alpha}=0$. Near $\gamma$, the first-order singular field (and hence $h^1_{\mu\nu}$) behaves as
\beq
h^{\S1}_{\mu\nu} = \frac{2m}{r}(g_{\mu\nu}+2u_{\mu}u_\nu)+\O(r^0),\label{hS1-local}
\eeq
where $r$ is a geodesic spatial distance (proper in $g_{\mu\nu}$) from $\gamma$. The second-order field behaves as
\begin{subequations}\label{hS2-local}
\begin{align}
h^{\rm s2}_{\mu\nu} &= \frac{\delta m_{\mu\nu}}{r}+\O(r^0),\\
h^{\rm Snl}_{\mu\nu} &\sim \frac{m^2}{r^2}+\O(r^{-1}),\label{hSnl-local}
\end{align}
\end{subequations}
where $\delta m_{\mu\nu}$ is given explicitly by Eq.~(133) of Ref.~\cite{Pound-Miller:14}. Later sections will refer to these local forms to elucidate several key ideas. For now, note their essential characteristic: they diverge on the object's self-accelerated center-of-mass worldline $\gamma$, rather than on a background geodesic. This distinguishes them from the singular field of the Gralla-Wald approximation described below.

Let me summarize. The self-consistent approximation described in this section is based on splitting the Einstein equation into a sequence of hyperbolic equations together with a constraint. The constraint determines the equation of motion of the object's center-of-mass worldline $\gamma$, and it constrains the evolution of the object's multipole moments, which are tensors on $\gamma$. From the analysis in Sec.~II of Ref.~\cite{Pound:12b}, one can expect that at a formal level, this is the only required input from the constraint: if the Cauchy data satisfies the constraint, then the constraint should be preserved by the coupled system comprising the hyperbolic equations, the equation of motion, and the evolution equations for the multipole moments. In practice, the set of coupled equations can be solved numerically with a puncture scheme~\cite{Barack-Golbourn:07, Vega-Detweiler:07, Pound:12a,Pound:12b}, in which one uses \emph{residual} field variables $h^{\res n}_{\mu\nu}:=h^n_{\mu\nu}-h^{\P n}_{\mu\nu}$ in place of $h^{\R n}_{\mu\nu}$. Here $h^{\P n}_{\mu\nu}$ is a \emph{puncture} field that mimics $h^{\S n}_{\mu\nu}$ near $\gamma$, such that  $h^{\res n}_{\mu\nu}=h^{\R n}_{\mu\nu}$ and $\partial_\alpha h^{\res n}_{\mu\nu}=\partial_\alpha h^{\R n}_{\mu\nu}$ on $\gamma$, allowing one to use $h^{\res n}_{\mu\nu}$ in the equation of motion~\eqref{2nd-geo}. 

No serious study has been made of constraint violation or numerical stability in these puncture schemes, particularly in the wide class of gauges considered here, but such issues are not of essential interest in this paper. The question of interest here is ``Suppose two researchers obtain asymptotic solutions of the form~\eqref{SC_expansion}, in whatever manner, in two different gauges. How are their two solutions related?'' This is the question addressed in the later sections of this paper.

\subsection{Gralla-Wald}
I next consider the Gralla-Wald approximation, named after the authors of Refs.~\cite{Gralla-Wald:08,Gralla:12}. This approximation consists of a strict Taylor expansion,
\beq\label{GW_expansion}
h_{\mu\nu}(x,\e)=\sum_{n>0}\e^n \GW{h}^n_{\mu\nu}(x);
\eeq
here, unlike Eq.~\eqref{SC_expansion}, the coefficients $ \GW{h}^n_{\mu\nu}$ \emph{do not depend on $\e$}. Since the metric depends on the object's motion, this expansion of $h_{\mu\nu}$ requires an expansion of the worldline itself,
\beq\label{z_expansion}
z^\mu(s,\e) = \sum_{n\geq0}\e^n \GW{z}_n^\mu(s),
\eeq
where $s$ is a parameter on the worldline. The leading-order term, $z^\mu_0(s):=\GW{z}^\mu_0(s)=z^\mu(s,0)$, is the coordinate description of a worldline $\gamma_0$, and the subleading terms $\GW{z}^\mu_{n>0}$ are vectors living on $\gamma_0$ that describe the deviation of the perturbed worldline $\gamma$ from $\gamma_0$. Hence, in this approximation, one treats the effect of the self-force as a small perturbation of the worldline itself, rather than as a small perturbation of the equation of motion for $z^\mu$. Hence, instead of seeking an equation of motion \a`a la~\eqref{SC_force_expansion} for $z^\mu$, one seeks evolution equations for each quantity $\GW{z}_n^\mu$: 
\begin{align}\label{GW_motion_expansion}
\frac{D^2\GW{z}^\mu_n}{d\tau_0^2}=\ldots, 
\end{align}
where $\tau_0$ is proper time (as measured in $g_{\mu\nu}$) on $\gamma_0$.

It is easy to intuit that one can derive expansions of the form~\eqref{GW_expansion} and \eqref{GW_motion_expansion} from the self-consistent approximation simply by substituting the expansion~\eqref{z_expansion} into the field~\eqref{SC_expansion} and equation of motion~\eqref{SC_force_expansion}. This intuition is borne out by the explicit expansion procedure of Sec.~\ref{worldline-expansion}. On the other hand, one cannot go the other way and obtain the self-consistent from the Gralla-Wald approximation (although Ref.~\cite{Gralla-Wald:08} contained a heuristic argument for how to make that leap at first order).

However, one can certainly work entirely in the Gralla-Wald expansion without ever relating it to the self-consistent one. The general procedure is very similar to that described in Sec.~\ref{SC_general_formulation}; because of the similarity, I shall forgo a general description and simply emphasize key points and concrete equations through second order. Starting from the expansion~\eqref{GW_expansion}, one obtains results very similar to those of the self-consistent expansion, but with a significant modification due to the different treatment of the worldline. In the self-consistent expansion, the fields $h^n_{\mu\nu}$ are functionals of $\gamma$ and of a set of multipole moments on $\gamma$; in the Gralla-Wald expansion, the fields $\GW{h}^n_{\mu\nu}$ are functionals of $\gamma_0$, not of $\gamma$, and of a \emph{different} set of multipole moments on $\gamma_0$. The multipole moments are modified by the deviation vectors $\GW{z}_{n>0}^\mu$, which themselves correspond to mass dipole moments and which alter the other moments. 

This is most easily understood by starting at the end. Consider Eq.~\eqref{hS1-local}, which shows that $h^1_{\mu\nu}$ behaves like a Coulomb field;  in flat space it would read $\propto \frac{\e m}{|x^i-z^i(t)|}$, where $(t,x^i)$ are inertial Cartesian coordinates. If we treat the worldline as in Eq.~\eqref{z_expansion}, then $\frac{\e m}{|x^i-z^i|}$ becomes $\frac{\e m}{|x^i-z_0^i|}+\frac{\e^2m\GW{z}_{1j}(x^j-z_0^j)}{|x^i-z_0^i|^3}+\O(\e^3)$. In words, the field in the self-consistent expansion, $\frac{\e m}{|x^i-z^i|}$, diverges at the object's perturbed center-of-mass position $z^i$; the field in the Gralla-Wald expansion, $\frac{\e m}{|x^i-z_0^i|}+\frac{\e^2m\GW{z}_{1j}(x^j-z_0^j)}{|x^i-z_0^i|^3}+\O(\e^3)$, diverges at the object's zeroth-order position $z_0^i$, and the deviation of the object away from that position alters the functional form of the metric,  introducing a mass dipole moment $M_j=m\GW{z}_{1j}$. Or as 4-vectors,
\beq\label{M=mz1}
M^\mu=m\GW{z}^\mu_{1\perp},
\eeq
where $\GW{z}_{1\perp}^{\mu}=(g^\mu{}_\nu+u_0^\mu u_{0\nu})\GW{z}_1^\nu$ is the projection of $\GW{z}_1^\nu$ orthogonal to $\gamma_0$; here $u_0^\mu:=\frac{dz^\mu_0}{d\tau_0}$, with $\tau_0$ being proper time (as measured in $g_{\mu\nu}$) on $\gamma_0$. Equation~\eqref{M=mz1} is in fact how we \emph{define} the object's deviation from $\gamma_0$: $\GW{z}^\mu_{1\perp}:=M^\mu/m$.

Concretely, the local behavior of the fields in the Lorenz gauge is no longer given by Eqs.~\eqref{hS1-local} and \eqref{hS2-local} but by
\begin{subequations}
\begin{align}
\GW{h}^{\S1}_{\mu\nu} &= \frac{2m}{r}(g_{\mu\nu}+2u_{0\mu}u_{0\nu})+\O(r^0),\\
\GW{h}^{\rm s2}_{\mu\nu} &= \frac{2m\GW{z}_{1i}x^i}{r^3}(g_{\mu\nu}+2u_{0\mu}u_{0\nu})+\frac{\GW{\delta m}_{\mu\nu}}{r}+\O(r^0),\label{hs2GW-local}\\
\GW{h}^{\rm Snl}_{\mu\nu} &\sim \frac{m^2}{r^2}+\O(r^{-1}),
\end{align}
\end{subequations}
where $x^i$ are local spatial coordinates centered at $\gamma_0$ and $r$ is a geodesic spatial distance from $\gamma_0$. The monopole moment $\GW{\delta m}_{\mu\nu}$ is modified by the presence of $\GW{z}_{1i}$, and it is now given explicitly by Eq.~(145) of Ref.~\cite{Pound-Miller:14}

In any gauge, the field satisfies equations analogous to Eqs.~\eqref{h1SC}, \eqref{h2SC}, \eqref{hRSC}, and \eqref{hSSC}:
\begin{subequations}\label{hGW}
\begin{align}
\delta R_{\mu\nu}[\GW{h}^1] &= 8\pi \GW{\bar T}^1_{\mu\nu}[z_0],\label{h1GW}\\
\delta R_{\mu\nu}[\GW{h}^2] &= -\delta^2 R_{\mu\nu}[\GW{h}^1,\GW{h}^1]  \quad (x\notin\gamma_0)\label{h2GW}
\end{align} 
\end{subequations}
for the full field,
\begin{subequations}\label{hRGW}
\begin{align}
\delta R_{\mu\nu}[\GW{h}^{\R1}] &= 0,\label{hR1GW}\\
\delta R_{\mu\nu}[\GW{h}^{\R2}] &= -\delta^2 R_{\mu\nu}[\GW{h}^{\R1},\GW{h}^{\R1}]\label{hR2GW}
\end{align}
\end{subequations}
for the regular field, and
\begin{subequations}\label{hSGW}
\begin{align}
\delta R_{\mu\nu}[\GW{h}^{\S1}] &= 8\pi \GW{\bar T}^1_{\mu\nu}[z_0],\label{hS1GW}\\
\delta R_{\mu\nu}[\GW{h}^{\rm s2}] &= 8\pi \GW{\bar T}^2_{\mu\nu}[z_0,\GW{\delta m},\GW{z}_1],\label{hs2GW}\\
\delta R_{\mu\nu}[\GW{h}^{\rm Snl}] &= -\delta^2 R_{\mu\nu}[\GW{h}^{\S1},\GW{h}^{\S1}]\nonumber\\
							&\quad -2\delta^2 R_{\mu\nu}[\GW{h}^{\S1},\GW{h}^{\R1}]\quad (x\notin\gamma_0)\label{hSnlGW}
\end{align}
\end{subequations}
for the singular field. The stress-energy terms are defined in analogy with Eq.~\eqref{TSC}, and they (in their trace-reversed form) are given by distributions on $\gamma_0$,
\begin{subequations}
\begin{align}
\GW{T}^1_{\mu\nu} &= \int_{\gamma_0} m u_{0\mu} u_{0\nu} \delta(x,z_0)d\tau_0,\\
\GW{T}^2_{\mu\nu} &= \int_{\gamma_0}\left[\tfrac{1}{4}\GW{\overline{\delta m}}_{\mu\nu} +mu_{0\mu}u_{0\nu}\GW{z}_{1\perp}^{\gamma'}\nabla_{\gamma'}\right]\delta(x,z_0)d\tau_0.
\end{align}
\end{subequations}

Just as in the self-consistent approximation, we can also write the Einstein equation in distributional form,
\begin{align}\label{distributional-EFE-GW}
\delta R_{\mu\nu}[\e \GW{h}^1+&\e^2 \GW{h}^2]+\e^2\delta^2R_{\mu\nu}[\GW{h}^1,\GW{h}^1] \nonumber\\
&= 8\pi\e\GW{\bar T}^1_{\mu\nu}+8\pi\e^2\GW{\bar T}^2_{\mu\nu}+\O(\e^3).
\end{align}

The second-order approximation is completed by the evolution equations~\eqref{GW_motion_expansion} for $\GW{z}_n^\mu$. As in the self-consistent case, for a nonspinning object with vanishing quadrupole moment, the evolution equations can be written purely in terms of the regular field: in place of Eq.~\eqref{2nd-geo}, we here have the evolution equations~\eqref{z0}, \eqref{z1_generic}, and \eqref{z2_generic}, with Eqs.~\eqref{F1_generic} and \eqref{F2_generic}. The Riemann terms in these equations of motion are geodesic-deviation terms; they correspond to the fact that even in the absence of a force, two neighbouring curves $z^\mu$ and $z^\mu_0$ will deviate from one another due to the background curvature.


\section{From self-consistent to Gralla-Wald}\label{worldline-expansion}
I now show  how to obtain the Gralla-Wald approximation from the self-consistent one. This means two things: how to obtain equations of motion for $\GW{z}_n^\mu$ given an equation of motion for $z^\mu$, and how to substitute the expansion of $z^\mu$ into functionals such as $h^n_{\mu\nu}[z]$.

Two main tools are required in these derivations: the ordinary Lie derivative, $\Lie$, and a variant of it, $\varLie$, to be described below.

\subsection{Expanded forms of the equation of motion}\label{motion_expansions}
The difference between the self-consistent and Gralla-Wald approximations is most conspicuous in their equations of motion. In this section, I delineate that difference. I start from the geodesic equation in a smooth metric $\tilde{g}_{\mu\nu}$, and in Sec.~\ref{geodesic_expansion_in_h} I derive its self-consistent expansion, given by Eq.~\eqref{2nd-geo} below. In Sec.~\ref{geodesic_expansion_in_h_and_dz}, I start from the self-consistent result and by expanding the worldline, I derive the Gralla-Wald expansion of the geodesic equation, given by Eqs.~\eqref{z0}, \eqref{z1_generic}, and \eqref{z2_generic}, with Eqs.~\eqref{F1_generic} and \eqref{F2_generic}.

\subsubsection{Self-consistent expansion}\label{geodesic_expansion_in_h}
Generically, the geodesic equation~\eqref{geodesic-eff} reads 
\beq
\frac{d\dot z^\mu}{ds}+\tilde\Gamma^\mu_{\nu\rho}\dot z^\nu \dot z^\rho=\tilde\kappa \dot z^\mu,
\eeq
where $s$ is a potentially non-affine parameter on $z^\mu$, $\dot z^\mu:=\frac{dz^\mu}{ds}$ is its tangent vector field, $\tilde\Gamma^\mu_{\nu\rho}$ is the Christoffel symbol corresponding to $\tilde g_{\mu\nu}$, and $\tilde\kappa(s):=\frac{d}{ds}\ln\sqrt{-\tilde g_{\mu\nu}\dot z^\mu \dot z^\nu}$. Here for simplicity I have dropped the unknown $\O(\e^3)$ error term in Eq.~\eqref{geodesic-eff}; this also serves to emphasize that the derivation applies to the more general problem of expanding the exact geodesic equation in \emph{any} smooth metric.

Now, if we write the metric as the sum of two pieces, $\tilde g_{\mu\nu}=g_{\mu\nu}+h^\R_{\mu\nu}$, and if we take $s=\tau$, the proper time on $z^\mu$ as measured in $g_{\mu\nu}$, and if we rewrite the geodesic equation in terms of covariant derivatives compatible with $g_{\mu\nu}$, we find
\begin{equation}
\frac{D^2z^\mu}{d\tau^2} = -C^\mu{}_{\nu\rho}u^\nu u^\rho+\tilde\kappa u^\mu=:F^\mu(\tau,\e),\label{exact_geodesic}
\end{equation}
where 
\begin{subequations}
\begin{align}
C^\alpha{}_{\beta\gamma} &:= \tilde\Gamma^\alpha_{\beta\gamma}-\Gamma^\alpha_{\beta\gamma}
			= \frac{1}{2}\tilde g^{\alpha\delta}\left(2h^\R_{\delta(\beta;\gamma)}-h^\R_{\beta\gamma;\delta}\right)
\end{align}
\end{subequations}
is the difference between the Christoffel symbol associated with $\tilde g_{\mu\nu}$ and the one associated with $g_{\mu\nu}$. With $\tau$ as a parameter, $\tilde\kappa$ becomes
\begin{equation}
\tilde\kappa(\tau) = \frac{\frac{d}{d\tau}\sqrt{1-h^\R_{\mu\nu}u^\mu u^\nu}}{\sqrt{1-h^\R_{\mu\nu}u^\mu u^\nu}}.
\end{equation}
While $z^\mu$ is a geodesic of $\tilde{g}_{\mu\nu}$, Eq.~\eqref{exact_geodesic} tells us it is accelerated in $g_{\mu\nu}$.

So far no approximation has been made; Eq.~\eqref{exact_geodesic} is exact. If we now expand $C^\mu{}_{\nu\rho}$ and $\tilde\kappa$ in powers of $h^\R_{\mu\nu}$, we find
\begin{align}
\frac{D^2z^\alpha}{d\tau^2} &= -\frac{1}{2}(g^{\alpha\delta}-h^{\R\alpha\delta})\!\left(2h^\R_{\delta(\beta;\gamma)}-h^\R_{\beta\gamma;\delta}\right)\!u^\beta u^\gamma\nonumber\\
		&\quad-h^\R_{\beta\gamma}u^\alpha \frac{D^2z^\beta}{d\tau^2} u^\gamma	-\frac{1}{2}h^\R_{\beta\gamma;\delta}u^\alpha u^\beta u^\gamma u^\delta\nonumber\\
		&\quad	-\frac{1}{2}h^\R_{\mu\nu}h^\R_{\beta\gamma;\delta}u^\alpha u^\beta u^\gamma u^\delta u^\mu u^\nu +\O[(h^\R)^3]. \label{expanded_geodesic}
\end{align}
Here already is the spirit of the self-consistent expansion: by expanding in powers of $h^\R_{\mu\nu}$ rather than powers of $\e$, I avoid expanding $z^\mu(\tau,\e)$ itself. However, Eq.~\eqref{expanded_geodesic} is complicated by the fact that the acceleration $\frac{D^2z^\alpha}{d\tau^2}$ appears in a nontrivial way on the right-hand side. To disentangle it, I now explicitly substitute the self-consistent expansion~\eqref{SC_force_expansion}.\footnote{One could instead assume that $\frac{D^2z^\alpha}{d\tau^2}$ possesses an expansion in powers of $h^\R_{\mu\nu}$.} After simple rearrangements, one finds
\begin{align}
F^\alpha_0 &=0, \\
F^\alpha_1 &= -\frac{1}{2}P^{\alpha\delta}\!\left(2h^{\R1}_{\delta(\beta;\gamma)}-h^{\R1}_{\beta\gamma;\delta}\right)\!u^\beta u^\gamma,\label{F1}\\
F^\alpha_2 &= -\frac{1}{2}P^{\alpha\delta}\!\left(2h^{\R2}_{\delta(\beta;\gamma)}-h^{\R2}_{\beta\gamma;\delta}\right)\!u^\beta u^\gamma\nonumber\\
&\quad+\frac{1}{2}P^{\alpha\mu}h^{\R1}_\mu{}^\delta\!\left(2h^{\R1}_{\delta(\beta;\gamma)}-h^{\R1}_{\beta\gamma;\delta}\right)\!u^\beta u^\gamma,\label{F2}
\end{align}
and hence, 
\beq\label{2nd-geo-v1}
\frac{D^2 z^\mu}{d\tau^2} = \e F_1^\mu(\tau;z)+\e^2 F_2^\mu(\tau;z)+\O(\e^3),
\eeq
where $P^{\alpha\mu}:= g^{\alpha\mu}+u^\alpha u^\mu$ projects orthogonally to $u^\mu$. This result may also be written in the more compact form
\begin{align}
\frac{D^2z^\alpha}{d\tau^2} &=  -\frac{1}{2}P^{\alpha\mu}
		\left(g_\mu{}^\delta-h^{\R}_\mu{}^\delta\right)\!\left(2h^\R_{\delta\beta;\gamma}-h^\R_{\beta\gamma;\delta}\right)\!u^\beta u^\gamma \nonumber\\
		&\quad+\O(\e^3).\label{2nd-geo}
\end{align}

Equation~\eqref{2nd-geo-v1} is the self-consistent approximation to the geodesic equation. It is defined by expanding Eq.~\eqref{exact_geodesic} in powers of $\e$ \emph{while holding the solution $z^\mu(\tau,\e)$ to that equation fixed}.

\subsubsection{Gralla-Wald expansion}\label{geodesic_expansion_in_h_and_dz}
In the previous section, I expanded the geodesic equation while holding $z^\mu(\tau,\e)$ fixed. I now expand $z^\mu(\tau,\e)$ as well. This procedure yields a sequence of equations for the vectors $\GW{z}_n^\mu$ that measure the deviation away from the zeroth-order worldline in the Gralla-Wald approximation. 

In previous work~\cite{Gralla-Wald:08,Gralla:12,Pound:14c}, this expansion has always been performed after choosing some set of coordinates and some parameter $s$ (which may differ from $\tau$) on the worldline. Here I wish to develop a more general and more geometrical picture. First, let me set the stage. Consider the two smooth metrics $g_{\mu\nu}$ and $\tilde g_{\mu\nu}$ on a smooth manifold $\man$.\footnote{In Sec.~\ref{gauge} I will put the perturbed and background metric on different manifolds, but here it is simplest to consider a single manifold.} Now consider a family of worldlines $\gamma_\e:s\in\mathbb{R}\mapsto\gamma_\e(s)\subset\man$ with parameter $s$ and coordinates $z_\e^\mu(s)=x^\mu(\gamma_\e(s))$. Each family member satisfies the self-consistent equation of motion~\eqref{2nd-geo}, which I now write as
\begin{equation}\label{family_acceleration}
a^\mu_\e=F^\mu_\e(\tau_\e),
\end{equation}
where $\tau_\e=\tau_\e(s)$ is proper time on $\gamma_\e$, $a^\mu_\e:=\frac{D^2 z^\mu_\e}{d\tau_\e^2}$ is the acceleration of $\gamma_\e$, and $F^\mu_\e(\tau_\e)$ is given by the right-hand side of Eq.~\eqref{2nd-geo}. The family generates a two-dimensional timelike surface $\mathcal{S}\subset\man$ with coordinates $(s,\e)$, embedded in $\man$ according to $x^\mu(s,\e)=z_\e^\mu(s)$, which I will also write simply as $z^\mu(s,\e)$.

Now if we consider an expansion in powers of $\e$, we immediately run up against two apparent ambiguities. We can write the expansion as 
\begin{equation}\label{z_expansion_fixed-s}
z^\mu(s,\e) = z^\mu_0(s) +\e \GW{z}^\mu_1(s) + \e^2 \GW{z}_2^\mu(s)+ O(\e^3),
\end{equation}
where
\begin{equation}\label{zn_fixed-s}
\GW{z}^\mu_n(s) = \frac{1}{n!}\e^n\frac{\partial^n z^\mu}{\partial\e^n}(s,0).
\end{equation}
The first ambiguity in this expansion stems from the fact that we are expanding not a vector but a scalar field, equal to the $\mu$th coordinate field on the worldline. Because of this, the expansion depends on the choice of coordinates. How does this play out? The zeroth-order term in the expansion is $z_0^\mu(s)=z^\mu(s,0)$, which, while it is a set of coordinates along a curve $\gamma_0$, is invariant in the sense of uniquely identifying that curve. The first-order term, $\GW{z}_1^\mu(s)$, is a derivative along a curve (of increasing $\e$ and fixed $s$) in $\mathcal{S}$. This derivative is evaluated on $\gamma_0$, making it a vector on $\gamma_0$; again, this is a covariant quantity. But at second order and beyond, the deviations $\GW{z}^\mu_n$ lose their vectorial character: unlike the first derivative along a curve, second and higher derivatives are not vectors. The function $z^\mu(s,\e)$ describes a curve in a particular set of coordinates, and the corrections $\GW{z}^\mu_n$ depend on that choice of coordinates. 

The second ambiguity in the expansion stems from the fact that we are expanding at fixed $s$. Geometrically, we are taking the limit $\e\to0$ along curves of fixed $s$ in $\mathcal{S}$. Under a reparametrization $s\to s'(s,\e)$,  the terms $\GW{z}_{n>0}^\mu$ are altered, becoming derivatives along a different curve, a curve of increasing $\e$ along which $s'$, not $s$, is constant. For a general ``small'' reparametrization $s\to s'=s+\O(\e)$, this is a type of gauge freedom, similar to but distinct from the usual gauge freedom of perturbation theory: rather than a small transformation of the extrinsic coordinates $x^\mu$, it is a small transformation of the intrinsic coordinates $(s,\e)$ on $\mathcal{S}$. The effect of reparametrization is displayed explicitly in Appendix~\ref{reparametrization}. 

To avoid a mire of coordinate- and parametrization-dependent results, I will isolate the coordinate- and parametrization-independent content of each $\GW{z}^\mu_n$. These are the quantities I will derive evolution equations for, and as will become apparent in Sec.~\ref{field_expansions}, they are the only quantities necessary for a practical Gralla-Wald approximation.

We can get at these desirable quantities through more obviously geometrical ones. First note that the surface $\mathcal{S}$ can be generated by the two vector fields $\dot z^\mu(s,\e)=\frac{\partial z^\mu}{\partial s}$ and $v^\mu(s,\e):=\frac{\partial z^\mu}{\partial\e}$. The field $v^\mu$ describes the deviation between neighbouring curves $\gamma_\e$ and $\gamma_{\e+d\e}$, and the first-order term $\GW{z}^\mu_1$ is simply its restriction to $\gamma_0$, 
\beq
\GW{z}^\mu_1 = v^\mu\big|_{\gamma_0}.
\eeq
The direction of this quantity plainly depends on the choice of parameter $s$. As shown in Appendix~\ref{reparametrization}, with a reparametrization we can freely adjust the piece of $\GW{z}^\mu_1$ that lies parallel to $\gamma_0$, but we cannot alter the piece perpendicular to $\gamma_0$. Hence, a covariant and parametrization-invariant measure of deviation is
\beq
\GW{z}^\mu_{1\perp} := P^\mu{}_{\nu} v^\nu\big|_{\gamma_0}.
\eeq
No matter the coordinate system and parametrization in which the expansion~\eqref{z_expansion_fixed-s} is performed, the coordinate- and parametrization-invariant piece of $\GW{z}^\mu_1$ can be picked out using $\GW{z}^\mu_{1\perp}=P^\mu_{0\nu} \GW{z}^\nu_1$, where $P_0^{\alpha\mu}:= g^{\alpha\mu}+u_0^\alpha u_0^\mu$.

For the second-order term, we can construct a ``deviation of the deviation'' from a derivative of $v^\mu$. First consider the vector 
\beq
w^\alpha:=\frac{1}{2}\frac{Dv^\alpha}{d\e} = \frac{1}{2}v^\beta\nabla_\beta v^\alpha
\eeq
and its restriction to $\gamma_0$, 
\beq\label{z2N}
\GW{z}^\alpha_{2{\rm N}}:= w^\alpha|_{\gamma_0}.
\eeq
The subscript ``N'' stands for ``normal'': $\GW{z}^\alpha_{2{\rm N}}$ is the second-order term in the expansion~\eqref{z_expansion_fixed-s} when that expansion is performed in locally inertial (i.e., normal) coordinates centered on $\gamma_0$. In those coordinates, $\frac{1}{2}v^\beta\nabla_\beta v^\alpha|_{\gamma_0}= \frac{1}{2}v^\beta\partial_\beta v^\alpha|_{\gamma_0}= \frac{1}{2}\frac{\partial^2 z^\mu}{\partial\e^2n}(s,0)=\GW{z}^\alpha_{2}(s)$. Unfortunately, as shown in Appendix~\ref{reparametrization}, not only $\GW{z}^\alpha_{2{\rm N}}$ but also $\GW{z}^\mu_{2{\rm N}\perp}=P_0^\mu{}_\nu\GW{z}^\nu_{2{\rm N}}$ depends on one's choice of parametrization. 

Fortunately, we can easily construct a related quantity that does not depend on parametrization. This is done by projecting out all information about the flow parallel to $\gamma$. Replacing $v^\mu$ with its orthogonal projection $v^\mu_\perp:=P^\mu{}_\nu v^\nu$, and then taking the orthogonal piece of the result, we obtain a quantity
\beq
\GW{z}^\alpha_{2\ddagger}:=  \frac{1}{2} P^\alpha{}_\gamma v_\perp^\beta\nabla_\beta v_\perp^\gamma|_{\gamma_0}
\eeq
that Appendix~\ref{reparametrization} shows is paramatrization-independent. $\GW{z}^\alpha_{2\ddagger}$ is simply related to $\GW{z}^\alpha_{2{\rm N}\perp}$ according to
\beq\label{z2cross-vs-z2perp}
\GW{z}^\alpha_{2\ddagger} = \GW{z}^\alpha_{2{\rm N}\perp}+\GW{z}_{1\parallel}\GW{u}^\beta_{1\perp},
\eeq
where $\GW{z}_{1\parallel}:=\GW{z}_{1}^\mu u_{0\mu}$, $\GW{u}_1^\mu:= \frac{D\GW{z}_1^\mu}{d\tau_0}$, and $\GW{u}_{1\perp}^\mu:=P^\mu_0{}_\nu\GW{u}_{1}^\nu=\frac{D\GW{z}_{1\perp}^\mu}{d\tau_0}$. If we construct $\GW{z}^\mu_n$ according to Eq.~\eqref{z_expansion_fixed-s} in some coordinate system and with some choice of parameter $s$, then we can extract the covariant and parametrization-invariant content of $\GW{z}^\mu_2$ using the relation
\beq\label{z2cross-vs-z2}
\GW{z}^\alpha_{2\ddagger} = P^\alpha_0{}_\beta\left(\GW{z}^\beta_2+\frac{1}{2}\Gamma^\beta_{\mu\nu}\GW{z}^\mu_1\GW{z}^\nu_1
														+\GW{z}_{1\parallel}\GW{u}^\beta_1\right),
\eeq
which follows from $\GW{z}^\alpha_{2{\rm N}} = \frac{1}{2}\frac{D^2z^\alpha}{d\e^2}\big|_{\gamma_0}= \GW{z}^\alpha_2+\frac{1}{2}\Gamma^\alpha_{\mu\nu}\GW{z}_1^\mu\GW{z}_1^\nu$ and Eq.~\eqref{z2cross-vs-z2perp}.

$\GW{z}^\alpha_{1\perp}$ and $\GW{z}^\alpha_{2\ddagger}$ will be my first- and second-order measures of the deviation of the accelerated worldline from the zeroth-order geodesic. They are the quantities I seek evolution equations for. In Sec.~\ref{field_expansions}, I will show that they suffice to obtain the Gralla-Wald approximation~\eqref{GW_expansion} in any coordinate system, despite the coordinate dependence of Eq.~\eqref{z_expansion_fixed-s}. 

Evolution equations for $\GW{z}^\alpha_{1\perp}$ and $\GW{z}^\alpha_{2\ddagger}$ can be derived in several ways, but here I wish to use a technique that will also apply directly to the discussion of gauge transformations in later sections. And as will become clear in those later sections, that means performing an expansion along a flow generated by a vector field. Let $\varphi_\e$ be the diffeomorphism describing the flow generated by $v^\mu$, and let $A$ be a smooth tensor field on $\mathcal{S}$ (suppressing indices on $A$ for compactness). The expansion of $A$ along a flow line beginning at $z^\mu_0(s)$ reads 
\begin{subequations}
\begin{align}
(\varphi^*_\e A)(z_0(s)) &= (e^{\e \Lie_v}A)(z_0(s)) \\
						&=  A(z_0(s)) +\e\Lie_vA(z_0(s))\nonumber\\
						&\quad+\frac{1}{2}\e^2\Lie_v^2A(z_0(s))+\O(\e^3),
\end{align}
\end{subequations}
where $\varphi^*_\e$ is the pullback.

We want to apply this expansion to the equation~\eqref{family_acceleration}, and from sequential orders obtain the evolution equations for $z^\mu_0$, $\GW{z}^\mu_{1\perp}$, and $\GW{z}^\mu_{2\ddagger}$. For convenience in performing the expansion, I choose the parameter $s$ such that it reduces to $\tau_0$ on $\gamma_0$. This does not imply any loss of generality, since it does not restrict the direction of $v^\mu$; it is analogous to choosing a particular set of background coordinates in perturbation theory, which does not restrict the choice of gauge.

Now, the zeroth-order term in the expansion is simply $a^\mu(s,0)=F^\mu(s,0)$. Since $F^\mu(s,0)=0$, this gives us the geodesic equation in the background metric,
\beq\label{z0}
\frac{D^2z_0^\mu}{d\tau_0^2}=0. 
\eeq 

The first-order term is 
\beq\label{Liea=LieF}
\Lie_v a^\mu(s,0)=\Lie_v F^\mu(s,0). 
\eeq
To evaluate the left-hand side, I write the acceleration explicitly as
\beq\label{a(s)}
a^\mu(s,\e) := \frac{D^2z^\mu}{d\tau^2}=\left(\!\frac{ds}{d\tau}\!\right)^{\!\!2}\left[\ddot z^\mu-\kappa\dot z^\mu\right],
\eeq
where $\dot z^\mu:=\frac{dz^\mu}{ds}$, $\ddot z^\mu:=\frac{D\dot z^\mu}{ds}$, and $\kappa(s):=\frac{d}{ds}\ln\sqrt{-g_{\mu\nu}\dot z^\mu \dot z^\nu}$. I then apply Eqs.~\eqref{Liedsdtau}, \eqref{Liekappa}, \eqref{vDzddot}, and the Ricci identity. The result is
\begin{subequations}
\begin{align}
\Lie_v a^\mu\big|_{\gamma_0} &=\left(v^\nu a^\mu{}_{;\nu}-a^\nu v^\mu{}_{;\nu}\right)\big|_{\gamma_0}\\
					&= \frac{D^2\GW{z}^\mu_{1\perp}}{d\tau_0^2}+R^\mu{}_{\alpha\nu\beta}u_0^\alpha \GW{z}^\nu_{1\perp}u_0^\beta.
\end{align}
\end{subequations}

To evaluate the right-hand side of Eq.~\eqref{Liea=LieF}, I first substitute the expansion~\eqref{GW_expansion} into the right-hand side of Eq.~\eqref{2nd-geo} to get the force in the form
\beq
F^\mu(s,\e;\gamma_\e) = \e f_1^\mu(s,\e) + \e^2 f_2^\mu(s,\e) +\O(\e^3), 
\eeq
where $f_1^\mu$ is given by Eq.~\eqref{F1} with  the replacement $h^{\R1}_{\mu\nu}\to\GW{h}^{\R1}_{\mu\nu}$, and $f_2^\mu$ by Eq.~\eqref{F2} with $h^{\R n}_{\mu\nu}\to\GW{h}^{\R n}_{\mu\nu}$. The forces $f^\mu_n$ are ordinary tensors on $\gamma_\e$, whereas the forces $F^\mu_n$ are tensor-valued functionals. I leave it to the next section to show how the expansion~\eqref{GW_expansion} is explicitly obtained from the coefficients $h^{\R n}_{\mu\nu}(x;z_\e)$ of the self-consistent expansion. So now noting that $\Lie_v\e=1$, we find
\begin{subequations}
\begin{align}
\Lie_v F^\mu\big|_{\gamma_0} &= -\frac{1}{2}P_0^{\mu\delta}\!\left(2\GW{h}^{\R1}_{\delta\beta;\gamma}-\GW{h}^{\R1}_{\beta\gamma;\delta}\right)\!u_0^\beta u_0^\gamma\\
&=:\GW{F}^\mu_1.\label{F1_generic}
\end{align}
\end{subequations}
Putting these results together, we get
\begin{equation}\label{z1_generic}
\frac{D^2\GW{z}^\alpha_{1\perp}}{d\tau_0^2} = \GW{F}_1^\alpha - R^\alpha{}_{\mu\beta\nu}u_0^\mu \GW{z}^\beta_{1\perp} u_0^\nu.
\end{equation}
Note that this result is independent of the choice of parameter $s$. For any choice, the expansion of the equation of motion along the flow of $v^\mu$ \emph{only} determines the evolution of $\GW{z}^\alpha_{1\perp}$; the piece of $\GW{z}^\alpha_{1}$ parallel to $u^\mu_0$, which can be arbitrarily altered by a reparametrization, is not determined by the equation of motion.

The second-order term in the equation of motion~\eqref{family_acceleration} is 
\beq\label{LieLiea=LieLieF}
\frac{1}{2}\Lie^2_v a^\mu(s,0)=\frac{1}{2}\Lie^2_v F^\mu(s,0). 
\eeq
To evaluate and simplify the left-hand side, I again begin with Eq.~\eqref{a(s)} and then repeatedly apply Eqs.~\eqref{Liedsdtau}, \eqref{Liekappa}, \eqref{vDzddot}, and the Ricci identity. The result is 
\begin{align}\label{LieLiea}
\frac{1}{2}\Lie^2_v a^\alpha\big|_{\gamma_0} &= \frac{D^2\GW{z}^\alpha_{2\rm N\perp}}{d\tau_0^2}
			+P^\alpha_0{}_{\!\rho} R^\rho{}_{\mu\beta\nu}\left(\!u_0^\mu \GW{z}_{2{\rm N}}^\beta u_0^\nu
			+2\GW{u}_1^\mu \GW{z}_1^\beta u_0^\nu\right)\nonumber\\
			&\quad -2P_0^\alpha{}_{\!\rho} R^\rho{}_{\mu\beta\nu;\gamma}\GW{z}_1^{(\mu} u_0^{\beta)} \GW{z}_1^{[\nu} u_0^{\gamma]} 
			+u_0^\alpha \GW{a}_{1\parallel}\GW{u}_{1\parallel}\nonumber\\
			&\quad + \GW{u}^\alpha_1\GW{a}_{1\parallel}  + u_0^{\alpha}\GW{F}_1^{\mu} \GW{u}_{1\mu}+2\GW{F}_1^{\alpha} \GW{u}_{1\parallel}
			\nonumber\\
			&\quad-\GW{F}_1^\beta v^\alpha{}_{;\beta}\big|_{\gamma_0},
\end{align}
where $\GW{u}_{1\parallel}:=u_{0\mu}\frac{D\GW{z}^\mu_1}{d\tau_0}$ and $\GW{a}_{1\parallel}:=u_{0\mu}\frac{D^2\GW{z}^\mu_1}{d\tau_0^2}$. On the left-hand side of Eq.~\eqref{LieLiea=LieLieF} we have
\begin{subequations}\label{LieLieF}
\begin{align}
\frac{1}{2}\Lie^2_v F^\mu\big|_{\gamma_0} &= f_2^\mu(s,0) +  \Lie_v f_1^\mu(s,0)\\
																&= f_2^\mu(s,0) + \GW{z}_1^\alpha f_{1;\alpha}^\mu(s,0) \nonumber\\
																&\quad - \GW{F}_1^{\alpha} v^{\mu}{}_{;\alpha}\big|_{\gamma_0}.
\end{align}
\end{subequations}
Combining Eqs.~\eqref{LieLiea} and \eqref{LieLieF} in Eq.~\eqref{LieLiea=LieLieF}, and rewriting the result in terms of the variables $\GW{z}^\mu_{1\perp}$ and $\GW{z}^\mu_{2\ddagger}$, we arrive at
\begin{align}\label{z2_generic}
\frac{D^2\GW{z}_{2\ddagger}^\alpha}{d\tau_0^2} &= \GW{F}^\alpha_2  - P_0^\alpha{}_{\!\rho} R^\rho{}_{\mu\beta\nu}\left(u_0^\mu \GW{z}_{2\ddagger}^\beta u_0^\nu+2\GW{u}_{1\perp}^\mu \GW{z}_{1\perp}^\beta u_0^\nu\right)\nonumber\\
			&\quad				+2P_0^\alpha{}_{\!\rho} R^\rho{}_{\mu\beta\nu;\gamma}\GW{z}_{1\perp}^{(\mu} u_0^{\beta)} \GW{z}_{1\perp}^{[\nu} u_0^{\gamma]}.
\end{align}
The force $\GW{F}_2^\mu$ appearing in this equation of motion is given by 
\begin{subequations}
\begin{align}
\GW{F}_2^\mu &:= f_2^\mu(s,0)  + \GW{z}_{1\perp}^\alpha f_{1;\alpha}^\mu(s,0)-u^\mu_0\GW{F}_{1\nu}\GW{u}^\nu_{1\perp}\\
&= -\frac{1}{2}P_0^{\mu\rho}\left(2\GW{h}^{\R2}_{\rho\sigma;\lambda}-\GW{h}^{\R2}_{\sigma\lambda;\rho}\right)u^\sigma_0 u^\lambda_0 \nonumber\\
		&\quad-\frac{1}{2}P_0^{\mu\rho}\left(2\GW{h}^{\R1}_{\rho\sigma;\lambda\delta}-\GW{h}^{\R1}_{\sigma\lambda;\rho\delta}\right)u^\sigma_0 u^\lambda_0\GW{z}_{1\perp}^\delta\nonumber\\
		&\quad 	 -\left(2\GW{h}^{\R1}_{\nu\sigma;\lambda}-\GW{h}^{\R1}_{\sigma\lambda;\nu}\right)\!
		\left(\tfrac{1}{2}\GW{u}^{\mu}_{1\perp} u^{\nu}_0u^\sigma_0u_0^\lambda+P_0^{\mu\nu}\GW{u}^{(\sigma}_{1\perp} u^{\lambda)}_0\right)\nonumber\\
		&\quad +\frac{1}{2}P_0^{\mu\nu}\GW{h}^{\R1}_\nu{}^\rho\left(2\GW{h}^{\R1}_{\rho\sigma;\lambda}-\GW{h}^{\R1}_{\sigma\lambda;\rho}\right)u^\sigma_0 u^\lambda_0.
		\label{F2_generic}
\end{align}
\end{subequations}

The sequence of equations~\eqref{z0}, \eqref{z1_generic}, and \eqref{z2_generic}, with Eqs.~\eqref{F1_generic} and \eqref{F2_generic}, are the Gralla-Wald expansion of the geodesic equation~\eqref{2nd-geo}. They are covariant and reparametrization-invariant evolution equations for the deviation of the accelerated worldline $\gamma_\e$ from the geodesic $\gamma_0$. As such, they represent a refinement of the coordinate- and parametrization-specific expansion in Ref.~\cite{Gralla:12} and of the covariant but parametrization-dependent one in~\cite{Pound:15a}. Other covariant expansion methods have been used for somewhat similar purposes~\cite{Vines:14}, but the analysis here has the advantage of clearly excising the parametrization dependence of the Gralla-Wald expansion. More importantly for the purposes of this paper, its use of Lie derivatives exactly parallels their use in gauge transformations, which in later sections will help to clarify an essential notion: the gauge-dependence of the worldline.

\subsection{Expansions of the metric and field equations}\label{field_expansions}
The preceding section illuminated the relationship between the equations of motion in our two formalisms. But of course, these equations of motion are largely meaningless in themselves; to fully understand the relationship between the formalisms, one must know the relationship between their metric perturbations $h^n_{\mu\nu}$ and $\GW{h}^n_{\mu\nu}$. In this section, I show explicitly how the metric perturbations (and their field equations) in the Gralla-Wald  approximation can be obtained from the self-consistent approximation by expanding the worldline.

\subsubsection{Expansion of functionals of $z^\mu$}
We are ultimately interested in expanding the metric perturbations and stress-energy tensor. But first consider the expansion of a generic tensor-valued functional $A(x;z)$ of arbitrary rank $(p,q)$ (suppressing indices for generality). The functionals of interest to us may be written in the form\footnote{It is not clear whether the perturbations $h^{n\geq2}_{\mu\nu}$ can be written in such a simple integral form, but they can be written in a more complicated, implicit integral form~\cite{Pound:10a}, which also allows an explicit (if more laborious) expansion of the worldline.}
\beq
A(x;z) = \int_\gamma B(x,z(s))\sqrt{-g_{\mu'\nu'}\dot z^{\mu'}\dot z^{\nu'}}ds,
\eeq
where $B(x,z)$ is a bitensor that behaves as a scalar at $z^\mu(s)$ and as a tensor of rank $(p,q)$ at $x^\mu$, and primed indices indicate evaluation at $x^{\mu'}=z^\mu(s)$. (See Ref.~\cite{Poisson-Pound-Vega:11} for a pedagogical introduction to bitensors.) Concrete examples of such a quantity are $T_1^{\mu\nu}(x;z)$ and $h^1_{\mu\nu}(x;z)$, shown below in Eqs.~\eqref{T1_param-invar} and \eqref{h1_z}. We can also write the functional as $A(x;z) = \int_\gamma \tilde B(x,z(s))ds$, where $\tilde B(x,z):=B(x,z(s))\sqrt{-g_{\mu'\nu'}\dot z^{\mu'}\dot z^{\nu'}}$ is a tensor at $x^\mu$ and a scalar density with respect to reparametrizations of $\gamma$. 

To facilitate the expansion of $A$, I introduce a Lie derivative $\varLie$ that acts on the functional dependence on $z^\mu$:
\begin{subequations}\label{varLie2}
\begin{align}
\varLie_\xi A(x;z) &:= \frac{d}{d\lambda}A(x;z+\lambda\xi)\big|_{\lambda=0}\\
 &=  \int_\gamma \Lie_{\xi'} \tilde B(x,z(s))ds.
\end{align}
\end{subequations}
This Lie derivative is closely related to the ordinary one that acts on the dependence on $x^\mu$. $\Lie$ drags fields along a flow of the field point $x^\mu$ relative to the worldline $z^\mu$; $\varLie$ drags fields along a flow of the worldline $z^\mu$ relative to the field point $x^\mu$. 

An expansion of $A(x;z(s,\e))$ in the limit $\e\to0$ is an expansion along a flow of increasing $\e$ in $\mathcal{S}$. We can write this in terms of $\varLie$ as 
\begin{align}\label{A_expansion}
A(x;z_\e) &= \sum_{n\geq0} \e^n\vardelta^n\! A(x;z_0), 
\end{align}
where
\beq\label{dnA}
\vardelta^n\! A(x;z_0) := \frac{1}{n!}\varLie^n_v A(x;z_0).
\eeq
With simple manipulations, one may reexpress this in terms of $z^\mu_1$ and $z^\mu_{2\rm N}$ to find
\begin{align}
\vardelta A(x;z_0) &= \int_{\gamma_0}\GW{z}^{\mu'}_1 \nabla_{\mu'}\tilde B(x,z_0)ds,\label{dA}\\
\vardelta^2\! A(x;z_0) &= \int_{\gamma_0}\big[\GW{z}_{2\rm N}^{\mu'}\nabla_{\mu'}\tilde B(x,z_0) \nonumber\\
												&\quad + \tfrac{1}{2}\GW{z}^{\mu'}_1 \GW{z}^{\nu'}_1\nabla_{\mu'}\nabla_{\nu'}\tilde B(x,z_0)\big]ds, \label{d2A}
\end{align}
where primed indices now refer to the point $z^\mu_0(s)$. 

We can also go one step further and express $\vardelta A(x;z_0)$ and $\vardelta^2\! A(x;z_0)$ in terms of the reparametrization-invariant quantities $z^\mu_{1\perp}$ and $z^\mu_{2\ddagger}$. After expressing $z^\mu_1$ and $z^\mu_{2\rm N}$ in terms of these quantities [using Eq.~\eqref{z2cross-vs-z2perp}] and writing $\tilde B(x,z)=B(x,z(s))\sqrt{-g_{\mu'\nu'}\dot z^{\mu'}\dot z^{\nu'}}$, one may eliminate all parametrization-dependent content by repeatedly applying Eq.~\eqref{commutation} and integrating by parts, utilizing the Ricci identity, and appealing to the equation of motion~\eqref{z1_generic}. The result is
\begin{align}
\vardelta A(x;z_0) &= \int_{\gamma_0}\GW{z}^{\mu'}_{1\perp}\nabla_{\mu'}B(x,z_0)d\tau_0,\label{dAinv}\\
\vardelta^2\! A(x;z_0) &= \int_{\gamma_0}\left[\GW{z}_{2\rm \ddagger}^{\mu'}\nabla_{\mu'}B(x,z_0)
												+\tfrac{1}{2}z_{1\perp}^{\mu'}\GW{F}_{1\mu'}B(x,z_0)\right. \nonumber\\
												&\quad +\left. \tfrac{1}{2}\GW{z}^{\mu'}_{1\perp} \GW{z}^{\nu'}_{1\perp}\nabla_{\mu'}\nabla_{\nu'}B(x,z_0)\right]d\tau_0. \label{d2Ainv}
\end{align}
Here I have expressed the final results in terms of the parameter $\tau_0$ on $\gamma_0$, but as discussed in the previous section, this choice of parameter on $\gamma_0$ represents no loss of generality. We see explicitly in these results that $\GW{z}^\mu_{1\perp}$ and $\GW{z}^\mu_{2\ddagger}$ are the only required measures of first- and second-order deviation.

\subsubsection{Stress-energy tensor}\label{T1_expansion}
We are interested in applying the above expansion to two quantities: the metric perturbation and the stress-energy tensor. Let us first consider the stress-energy. 

According to Eq.~\eqref{A_expansion}, the $n$th-order stress-energy can be expanded as
\begin{align}
\e^n T_n^{\mu\nu}(x;z) &= \e^n T_n^{\mu\nu}(x;z_0)+\e^{n+1} \vardelta T_n^{\mu\nu}(x;z_0,\GW{z}_1)\nonumber\\
											&\quad +\O(\e^{n+2}).
\end{align}
and so the first- and second-order stress-energies in the Gralla-Wald formalism are
\begin{subequations}\label{GWT}
\begin{align}
\GW{T}_1^{\mu\nu}(x;z_0) &= T_1^{\mu\nu}(x;z_0),\label{GWT1}\\
\GW{T}_2^{\mu\nu}(x;z_0,\delta m, \GW{z}_1) &= T_2^{\mu\nu}(x;z_0,\delta m) \nonumber\\
															&\quad +\vardelta T_1^{\mu\nu}(x;z_0,\GW{z}_1),\label{GWT2}
\end{align}
\end{subequations}
where $T^{\mu\nu}_1$ and $T^{\mu\nu}_2$ are given by Eq.~\eqref{TSC-1,2}.

The unknown term at this stage is $\vardelta T_1^{\mu\nu}(x;z_0,\GW{z}_1)$. To calculate it, I first write $T_1^{\mu\nu}(x;z)$ in the reparametrization-invariant form~\cite{Poisson-Pound-Vega:11}
\begin{align}
T^{\alpha\beta}_1(x;z) &= m\int_\gamma g^\alpha_{\alpha'}(x,z)g^\beta_{\beta'}(x,z) \dot z^{\alpha'}\dot z^{\beta'} \nonumber\\
								&\quad	\times\frac{\delta(x,z)ds}{\sqrt{-g_{\mu'\nu'}(z)\dot z^{\mu'}\dot z^{\nu'}}},\label{T1_param-invar}
\end{align}
where $g^\alpha_{\alpha'}(x,z)$ is a parallel propagator from the source point $x^{\mu'}=z^\mu(s,\e)$ to the field point $x^\mu$. One may then derive $\vardelta T_1^{\mu\nu}$ from Eq.~\eqref{dAinv}. Simplifying the result using Eq.~\eqref{commutation} and the distributional identities $\nabla_{\mu'}\delta(x,z)=-g^\mu_{\mu'}\nabla_\mu\delta(x,z)$ and $g^\alpha_{\alpha';\beta'}\delta(x,z)=0$~\cite{Poisson-Pound-Vega:11}, one finds
\begin{align}\label{dT_zperp}
\vardelta T^{\alpha\beta}_1 &= m\!\int_{\gamma_0}\! g^\alpha_{\alpha'}g^\beta_{\beta'}
					\Big[2u_0^{(\alpha'}\GW{u}_{1\perp}^{\beta')}\delta(x,z_0)\nonumber\\
					&\quad -u_0^{\alpha'}u_0^{\beta'}\GW{z}_{1\perp}^{\gamma'}g^\gamma_{\gamma'}\nabla_{\gamma}\delta(x,z_0)\Big]d\tau_0.
\end{align}

We can now plainly identify the content of $\vardelta T^{\alpha\beta}_1$. It contains two types of terms: a $\nabla_\mu\delta$ term and a $\delta$ term. The first simply represents the displacement of the point particle mass away from $\gamma_0$. The second is proportional to $mu_0^{(\alpha'}\GW{u}_{1\perp}^{\beta')}$; under the usual interpretation of components of the stress-energy tensor,  this represents the particle's density of linear momentum in the direction perpendicular to $\gamma_0$, arising from the fact that the particle is not just displaced from $\gamma_0$, but moving away from it. 

\subsubsection{Expansion of the first-order field}
Next I turn to the expansion of the metric perturbation. In analogy with Eq.~\eqref{GWT}, we have
\begin{subequations}
\begin{align}
\GW{h}^1_{\mu\nu}(x;z_0) &= h^1_{\mu\nu}(x;z_0),\\
\GW{h}^2_{\mu\nu}(x;z_0,\GW{z}_1) &= h^2_{\mu\nu}(x;z_0) + \vardelta h^1_{\mu\nu}(x;z_0,\GW{z}_1).
\end{align}
\end{subequations}
Likewise, the singular and regular fields are expanded as
\begin{subequations}
\begin{align}
\GW{h}^{\S/\R1}_{\mu\nu} &= h^{\S/\R1}_{\mu\nu}(x;z_0),\\
\GW{h}^{\S/\R2}_{\mu\nu} &= h^{\S/\R2}_{\mu\nu}(x;z_0) + \vardelta h^{\S/\R1}_{\mu\nu}(x;z_0,\GW{z}_1).
\end{align}
\end{subequations}

As in the case of the stress-energy, the unknown terms here are the $\vardelta$ terms. I express them concretely by assuming the existence of a Green's function for Eq.~\eqref{h1SC}. Given a Green's function satisfying \footnote{To make the index structure clear, I write $E_1^{\mu\nu}[G]$ as $E^{\mu\nu\rho\sigma} G_{\rho\sigma\mu'\nu'}$.}
\beq
E^{\mu\nu\rho\sigma} G_{\rho\sigma\mu'\nu'} = 8\pi \left(g^\mu_{(\mu'}g^{\nu}_{\nu')}-\tfrac{1}{2}g^{\mu\nu}g_{\mu'\nu'}\right)\delta(x,x')
\eeq
and some boundary conditions, the first-order self-consistent field satisfying the same boundary conditions is 
\begin{subequations}\label{h1_z}%
\begin{align}
h^1_{\mu\nu}(x;z) &= \int G_{\mu\nu\mu'\nu'}(x,x') T^{\mu'\nu'}_1(x';z) dV'\\
							&= m\int_\gamma G_{\mu\nu\mu'\nu'}\frac{\dot z^{\mu'}\dot z^{\nu'}}{\sqrt{-g_{\alpha'\beta'}\dot z^{\alpha'}\dot z^{\beta'}}}ds.
\end{align}
\end{subequations}
Applying Eq.~\eqref{dAinv} and simplifying using Eq.~\eqref{commutation}, we get
\begin{align}\label{dh_zperp}
\vardelta h^1_{\mu\nu}(x;z_0,\GW{z}_1) &= m\!\int_{\gamma_0}\!\! \Big(2 G_{\mu\nu\mu'\nu'}u_0^{(\mu'}u_{1\perp}^{\nu')}\nonumber\\
													&\quad + G_{\mu\nu\mu'\nu';\gamma'}u^{\mu'}_0u^{\nu'}_0\GW{z}_{1\perp}^{\gamma'}\Big)d\tau_0.
\end{align}

Our interpretation of the two terms in $\vardelta T^{\mu\nu}_1$ informs our interpretation of this result: the first term in $\vardelta h^1_{\mu\nu}$ is the contribution to the metric from the particle's momentum directed away from $\gamma_0$, while the second term is the contribution from the displacement of the particle away from $\gamma_0$. In the Lorenz gauge, one can use standard local expansions of the Green's function~\cite{Poisson-Pound-Vega:11} to establish this more directly. Although I omit the details here, the result of that calculation~\cite{Pound:15a} is that the $G_{\mu\nu\mu'\nu';\gamma'}$ term in Eq.~\eqref{dh_zperp} provides the explicit $\GW{z}_1^\mu$ term  in the local expression~\eqref{hs2GW-local}, while the $G_{\mu\nu\mu'\nu'}$ term provides the $\GW{z}_1^\mu$-modfication to $\GW{\delta m}_{\mu\nu}$ mentioned below that equation.

\subsubsection{Field equations}
Finally, we can now immediately obtain the field equations~\eqref{hGW}--\eqref{hSGW} of the Gralla-Wald approximation by substituting the expansions of $h^n_{\mu\nu}$ and $T_n^{\mu\nu}$ into the field equations~\eqref{h1SC}, \eqref{h2SC}, and \eqref{hRSC}--\eqref{hSSC} of the self-consistent approximation and regrouping the results according to powers of $\e$. 

The analysis in this section has accomplished two things. First, it has shown explicitly how to implement an expansion of the worldline in the equations of motion and field equations, and the relationship between the solutions. Second, it has shown how to do this in a way that is covariant and reparametrization invariant. Since the quantities $\vardelta T^{\mu\nu}_1$ and $\vardelta h^{\mu\nu}_1$ depend only on the perpendicular deviation $\GW{z}^\mu_{1\perp}$, there is no influence from the arbitrarily specified parallel deviation. 

In the explicit expansions I have shown, I have not used the second variation~\eqref{d2A} of any quantity; they would first become involved via terms $\vardelta^2 T^1_{\mu\nu}$ and $\vardelta^2h^1_{\mu\nu}$ in the third-order Gralla-Wald approximation. However, Eq.~\eqref{d2Ainv} shows that these quantities will again depend only on the covariant and reparametrization-invariant content of the second deviation. More generally, it should be the case that at any order, from the equation of motion for $z^\mu$, one can obtain equations of motion for covariant, parametrization-invariant deviation vectors on $z_0^\mu$, and one can then write the expansion~\eqref{A_expansion} in terms of those vectors.

Perhaps most importantly, as I mentioned in the introduction, these tools and results are not confined to relating the self-consistent to the Gralla-Wald approximation. They can be utilized in any scenario in which one wishes to expand the self-consistent worldline around some neighbouring (possibly non-geodesic) worldline.


\section{Gauge transformations in the self-consistent and Gralla-Wald approximations}\label{gauge}

In the preceding sections, I have elucidated the two approximations' treatments of perturbative motion and the relation between those treatments. I now describe the effects of smooth gauge transformations within these approximations. As we shall see, the differing treatments of the worldline entail differing treatments of gauge transformations.

Section~\ref{gauge_review} opens the discussion with a review of gauge freedom in perturbation theory,  following Refs.~\cite{Geroch:69, Stewart-Walker:74,Bruni-etal:96}. The material is standard, but it establishes my notation and reminds the reader of the basics, which are well known at linear order but less familiar at nonlinear orders. Sections~\ref{gauge_SC} and \ref{gauge_GW} then focus in turn on the self-consistent and Gralla-Wald approximations.

In Sec.~\ref{third-freedom} I discuss some general issues related to the freedom of choosing what to hold fixed in the limit $\e\to0$, which is distinct from the usual gauge freedom discussed in Secs.~\ref{gauge_review}--\ref{gauge_GW}.

\subsection{Review of gauge freedom in perturbation theory}\label{gauge_review}

\subsubsection{Active view}
In perturbation theory, we consider a family of metrics ${\sf g}_{\mu\nu}(x,\e)$, or simply ${\sf g}$ in the absence of a chart. (It will be convenient in this section to adopt index-free notation for tensors.) The family of metrics lives on a family of manifolds $\man_\e$, and a given choice of gauge refers to an identification map $\phi^X_\e:\man_0\to\man_\e$. The identification map induces a flow through the family down to the base manifold $\man_0$ where the background metric $g:={\sf g}_0$ lives.  

 Call the generator of this flow $X:=\frac{d\phi^X_\e}{d\e}$. We wish to approximate a tensor ${\sf A}$ at a point $q=\phi^X_\e(p)\in\man_\e$ in the limit of small $\e$. In ordinary (i.e., not singular~\cite{Kevorkian-Cole:96,Kates:81,Pound:10b}) perturbation theory, ${\sf A}$ is well approximated by a Taylor expansion around its value at $\e=0$, given by 
\beq
(\phi^X_\e{}^*{\sf A})(p) = (e^{\e\Lie_X}{\sf A})(p) = \sum_{n\geq0}\frac{\e^n}{n!} (\Lie^n_X{\sf A})(p),\label{X-expansion}
\eeq
where $p\in\man_0$. This expansion depends on the choice of gauge $X$, and we define the $n$th-order perturbation $A^X_n$ to be
\beq
A^X_n(p) := \frac{1}{n!} (\Lie^n_X{\sf A})(p).
\eeq

Now say we work in a different gauge. This corresponds to a different choice of identification map $\phi^Y_\e:\man_0\to\man_\e$ or flow generator $Y:=\frac{d\phi^Y_\e}{d\e}$. The approximation of the tensor ${\sf A}$ in terms of tensors at the point $p\in\man_0$ is now given by 
\beq
(\phi^Y_\e{}^*{\sf A})(p) = (e^{\e \Lie_Y}{\sf A})(p),\label{Y-expansion}
\eeq
and the $n$th-order perturbation is
\beq
A^Y_n(p) := \frac{1}{n!} (\Lie^n_Y{\sf A})(p).
\eeq

\begin{figure}[t]
\begin{center}
\includegraphics[width=0.85\columnwidth]{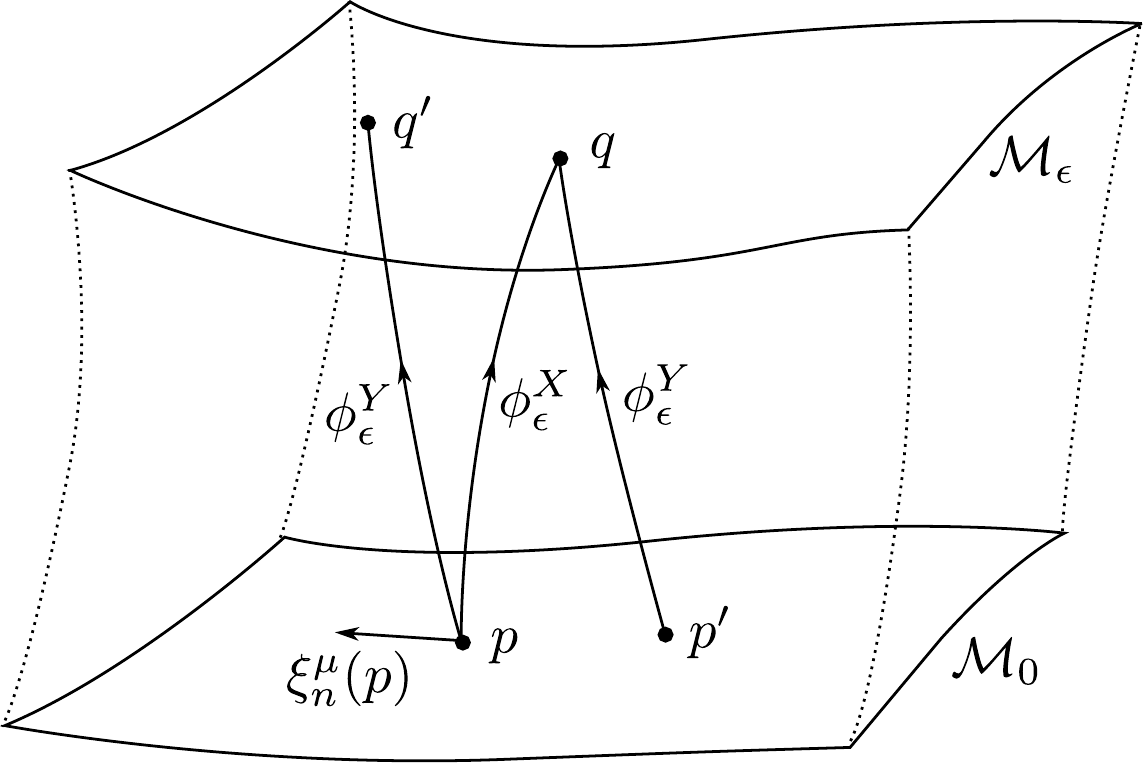}
\caption{\label{Fig:points}Relationship between points in the perturbed and background spacetimes in two different gauges. In the two gauges $X$ and $Y$, the point $p\in\man_0$ is identified with the two different points $q$ and $q'$ in $\man_\e$, respectively. Likewise, a given point $q$ in $\man_\e$ is identified with the two different points $p$ and $p'$ in $\man_0$. Very roughly speaking, the transformation generators $\xi_n^\mu$ point from $q$ to $q'$, and their opposites, $-\xi^\nu_n$, from $p$ to $p'$.}
\end{center}
\end{figure}

The two expansions~\eqref{X-expansion} and \eqref{Y-expansion} approximate the tensor's value at two different points in $\man_\e$, $q=\phi^X_\e(p)$ and $q'=\phi^Y_\e(p)$, but in both cases the terms $A^X_n$ and $A^Y_n$ are evaluated at the same point $p$ in $\man_0$; the situation is illustrated in Fig.~\ref{Fig:points}. We now ask how the quantities $A^X_n$ and $A^Y_n$ differ when evaluated at this point $p$. Their difference is
\beq
\Delta A_n(p) = \frac{1}{n!} (\Lie^n_Y{\sf A})(p)-\frac{1}{n!} (\Lie^n_X{\sf A})(p).
\eeq
The first- and second-order terms are easily expressed in the usual form 
\begin{subequations}\label{DeltaA}
\begin{align}
\Delta A_1 &= \Lie_{\xi_1} A_0,\label{DeltaA1}\\
\Delta A_2 &= \Lie_{\xi_2} A_0 + \frac{1}{2}\Lie^2_{\xi_1}A_0 + \Lie_{\xi_1} A_1,\label{DeltaA2}
\end{align}
\end{subequations}
where $\xi_1:= Y-X$ and $\xi_2:= \frac{1}{2}[X,Y]$ are the usual gauge vectors. 
Higher-order terms can be easily worked out. 

For compactness, in later sections I will adopt a more familiar notation, writing the gauge transformation as $A_n \to A_n'=A_n+\Delta A_n$, where the primed tensor refers to the $Y$ gauge and the unprimed to the $X$ gauge.

\subsubsection{Passive view}

The notion of gauge described in the previous section does not utilize coordinates. However, gauge transformations are sometimes more conveniently thought of as small coordinate transformations. 

Let us introduce a coordinate system $x^\mu$ on $\man_0$ and use the identification maps $\phi^X_\e$ and $\phi^Y_\e$ to carry the coordinates up to $\man_\e$. In gauge $X$, keep the name $x^\mu$ for the coordinates on $\man_\e$, such that the point $q=\phi^X_\e(p)$ has a coordinate label $x^\mu(q)=x^\mu(p)$. In gauge $Y$, use the name $x'^\mu$ for the coordinates on $\man_\e$, such that the point $q'=\phi^Y_\e$ has the coordinate label $x'^\mu(q')=x^\mu(p)$. The two choices of gauge then become expansions at fixed coordinate values $x^\mu$ and $x'^\mu$, respectively. So we may do away with the differing identification maps, adopt a fixed identification (say, $X$), and think of the gauge transformation as the coordinate transformation $x^\mu\to x'^\mu$. For small $\e$, this is a near-identity transformation. To figure out its small-$\e$ expansion, first consider the differing coordinate values at the points $q$ and $q'$ in the chart $x^\mu$. The chart consists of four scalar fields, to which we can apply the general expansion~\eqref{Y-expansion} to find
\begin{align}\label{qtoq'}
x^\mu(q') &= x^\mu(q) + \e \xi^\mu_1(x(q)) \nonumber\\&\quad+ \e^2\!\!\left[\xi^\mu_2(x(q))+\frac{1}{2}\xi^\nu_1(x(q))\partial_\nu\xi^\mu_2(x(q))\right]+\O(\e^3),
\end{align}
where I have expressed $Y$ derivatives as $\xi$ derivatives using $\Lie_X x^\mu=0$, and I have expressed the components on the right-hand side as functions of $x^\mu(q)$ using the fact that $x^\mu(p)=x^\mu(q)$. Now, rather than an active diffeomorphism on $\man_\e$, let us consider this as the passive change in chart $x^\mu(q)\mapsto x'^\mu(x(q))$. From the definitions above, we have $x'^\mu(q')=x^\mu(q)$. Rewriting Eq.~\eqref{qtoq'} as an equation for $x^\mu(q)$, we then arrive at
\begin{align}
x'^\mu(q') &= x^\mu(q')-\epsilon \xi^\mu_1(x(q'))\nonumber\\&\quad-\epsilon^2\!\! \left[\xi^\mu_2(x(q'))-\frac{1}{2}\xi^\nu_1(x(q'))\partial_\nu\xi_1^\mu(x(q'))\right]\nonumber\\
 &\quad +\O(\epsilon^3).\label{coord_transformation}
\end{align}

One may now use this transformation to obtain gauge transformation rules. For example, the components of ${\sf g}$ transform as
\begin{subequations}\label{component_transformation}%
\begin{align}
{\sf g}'_{\mu\nu}(x',\e) &= \frac{\partial x^\alpha}{\partial x'^\mu}\frac{\partial x^\beta}{\partial x'^\nu}{\sf g}_{\alpha\beta}(x(x'),\e)\\
								&= {\sf g}_{\mu\nu}(x',\e) +\Lie_\xi{\sf g}_{\mu\nu}(x',\e)+\frac{1}{2}\Lie^2_\xi{\sf g}_{\mu\nu}(x',\e) \nonumber\\&\quad+\O(\e^3),  
\end{align}
\end{subequations}
where $\xi^\mu=\e\xi_1^\mu+\e^2\xi^\mu_2+\O(\e^3)$. We see that in general, the tensor transformation is an expansion along the flow generated by $\xi^\mu$. Recall that we are now using a single identification map, say $\phi^X_\e$, and with that identification, component values of ${\sf g}$ are identical to component values of $(\phi_\e{}^*{\sf g})$. Also note that Eq.~\eqref{component_transformation} applies even if ${\sf g}_{\mu\nu}(x',\e)$ is not expanded for small $\e$. But if it is so expanded, then the passive transformation law~\eqref{component_transformation} agrees with Eq.~\eqref{DeltaA}.

\subsection{Gauge in the self-consistent approximation}\label{gauge_SC}

I now move to the self-consistent expansion. This is \emph{not} an ordinary expansion of the sort described by Eq.~\eqref{X-expansion}. To understand what sort it is, consider its form in a chart $x^\mu$. In a gauge in which no logarithms of $\e$ appear in the metric, it is plain to see from Eq.~\eqref{SC_expansion} that the self-consistent expansion has the form of a Taylor expansion at not only fixed coordinate values, but also at a fixed coordinate description of the worldline, $z^\mu=x^\mu(\gamma)$. Hence, if we imagine working with the expansion~\eqref{SC_expansion} prior to imposing the gauge condition, leaving $\gamma_\e$ freely specifiable, then when we take the limit $\e\to0$ and move down through our family of spacetimes, we shrink the object toward zero size and mass on the worldline $\gamma_\e^X:=(\phi^X_\e)^{-1}(\gamma)$ in $\man_0$. 

After we impose the gauge condition and fix $\gamma_\e$'s dependence on $\e$, the situation becomes more complex. In this case, the limit $\e\to0$ takes  $\gamma_\e$ to $\gamma_0$, and it may seem that we can no longer imagine the ``fixed $z^\mu$'' limit. However, this is slightly misleading. The essential aspect of the approximation is that for a given value of $\e$, the object's worldline, as identified by the location of the metric's $\sim 1/r^n$ behavior, \emph{is at the same coordinate position in $\man_0$ as in $\man_\e$}.\footnote{Strictly speaking, $\gamma_\e$ lies in $\widetilde\man_\e$, the manifold of the effective spacetime $(\tilde g_{\mu\nu},\widetilde\man_\e)$. This technicality is required because $\man_\e$ may not be be diffeomorphic to $\man_0$ in a region including $\gamma_0$.} What is happening here is that we map from $\man_\e$ to $\man_0$ holding $\gamma_\e$ fixed, shrinking the object to zero size on $\gamma_\e^X$ and obtaining the perturbations $h^n_{\mu\nu}(x;z)$ that live on $\man_0$. By enforcing the gauge condition, we then ensure that those fields on $\man_0$ provide an asymptotic solution to the Einstein equation.

\begin{figure}[t]
\begin{center}
\includegraphics[width=0.85\columnwidth]{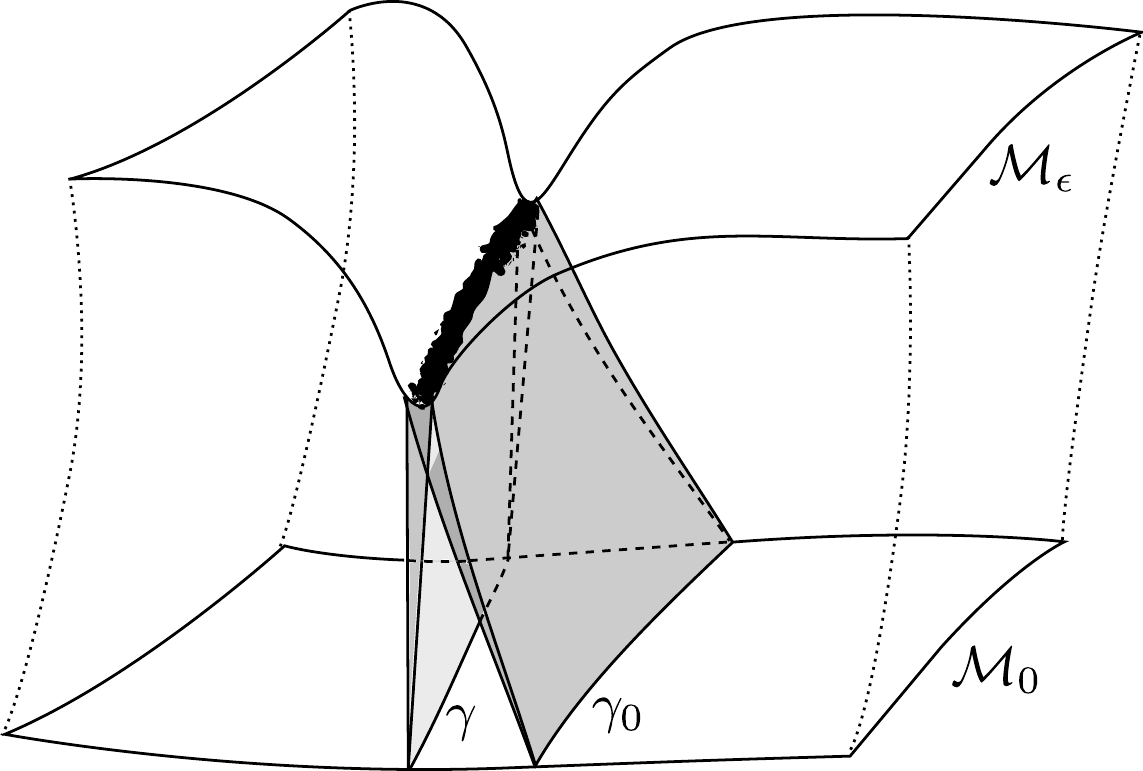}
\caption{\label{Fig:limits}Gralla-Wald and self-consistent limits, shown superimposed. A family of spacetimes is stacked vertically, with $\e$ running upwards. The top surface is the specific spacetime we seek to approximate, with a specific value of $\e$. In that spacetime, the small object is shown in black. In the limit toward $\e\to0$, this object shrinks toward zero size, and the strong curvature due to the object likewise decreases toward zero. The Gralla-Wald approximation shrinks the object down to a background geodesic $\gamma_0$. The self-consistent approximation shrinks the object down to a worldline $\gamma$ that is accelerated in the background.}
\end{center}
\end{figure}

This setup is illustrated in Fig.~\ref{Fig:limits}, and it remains true even in the case that logarithms appear.

\begin{figure}[t]
\begin{center}
\includegraphics[width=0.85\columnwidth]{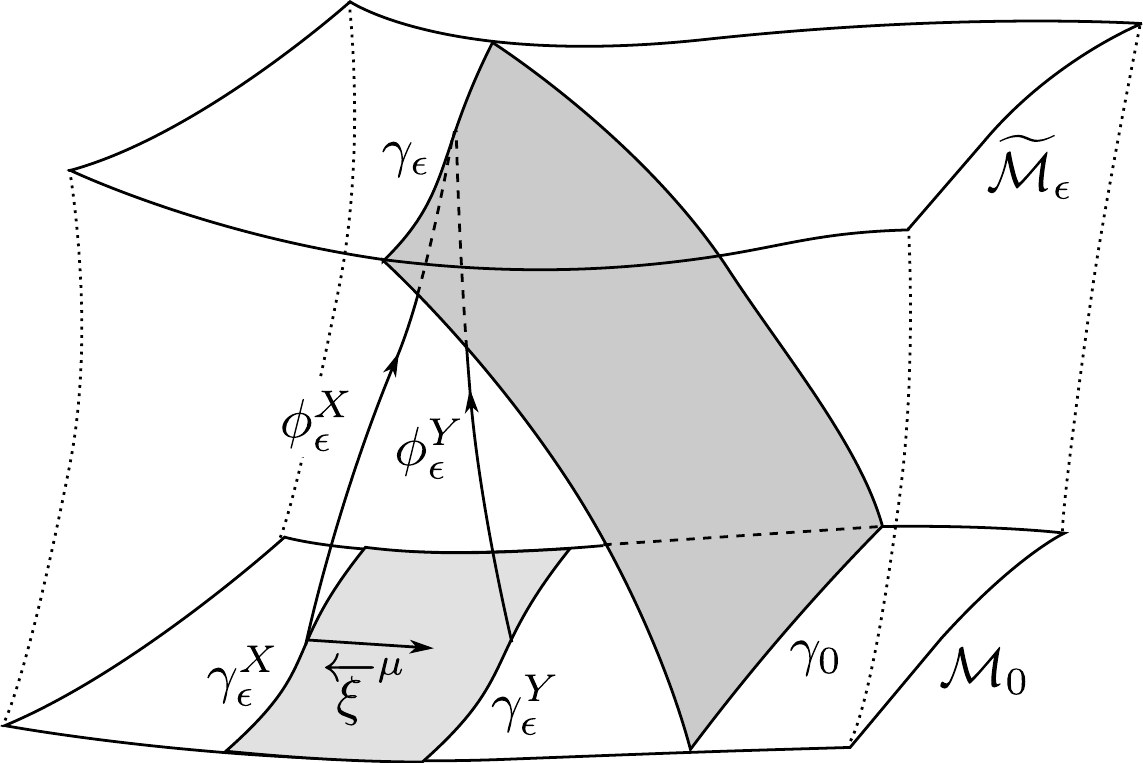}
\caption{\label{Fig:worldlines} Worldlines in different gauges in the self-consistent and Gralla-Wald pictures. The family of perturbed worldlines $\{\gamma_\e\}_\e$ generate a surface that runs down through the family of manifolds $\widetilde\man_\e$, terminating at the zeroth-order worldline $\gamma_0$; when this surface is mapped down to $\man_0$ using a particular identification map, it becomes the surface $\mathcal{S}$ of Sec.~\ref{geodesic_expansion_in_h_and_dz}. In gauge $X$, $\gamma_\e$ is identified with $\gamma_\e^X$ in $\man_0$; in gauge $Y$, with $\gamma_\e^Y$. The two gauge-related worldlines in $\man_0$ are connected by a surface generated by the vector field $\protect\overleftarrow{\xi}^\mu = -\e\xi_1^\mu-\e^2\xi_2^\mu+\O(\e^3)$. The self-consistent approximation works directly with $\gamma_\e^X$ or $\gamma_\e^Y$. The Gralla-Wald approximation uses expansions of those worldlines around $\gamma_0$.}
\end{center}
\end{figure}

Under a gauge transformation, we not only change the coordinates $x^\mu\to x'^\mu$, but also the coordinate description of the worldline, $z^\mu\to z'^\mu$, and that change must be accounted for in the transformation law for the perturbations $h^n_{\mu\nu}$. In coordinate-independent language, we say that in the two gauges we shrink the object toward zero size and mass on two different worldlines in $\man_0$: $\gamma_\e^X:=(\phi^X_\e)^{-1}(\gamma)$ in gauge $X$; $\gamma_\e^Y:=(\phi^Y_\e)^{-1}(\gamma)$ in gauge $Y$. Figure~\ref{Fig:worldlines} shows the relationship between these two worldlines.

In the course of this section, I use these ideas to work through transformation laws for the worldline, the metric perturbations, the singular and regular fields, and finally, tensors that live only on the worldline, such as the self-force.

\subsubsection{Transformation of the worldline}
The transformation of the worldline, $\gamma^X_\e\to\gamma^Y_\e$, is an immediate consequence of the coordinate transformation law~\eqref{coord_transformation}. Under a gauge transformation, the coordinates $z^\mu(s,\e)=x^\mu(\gamma_\e(s))$ become
\begin{align}
z'^{\mu}(s,\e) &= z^\mu(s,\e)-\epsilon \xi^\mu_1(z)\nonumber\\
&\quad-\epsilon^2\!\! \left[\xi^\mu_2(z)-\frac{1}{2}\xi^\nu_1(z)\partial_\nu\xi_1^\mu(z)\right]+O(\epsilon^3),\label{z_transformation}
\end{align}
where functions of $z^\mu$ are evaluated at $z^\mu(s)$. 

Since it is easy to get lost in the sign conventions, it is worth verifying we have the correct signs. Going back to Fig.~\ref{Fig:points} and Eq.~\eqref{qtoq'}, we see that the transformation from $q$ to $q'$ maps to the left in the figure, in the direction of $Y-X=\xi_1$. This is the direction of the transformation between the two points in $\man_\e$. Now look instead at how the point $q$ is mapped to two different points in $\man_0$: $p=(\phi_\e^X)^{-1}(q)$ and $p'=(\phi_\e^Y)^{-1}(q)$. These two points are related by a map to the right, in the direction of $X-Y=-\xi_1$. Comparing the appearance of $p$ and $p'$ in Fig.~\ref{Fig:points} to that of $\gamma^X_\e$ and $\gamma^Y_\e$ in Fig.~\ref{Fig:worldlines}, we see that we have the correct (minus) sign in front of $\xi^\mu_1$ in Eq.~\eqref{z_transformation}.

\subsubsection{Transformation of the metric perturbations}\label{h-transform-SC}
Now we may consider the transformation of the metric perturbations. As described above, the expansions in the two gauges are performed by writing ${\sf g}_{\mu\nu}$ in the coordinates $x^\mu$ or $x'^\mu$ as ${\sf g}_{\mu\nu}(x,\e;z)$ or ${\sf g}'_{\mu\nu}(x',\e;z')$ and then expanding for small $\e$ at fixed $x^\mu$ and $z^\mu$ or $x'^\mu$ and $z'^\mu$. In terms of $h_{\mu\nu}:= {\sf g}_{\mu\nu}-g_{\mu\nu}$ and $h'_{\mu\nu}:= {\sf g}'_{\mu\nu}-g_{\mu\nu}$, the two expansions read
\begin{align}
h_{\mu\nu}(x,\e;z) &= \sum \e^n {h}^{n}_{\mu\nu}(x;z),\label{h-exp}\\
h'_{\mu\nu}(x',\e;z') &= \sum \e^n {h'}^{n}_{\mu\nu}(x';z').\label{h'-exp} 
\end{align}
Transforming between the two coordinate systems, we find that ${\sf g}_{\mu\nu}(x,\e;z)$ and ${\sf g}'_{\mu\nu}(x',\e;z')$ are related by
\begin{subequations}
\begin{align}
{\sf g}'_{\mu\nu}(x',\e;z') &= \frac{\partial x^\alpha}{\partial x'^\mu}\frac{\partial x^\beta}{\partial x'^\nu}{\sf g}_{\alpha\beta}(x(x'),\e;z(z')),\\
			&= {\sf g}_{\mu\nu}(x',\e;z(z')) + \Lie_\xi {\sf g}_{\mu\nu}(x',\e;z(z')) \nonumber\\&\quad+ \frac{1}{2}\Lie^2_\xi g_{\mu\nu}(x',\e;z(z')) + \O(\e^3).
\end{align}
\end{subequations}
Now expanding ${\sf g}_{\mu\nu}(x',\e;z(z'))$, we get 
\begin{align}
h'_{\mu\nu}(x',\e) &=   \e\big[h^1_{\mu\nu}(x';z) + \Lie_{\xi_1} g_{\mu\nu}(x')\big]+ \e^2 \big[ h^2_{\mu\nu}(x';z) \nonumber\\
			&\quad + \Lie_{\xi_2} g_{\mu\nu}(x')  + \Lie_{\xi_1} h^1_{\mu\nu}(x';z) \nonumber\\
			&\quad  + \tfrac{1}{2}\Lie^2_{\xi_1} g_{\mu\nu}(x') \big] +\O(\e^3).\label{h'z}
\end{align}
In Eq.~\eqref{h'z} I have not yet expanded $z^\mu(z')$ around $z^\mu=z'^\mu$. After we perform that expansion, the perturbation reads 
\begin{align}
h'_{\mu\nu}(x',\e) &= \e\big[h^1_{\mu\nu}(x';z') + \Lie_{\xi_1} g_{\mu\nu}(x')\big] + \e^2\big[h^2_{\mu\nu}(x';z')  \nonumber\\
			&\quad + \Lie_{\xi_2} g_{\mu\nu}(x') + (\Lie_{\xi_1}+\varLie_{\xi_1}) h^1_{\mu\nu}(x';z')\nonumber\\
			&\quad + \tfrac{1}{2}\Lie^2_{\xi_1} g_{\mu\nu}(x') \big] +\O(\e^3).\label{h'z'}
\end{align}
where $\varLie_{\xi_1}h^{1}_{\mu\nu}(x';z')=\vardelta h^{1}_{\mu\nu}(x';z',\xi)$ uses the Lie derivative introduced in Sec.~\ref{field_expansions}. 

The two Lie derivatives in $(\Lie_{\xi_1}+\varLie_{\xi_1}) h^1_{\mu\nu}(x';z')$ have an important interpretation. $\Lie_{\xi_1}$ moves points relative to the worldline, and  $\varLie_{\xi_1}$ moves the worldline relative to the points; hence, their sum has no net effect on the position of points relative to the worldline. More concretely, the term $\Lie_{\xi_1}h^1_{\mu\nu}$ introduces a mass dipole moment into the metric, $\varLie_{\xi_1}h^1_{\mu\nu}$ introduces an equal but opposite moment, and in the end $h'_{\mu\nu}$ contains no mass dipole moment when written as a functional of the transformed worldline $z'^\mu$. The fact that the dipole moment vanishes relative to $z'^\mu$ can be seen explicitly by acting with $(\Lie_{\xi_1}+\varLie_{\xi_1})$ on the leading term in $h^1_{\mu\nu}$, which is $\frac{2m\delta_{\mu\nu}}{r}\sim \frac{2m\delta_{\mu\nu}}{|x^i-z^i|}$: in $\Lie_{\xi_1} \frac{2m\delta_{\mu\nu}}{r}$ the derivative moves $x^i$ relative to $z^i$, contributing a mass dipole term $\frac{-2m\xi^i_1n_i}{r^2}\delta_{\mu\nu}$ with moment $M^i=-m\xi^i_1$, while in $\varLie_{\xi_1} \frac{2m\delta_{\mu\nu}}{r}$ the derivative moves $z^i$, counteracting the effect of $\Lie_{\xi_1}$. Note that though their effects on the mass dipole moment cancel in this way, the two derivatives do not \emph{precisely} cancel one another, because $h^1_{\mu\nu}$ has a different tensorial character at $x$ than at $z$. We can infer the net action of the derivatives from the result~\eqref{LieT+varLieT}.

Equation~\eqref{h'z'} provides the transformation $h_{\mu\nu}\to h'_{\mu\nu}$ of the total perturbation. Comparing it to Eq.~\eqref{h'-exp}, we read off the transformation $h^n_{\mu\nu}\to h'^n_{\mu\nu}$ of the coefficients of $\e^n$ in the two expansions~\eqref{h-exp} and \eqref{h'-exp}:
\begin{subequations}\label{DhSC}%
\begin{align}
h'^1_{\mu\nu}(x';z') &=  h^1_{\mu\nu}(x';z') + \Lie_{\xi_1} g_{\mu\nu}(x') ,\label{h1'}\\
h'^2_{\mu\nu}(x';z') &=  h^2_{\mu\nu}(x';z') +  (\Lie_{\xi_1}+ \varLie_{\xi_1})h^{1}_{\mu\nu}(x';z') \nonumber\\
			&\quad+ \Lie_{\xi_2} g_{\mu\nu}(x') + \frac{1}{2}\Lie^2_{\xi_1} g_{\mu\nu}(x').\label{h2'}
\end{align}
\end{subequations}
Note what might be the most essential characteristic of these transformation laws: the metric perturbations in any given gauge always diverge on the center-of-mass worldline in that particular gauge. So $h^n_{\mu\nu}[z]$ diverges on $z^\mu$, but $h'^n_{\mu\nu}[z']$ diverges on $z'^\mu$.

With little additional effort, we could extend the analysis in this section to allow functional gauge generators $\xi^\mu_n(x;z)$. Such transformations are necessary (and natural) because $\xi^\mu_n$ is often required to transform away from a specific metric perturbation $h_{\mu\nu}(x;z)$, making $\xi^\mu_n$ depend on $h_{\mu\nu}(x;z)$ and therefore on $z^\mu$. However, to avoid overburdening the discussion, I leave this extension to the reader.

\subsubsection{Transformation of the singular and regular fields}
Though we derived the transformation laws for $h^n_{\mu\nu}$ with relative ease, we must take care in doing the same for the singular and regular fields $h^{\S n}_{\mu\nu}$ and $h^{\R n}_{\mu\nu}$. Suppose that in the $X$ gauge the regular field satisfies the ``nice'' properties that $\tilde g_{\mu\nu}$ is a vacuum metric and $z^\mu$ is a geodesic of that metric. In principle, after transforming to the $Y$ gauge, we may split $h'^n_{\mu\nu}$ into any singular and regular piece we like. However, clearly we would like to do so in a way that preserves the nice properties of the split, such that $\tilde {g}'_{\mu\nu}=g_{\mu\nu}+h'^{\R}_{\mu\nu}$ is a vacuum metric and $z'^\mu$ is a geodesic of that metric.

Appropriate transformation laws can be found by recalling that for a smooth metric and geodesic, the geodesic equation and vacuum Einstein equation are invariant under a generic smooth coordinate transformation. In our present context, the worldline automatically transforms according to the standard transformation law. It follows that if we let the effective metric also transform according to the ordinary coordinate-transformation law, with $\tilde g'_{\mu\nu}(x',\e;z')=\frac{\partial x^\alpha}{\partial x'^\mu}\frac{\partial x^\beta}{\partial x'^\nu} \tilde g_{\alpha\beta}(x(x'),\e;z(z'))$, then we preserve the nice properties of the regular field. Applying this rule, we have
\begin{subequations}\label{DhRSC}%
\begin{align}
h'^{\R1}_{\mu\nu}(x';z') &=  h^{\R1}_{\mu\nu}(x';z') + \Lie_{\xi_1} g_{\mu\nu}(x') ,\label{DhR1SC}\\
h'^{\R2}_{\mu\nu}(x';z') &=  h^{\R2}_{\mu\nu}(x';z') +  (\Lie_{\xi_1}+\varLie_{\xi_1}) h^{\R1}_{\mu\nu}(x';z') \nonumber\\
			&\quad+ \Lie_{\xi_2} g_{\mu\nu}(x') + \frac{1}{2}\Lie^2_{\xi_1} g_{\mu\nu}(x'). \label{DhR2SC}
\end{align}
\end{subequations}
At the same time, the perturbations $h'^n_{\mu\nu}=h'^{\R n}_{\mu\nu}+h'^{\S n}_{\mu\nu}$ must satisfy Eqs.~\eqref{DhSC}. This forces the singular field to transform as
\begin{subequations}\label{DhSSC}%
\begin{align}
h'^{\S1}_{\mu\nu}(x';z') &=  h^{\S1}_{\mu\nu}(x';z') ,\\
h'^{\S2}_{\mu\nu}(x';z') &=  h^{\S2}_{\mu\nu}(x';z') +  (\Lie_{\xi_1}+\varLie_{\xi_1}) h^{\S1}_{\mu\nu}(x';z').\label{DhS2SC}
\end{align}
\end{subequations}

To reiterate, with these rules, the effective metric in all smoothly related gauges is always a smooth solution to the vacuum Einstein equation, and the center-of-mass worldline is always a geodesic of that effective metric. Other transformation laws could also do the trick, but these are the most natural.

\subsubsection{Governing equations in different gauges}\label{transformed-EFE-SC}
By design, the transformation laws~\eqref{z_transformation}, \eqref{DhRSC}, and \eqref{DhSSC} ensure that \emph{the governing equations of the self-consistent approximation are invariant under a smooth gauge transformation}.

By this I mean, first and foremost, that in all smoothly related gauges,  the regular field satisfies the vacuum equation 
\beq\label{EFE-hR}
\delta R_{\mu\nu}[\e h^{\R1}+\e^2 h^{\R2}]+\e^2\delta^2 R_{\mu\nu}[h^{\R1},h^{\R1}] = \O(\e^3)
\eeq
and the worldline satisfies the geodesic equation~\eqref{2nd-geo}. These facts are obvious from the transformation laws for $h^{\R n}_{\mu\nu}$.

Even beyond the regular field, we can establish that the equations of the self-consistent approximation have a certain invariant form. For example, we can work out the distributional equation satisfied by the full field. Starting from Eq.~\eqref{distributional-EFE-SC-dR}, using Eq.~\eqref{DhSC} to express $h^n_{\mu\nu}[z]$ in terms of $h'^n_{\mu\nu}[z']$, appealing to Eqs.~\eqref{Lie R}--\eqref{Lie dR}, and expanding $\bar T^n_{\mu\nu}[z]$ around $\bar T^n_{\mu\nu}[z']$, we arrive at
\begin{align}
\delta R_{\mu\nu}&[\e h'^1+\e^2 h'^2]+\e^2\delta^2 R_{\mu\nu}[h'^1,h'^1] \nonumber\\
 &= 8\pi\e\bar T^1_{\mu\nu}[z']+8\pi\e^2\bar T^2_{\mu\nu}[z']\nonumber\\
 &\quad+8\pi\e^2(\Lie_{\xi_1}+\varLie_{\xi_1})\bar T^1_{\mu\nu}[z']+\O(\e^3).\label{EFE-h}
\end{align}
This is simply Eq.~\eqref{distributional-EFE-SC-dR} with a transformed distributional source,
\begin{subequations}\label{DT}
\begin{align}
\bar T'^1_{\mu\nu}[z'] &= \bar T^1_{\mu\nu}[z'],\\
\bar T'^2_{\mu\nu}[z'] &= \bar T^2_{\mu\nu}[z']+\e^2(\Lie_{\xi_1}+\varLie_{\xi_1})\bar T^1_{\mu\nu}[z'].
\end{align}
\end{subequations}
Therefore the field equation~\eqref{distributional-EFE-SC-dR} is satisfied in all gauges, although the source must be appropriately transformed between them. This change in the stress-energy corresponds precisely to the transformation of the singular field in Eq.~\eqref{DhSSC}. It is examined in detail in Appendix~\ref{Delta T2}.

So far we have only seen that that the total field equations $R_{\mu\nu}[g+h^\R]=0$ and $R_{\mu\nu}[g+h]=\sum_n\e^n\bar T^n_{\mu\nu}$ hold true in any gauge (minding that  $\bar T^n_{\mu\nu}$ must be transformed along with $h^{n}_{\mu\nu}$). However, we may like a stronger result: that the individual fields $h'^{\R n}_{\mu\nu}[z']$ and $h'^{n}_{\mu\nu}[z']$ satisfy the separate field equations~\eqref{hRSC}, \eqref{h1SC}, and \eqref{h2SC-distributional}. This is nontrivial because we are not working with ordinary Taylor series. In Sec.~\ref{SC_general_formulation}, I defined the functionals $h^n_{\mu\nu}$ as the solutions to Eqs.~\eqref{EFE-wgauge}, enforcing a particular division of $h_{\mu\nu}$. In writing down Eq.~\eqref{DhSC} for $h'^{n}_{\mu\nu}[z']$, I have adopted another particular division of $h'_{\mu\nu}$ in Eq.~\eqref{h'z'}. A priori, these two divisions could disagree; $\e h'^1_{\mu\nu}[z']$ could differ by $o(\e)$ from a solution to an equation of the form $E'^1_{\mu\nu}[\e h'^1]=8\pi\e\bar T^1_{\mu\nu}[z']$, for example.

To see that Eqs.~\eqref{hRSC}, \eqref{h1SC}, and \eqref{h2SC-distributional} do hold in all gauges, return to the idea that the self-consistent approximation corresponds to a Taylor series at fixed $z^\mu$. In accord with that idea, write the operators $E^n_{\mu\nu}$ in any gauge as
\begin{subequations}
\begin{align}
E^1_{\mu\nu}[h^1[z]]&=\frac{d}{d\e}R_{\mu\nu}\!\left[g+\e h^1[z]+\O(\e^2)\right]\!\!\big|_{\e=0},\\
E^2_{\mu\nu}[h^2[z]]&=\bigg\{\frac{1}{2}\frac{d^2}{d\e^2}R_{\mu\nu}\!\left[g+\e h^1[z]+\e^2h^2[z]+\O(\e^3)\right]\nonumber\\
								&\quad -\delta^2R_{\mu\nu}\!\!\left[h^1[z],h^1[z]\right]\!\!\bigg\}\bigg|_{\e=0},\label{E2}
\end{align}
\end{subequations}
where before evaluating the derivatives, the expansion~\eqref{SC_force_expansion} of the equation of motion is substituted, and the evaluation at $\e=0$ is performed while holding $z^\mu$ and $\dot z^\mu$ fixed.

With these definitions, let us first establish the desired result for $h'^1_{\mu\nu}[z']$:
\beq
E'^1_{\mu\nu}[h'^1[z']] = 8\pi \bar T^1_{\mu\nu}[z'].\label{E'=T'}
\eeq
The left-hand side may be rewritten as
\begin{subequations}
\begin{align}
E'^1_{\mu\nu}[h'^1[z']] &= \frac{d}{d\e}\delta R_{\mu\nu}[\e h'^1[z']]\big|_{\e=0}\\
 &= \frac{d}{d\e}\delta R_{\mu\nu}[\e h^1[z']]\big|_{\e=0}\\
 &= E^1_{\mu\nu}[h^1[z']],
\end{align}
\end{subequations}
where the second line is a consequence of $\delta R_{\mu\nu}[\Lie_{\xi_1}g]=0$. Equation~\eqref{E'=T'} then follows from $E^1_{\mu\nu}[h^1]=8\pi\bar T^1_{\mu\nu}$.

The desired result for $h'^2_{\mu\nu}[z']$,
\beq
E'^2_{\mu\nu}[h'^2[z']] +\delta^2R_{\mu\nu}[h'^1[z'],h'^1[z']] = 8\pi\bar T'^2_{\mu\nu}[z'], 
\eeq
follows from Eq.~\eqref{E2} in the same manner [repeating the operations that led to Eq.~\eqref{EFE-h}]. And we may likewise obtain the desired equations for $h'^{\R n}_{\mu\nu}[z']$.

Let me summarize: If the approximation scheme described in Sec.~\ref{SC_general_formulation} is applied in two slightly different coordinate systems related by an equation of the form~\eqref{coord_transformation}, then the terms in the two resulting approximations are related according to Eqs.~\eqref{DhSC}, \eqref{DhRSC}, and \eqref{DhSSC}. In all cases, the center-of-mass worldline is a geodesic of the effective metric, and the effective metric is a vacuum metric.

\subsubsection{Transformation of the self-force (and other fields on the worldline)}\label{transformation-on-z}
In the above sections, I described the effect of a gauge transformation on quantities evaluated at a given point in $\man_0$. However, in self-force theory, we are often interested in something quite different: the effect of a gauge transformation on quantities evaluated specifically on the worldline. This differs from the previous case because the position of the worldline itself changes under the transformation.

The correct treatment of this situation can be deduced from Fig.~\ref{Fig:worldlines}. Suppose that in gauge $X$, we calculate some tensor $A=A_0+\e A_1+\ldots$ on the worldline $\gamma_\e^X$, and in gauge $Y$ we calculate the same tensor as $A'=A_0+\e A'_1+\ldots$ on the worldline $\gamma_\e^Y$. How are the results related? The worldline $\gamma_\e^X$ is mapped to $\gamma_\e^Y$ via the transformation $\psi_\e:=(\phi_\e^Y)^{-1}\circ\phi_\e^X$. We can find the tangent field to this map, call it $\overleftarrow{\xi}^\mu$, by writing $e^{\Lie_{\overleftarrow{\xi}}}=\psi_\e^*=(\phi_\e^X)^*(\phi_\e^Y)^{-1*}=e^{\e\Lie_X}e^{-\e\Lie_Y}$. Expanding the exponentials, we find
\beq
\overleftarrow{\xi}^\mu = -\e\xi_1^\mu-\e^2\xi_2^\mu+\O(\e^3).\label{left-xi}
\eeq
We can then expand tensors at $\gamma_\e^Y$ around their values at $\gamma_\e^X$ using $\psi_\e^*=e^{\Lie_{\overleftarrow{\xi}}}$. Hence, the difference between the expansions in the two gauges on the two worldlines is
\beq\label{varDelta}
\varDelta A = \psi_\e^* A' - A.
\eeq
Here we find the change in the result not only due to the change of gauge, $A\to A'$, but also due to the change in evaluation point, $p\in\gamma_\e^X\to p'=\psi_\e(p)\in\gamma_\e^Y$. The comparison is made at the original point on $\gamma_\e^X$, and that is where the tensor $\varDelta A$ is defined. I use a slanted $\varDelta$ to distinguish this transformation from the ordinary transformation $A\to A'$.

Equation~\eqref{varDelta} becomes clearer when applied to a particular case. To demonstrate its use, I now work out the gauge transformation of the equation of motion $a^\mu=F^\mu$. 
 
 On the left-hand side, we have
\begin{align}\label{varDelta-a}
\varDelta a^\mu = -\e \Lie_{\xi_1}a^\mu -\e^2\left(\Lie_{\xi_2}a^\mu-\tfrac{1}{2}\Lie_{\xi_1}^2a^\mu\right)+\O(\e^3).
\end{align}
On the right-hand side, we have
\begin{align}\label{varDelta-F}
\varDelta F^\mu = \e \Delta F_1^\mu + \e^2 \left[\Delta F_2^\mu-\Lie_{\xi_1}(F_1^\mu+\Delta F_1^\mu)\right]+\O(\e^3).
\end{align}

To evaluate these expressions, I return to the type of calculations performed in Sec.~\ref{geodesic_expansion_in_h_and_dz}. In that section, I took Lie derivatives of the self-accelerated equation of motion to expand it around its value on a background geodesic. Now I expand its value on one self-accelerated worldline around its value on another self-accelerated worldline, with the two worldlines related by a gauge transformation. The role of the surface $\mathcal{S}$ in Sec.~\ref{geodesic_expansion_in_h_and_dz} is now played by the shaded surface connecting $\gamma_\e^X$ to $\gamma_\e^Y$ in Fig.~\ref{Fig:worldlines}, and the role of the vector field $v^\mu$ is now played by $\overleftarrow{\xi}^\mu$. However, the situations are not quite identical, partly because the acceleration does not vanish on either worldline, but more importantly because we cannot easily choose an arbitrary parameter on $\gamma_\e^Y$. If we parametrize $\gamma_\e^X$ with proper time, call it $\tau_X$, then the map $\psi_\e(\gamma_\e(\tau_X))$ induces a natural parameter $s(\tau_X)$ on $\gamma_\e^Y$, defined via $z'^\mu(s(\tau_X))=x^\mu(\psi_\e(\gamma_\e(\tau_X)))$. 

As in Sec.~\ref{geodesic_expansion_in_h_and_dz}, I rewrite the acceleration on $\gamma_\e^Y$ in terms of $s$,
\beq
a^\mu = \left(\frac{ds}{d\tau}\right)^2\left(\ddot z^\mu -\kappa \dot z^\mu\right),
\eeq
where $\tau$ is proper time on $\gamma_\e^Y$, and $s$ now refers specifically to the parameter induced by $\psi_\e$. With this specific parameter, just as we had $\Lie_{\dot z}v^\mu=\Lie_{v}\dot z^\mu=0$, we now have
\beq\label{commutation-xi}
\Lie_{\overleftarrow{\xi}}\dot z^\mu = \Lie_{\dot z}\overleftarrow{\xi}^\mu = 0,
\eeq
and so
\beq
\overleftarrow{\xi}^\nu\nabla_\nu\dot z^\mu = \dot z^\nu\nabla_\nu\overleftarrow{\xi}^\mu.
\eeq
If we did not use the naturally induced parameter, these useful relations would not hold.

Since $\overleftarrow{\xi}^\mu$ plays precisely the same role as $v^\mu$ did previously, we can apply all the results of Sec.~\ref{v-zdot-identities}. Concretely, Eqs.~\eqref{Liedsdtau}--\eqref{vDzddot} all hold true with the replacement $v^\mu\to\overleftarrow{\xi}^\mu$.


With those identities in hand, along with the expressions for $\Delta h^{\R n}_{\mu\nu}$, we can go ahead and calculate $\varDelta a^\mu$ and $\varDelta F^\mu$ according to Eqs.~\eqref{varDelta-a} and \eqref{varDelta-F}. Since $a^\mu=F^\mu$, we necessarily get $\varDelta a^\mu=\varDelta F^\mu$, and the explicit calculations bear this out. Either way, one finds the following transformation law for the self-force: 
\begin{align}
\varDelta F^\mu &= -\e\left(P^\mu{}_\nu\ddot\xi_1^\nu+R^\mu{}_{u\xi_1u}\right)\nonumber\\
				&\quad -\e^2 \left[P^\mu{}_\nu\ddot\zeta^\nu+R^\mu{}_{u\zeta u} +\xi^\mu_{1;\nu}\left(\ddot\xi_1^\nu+R^\nu{}_{u\xi_1 u}\right)\right.\nonumber\\
				&\quad -2R^\mu{}_{u\xi_1u}\dot\xi^\nu_1u_\nu+2R^\mu{}_{\dot\xi_1u\xi_1}+u^\mu R_{u\xi_1u\dot\xi_1}\nonumber\\
				&\quad +\tfrac{1}{2}\dot R^\mu{}_{\xi_1u\xi_1}+\tfrac{1}{2}u^\mu\dot R_{u\xi_1u\xi_1}-\tfrac{1}{2}R^\mu{}_{u\xi_1u|\xi_1}\nonumber\\
				&\quad -u^\mu\left(\ddot\xi^\nu_1\dot\xi_{1\nu}+4\ddot\xi^\nu_1 u_\nu\dot\xi_1^\rho u_\rho\right)-2\ddot\xi^\mu_1\dot\xi^\nu_1u_\nu+2F_1^\mu \dot\xi^\nu_1u_\nu\nonumber\\
				&\quad \left.+F^\nu_1\left(u^\mu\dot\xi_{1\nu}-\xi^\mu_{1;\nu}\right)\right]+\O(\e^3),\label{DFSC}
\end{align}
where $\zeta^\mu:=\xi^\mu_2-\frac{1}{2}\xi^\nu_1\xi^\mu_{1;\nu}$, overdots denote covariant differentiation with respect to $\tau$, and I have used the notation $\dot R^\mu{}_{\xi_1u\xi_1}:=R^\mu{}_{\alpha\beta\gamma;\nu}\xi_1^\alpha u^\beta\xi_1^\gamma u^\nu$ and $R^\mu{}_{u\xi_1u|\xi_1}:=R^\mu{}_{\alpha\beta\gamma;\nu}u^\alpha \xi_1^\beta u^\gamma \xi_1^\nu$ (for example). The first-order term is precisely the standard transformation law derived by Barack and Ori~\cite{Barack-Ori:01}. The second-order term appears here for the first time.

We can also apply the transformation law~\eqref{varDelta} to any other quantity of interest on the worldline. For example, applying it to the regular field evaluated on the worldline, we get
\begin{align}
\varDelta h^{\R}_{\mu\nu} &= \e \Lie_{\xi_1} g_{\mu\nu} + \e^2\big(\Lie_{\xi_2} g_{\mu\nu} - \tfrac{1}{2}\Lie^2_{\xi_1} g_{\mu\nu} \big) +\O(\e^3).
\end{align}
Here the $(\Lie_{\xi_1} +\varLie_{\xi_1})h^{\R 1}_{\mu\nu}$ terms in Eq.~\eqref{DhR2SC} are removed by the action of $\psi_\e^*$, leaving only a transformation of the background metric.

As a final observation in this section, I note that one could choose a gauge in which the ``accelerated'' worldline in the background is in fact the geodesic $\gamma_0$. This is easily seen from Fig.~\ref{Fig:worldlines}: in place of the vector fields $X$ or $Y$, simply choose a vector field $Z$ that is tangent to the shaded surface connecting $\gamma_0$ to $\gamma_\e$. This vector field will define an identification map $\phi^Z_\e$ satisfying $\phi^Z_\e(\gamma_0(s))=\gamma_\e(s)$. Hence, from this perspective \emph{the deviation of the self-accelerated worldline from the background geodesic is pure gauge}. This will be illustrated more explicitly in the next section. However, I will argue in Sec.~\ref{conclusion} that such a gauge choice is unacceptable on the physically interesting domains of size $\sim1/\e$; on such a domain, it forces the metric perturbation to become large.

\subsection{Gauge in the Gralla-Wald approximation}\label{gauge_GW}
I now consider the effect of a gauge transformation in the Gralla-Wald approximation. My treatment mirrors that of the previous section. However, since the coefficients $\GW{h}^n_{\mu\nu}$ in the Gralla-Wald expansion are independent of $\e$, gauge transformations require somewhat less care.

Unlike in the self-consistent case, where we imagine the object shrinking toward zero size on an accelerated worldline in $\man_0$, here the object shrinks toward zero size on a background geodesic $\gamma_0$, as illustrated in Fig.~\ref{Fig:limits}. 

\subsubsection{Transformation of the worldline}\label{zn-transformation}
In the self-consistent case, we had a transformation of the entire worldline $z^\mu$. In the Gralla-Wald case, we instead have transformations of each term $\GW{z}^\mu_n$ in the expansion of $z^\mu$. They can be found immediately by expanding Eq.~\eqref{z_transformation} in powers of $\e$, yielding
\begin{subequations}\label{DzGW}
\begin{align}
z_0'^\mu(s) &=z_0^\mu(s),\label{z0_transformation} \\
\GW{z}_1'^{\mu}(s) &= \GW{z}_1^\mu(s) - \xi_1^\mu(z_0), \label{z1_transformation}\\
\GW{z}_2'^{\mu}(s) &= \GW{z}_2^\mu(s) - \xi_2^\mu(z_0)+\frac{1}{2}\xi^\nu_1(z_0)\partial_\nu\xi_1^\mu(z_0)\nonumber\\
					&\quad-\GW{z}_1^\nu(s)\partial_\nu\xi^\mu_1(z_0),\label{z2_transformation}
\end{align}
\end{subequations}
where functions of $z_0^\mu$ are evaluated at $z^\mu_0(s)$. Note that the zeroth-order worldline is unchanged; the effect of the transformation is to alter the deviations relative to that worldline. 

These results depend on both the background coordinates and the parametrization of the accelerated worldline $z^\mu(s,\e)$. We are perhaps more interested in the covariant and parametrization-invariant quantities $\GW{z}^\mu_{1\perp}$ and $\GW{z}^\mu_{2\ddagger}$. Projecting out the parallel components in Eq.~\eqref{z1_transformation}, we get
\beq
\GW{z}_{1\perp}'^{\mu}(s) = \GW{z}_{1\perp}^\mu(s) - \xi_{1\perp}^\mu(z_0), \label{z1perp_transformation}
\eeq
where $\xi_{1\perp}^\mu:=P^\mu_{0\nu}\xi_{1}^\nu$. Evaluating Eq.~\eqref{z2_transformation} in a normal coordinate system centered on $\gamma_0$, we get
\begin{align}
\GW{z}_{2\rm N}'^{\mu}(s) &= \GW{z}_{2\rm N}^\mu(s) - \xi_2^\mu(z_0)+\frac{1}{2}\xi^\nu_1(z_0)\nabla_\nu\xi_1^\mu(z_0)\nonumber\\
					&\quad-\GW{z}_1^\nu(s)\nabla_\nu\xi^\mu_1(z_0).\label{z2N_transformation}
\end{align}
This result is covariant but parametrization-dependent. We can next apply Eq.~\eqref{z2cross-vs-z2perp} to get $\GW{z}_{2\ddagger}'^{\mu} = \GW{z}_{2\rm N\perp}'^{\mu} + \GW{z}'_{1\parallel}\GW{u}'^\mu_{1\perp}$. When combined with Eqs.~\eqref{z2N_transformation} and \eqref{z1_transformation}, this gives us (after some simplification)
\begin{subequations}\label{z2ddagger_transformation}%
\begin{align}
\GW{z}_{2\ddagger}'^{\mu} &= \GW{z}_{2\ddagger}^{\mu} - \xi^\mu_{2\ddagger} - P^\mu_{0\nu}\varsigma^\alpha\nabla_\alpha\xi^\nu_{1}\\
										&= \GW{z}_{2\ddagger}^{\mu} - \xi^\mu_{2\perp} +\xi_{1\parallel}P^\mu_{0\,\nu}\Lie_{\varsigma}u_0^\nu - P^\mu_{0\nu}\varsigma^\alpha\nabla_\alpha\xi^\nu_{1\perp},\label{z2ddagger_transformationB}
\end{align}
\end{subequations}
where $\xi^\mu_{2\ddagger}:=\xi^\mu_{2\perp}+\xi_{1\parallel}\frac{D\varsigma^\mu}{d\tau_0}$ and $\varsigma^\mu:=\GW{z}^\mu_{1\perp}-\tfrac{1}{2}\xi^\mu_{1\perp}$. Note that the second formula for the transformation, Eq.~\eqref{z2ddagger_transformationB}, involves derivatives of $u_0^\mu$ perpendicular to $\gamma_0$, requiring some extension of $u^\mu_0$ off $\gamma_0$; this extension is arbitrary, and it can be chosen to make the Lie derivative term vanish, for example.

\subsubsection{Transformation of the metric perturbations}
Now turn to the metric perturbations. Applying the transformation laws~\eqref{DeltaA}, we find 
\begin{subequations}\label{DhGW}
\begin{align}
\Delta \GW{h}^1_{\mu\nu}(x;\!z_0) &=\Lie_{\xi_1} g_{\mu\nu},\label{Deltah1GW}\\
\Delta \GW{h}^2_{\mu\nu}(x;\!z_0) &=\Lie_{\xi_2} g_{\mu\nu} +\! \frac{1}{2}\Lie^2_{\xi_1}g_{\mu\nu} + \Lie_{\xi_1} \GW{h}^1_{\mu\nu}(x;\!z_0). \label{Deltah2GW}
\end{align}
\end{subequations}
All the perturbations $\GW{h}'^n_{\mu\nu}$ in the new gauge, like all the perturbations $\GW{h}^n_{\mu\nu}$ in the old, diverge on the zeroth-order worldline $z^\mu_0$. This is an essential point of difference from the self-consistent approximation: In the self-consistent approximation, a gauge transformation shifts the curve on which the singular field diverges. In the Gralla-Wald approximation, the zeroth-order worldline, on which the singular field diverges, is invariant.

Instead of altering the curve on which the fields diverge, the gauge transformation alters the singularity on that curve, by altering the functions $\GW{z}_{n>0}^\mu$ that appear in $\GW{h}^{n>1}_{\mu\nu}$. It effects this change in singularity structure by altering the mass dipole moment. Recall the discussion in Sec.~\ref{h-transform-SC}, from which we can infer that the term $\Lie_{\xi_1} \GW{h}^1_{\mu\nu}$ generates a mass dipole moment $-m\xi^i_1$ relative to the worldline on which $\GW{h}^1_{\mu\nu}$ diverges, which  is $z^\mu_0$. In the self-consistent case, we began with no mass dipole moment in the unprimed gauge, and the dipole moment induced by $\Lie_{\xi_1}$ was exactly cancelled by the action of $\varLie_{\xi_1}$, leaving us again with no dipole moment. In the Gralla-Wald case, we instead generically begin with a mass dipole moment $M^\mu$, and the gauge transformation alters it by an amount $\Delta M^\mu=-m\xi^\mu_{1\perp}$.

I remind the reader that the first-order deviation in the Gralla-Wald approximation is defined as $z^\mu_{1\perp}:=M^\mu/m $. Hence, the transformation law $\Delta M^\mu=-m\xi^\mu_{1\perp}$, obtained from the transformation of the metric perturbation, precisely reproduces the transformation law~\eqref{z1perp_transformation}. In fact, at any order, the transformation laws for $\GW{z}_{n>0}^\mu$, as given at first and second order in Eq.~\eqref{DzGW}, can always be derived directly from the transformation laws for the metric perturbations. In principle, one need never appeal to Eq.~\eqref{z_transformation}. This becomes important in the case of transformations that are singular on the worldline, which spoils the application of Eq.~\eqref{z_transformation} but does not prevent the application of Eq.~\eqref{DhGW}~\cite{Gralla-Wald:08,Gralla:11,Pound-Merlin-Barack:14}.

\subsubsection{Transformation of the singular and regular fields}
We must now apportion  Eq.~\eqref{DhGW} into transformation laws for the singular and regular fields. In exact analogy with the self-consistent case, we make the natural choice of transforming the regular field like an ordinary smooth perturbation, which now means
\begin{subequations}\label{DhRGW}%
\begin{align}
\Delta \GW{h}^{\R1}_{\mu\nu}(x;\!z_0) &=\mathcal{L}_{\xi_1} g_{\mu\nu},\\
\Delta \GW{h}^{\R2}_{\mu\nu}(x;\!z_0) &=\mathcal{L}_{\xi_2} g_{\mu\nu} +\! \frac{1}{2}\mathcal{L}^2_{\xi_1} g_{\mu\nu}
				+\mathcal{L}_{\xi_1} \GW{h}^{\R1}_{\mu\nu}(x;\!z_0),
\end{align}
\end{subequations}
This leaves the singular field to transform as
\begin{subequations}\label{DhSGW}%
\begin{align}
\Delta \GW{h}^{\S1}_{\mu\nu}(x;z_0) &=0,\\
\Delta \GW{h}^{\S2}_{\mu\nu}(x;z_0) &=\mathcal{L}_{\xi_1} \GW{h}^{\S1}_{\mu\nu}(x;z_0).\label{DhS2}
\end{align}
\end{subequations}

As in the self-consistent case, these laws preserve all the nice properties of the singular-regular split.

\subsubsection{Governing equations in alternative gauges}
By construction, all the governing equations are invariant: the equations of motion~\eqref{z1_generic} and \eqref{z2_generic} apply in all gauges, and in all gauges the regular fields satisfy the vacuum Einstein equations~\eqref{hRGW}. 

As in the self-consistent case, the only thing to be mindful of is the transformation~\eqref{DhS2} of the singular field, or put another way, the transformation of the stress-energy. Repeating the analysis that led to Eq.~\eqref{EFE-h}, we find that the field equations~\eqref{hGW} for $\GW{h}^n_{\mu\nu}$ are invariant, but the stress-energy is altered between gauges. Specifically, $\GW{\bar T}'^1_{\mu\nu}=\GW{\bar T}^1_{\mu\nu}$, and $\GW{\bar T}'^2_{\mu\nu}=\GW{\bar T}^2_{\mu\nu}+\Delta\GW{\bar T}^2_{\mu\nu}$, where
\beq
\Delta \GW{\bar T}^2_{\mu\nu}= \Lie_{\xi_1}\GW{\bar T}^1_{\mu\nu}. 
\eeq
An explicit expression for $\Lie_{\xi_1}\GW{\bar T}^1_{\mu\nu}$ can be derived from one for $\Lie_{\xi_1}T_1^{\mu\nu}$, given in Eq.~\eqref{LieT}. From that equation, we see that the Lie derivative both alters $\delta m_{\mu\nu}$ and more notably, shifts the mass dipole moment by an amount $\Delta M^\mu=-m\xi^\mu_{1\perp}$, in agreement with the discussion above.


\subsubsection{Transformation of the self-force (and other fields on $\gamma_0$)}\label{DF-GW}
Working out gauge transformations of fields on the worldline is simultaneously simpler and subtler than it was in the self-consistent case. I first present the simpler viewpoint; at the end of the section I turn to the subtle point.

From the simple viewpoint, transformations of tensors on $\gamma_0$ are straightforward for the obvious reason that $\gamma_0$ is unmoved by the transformation. Hence, we can use the same transformation laws for tensors at a point on $\gamma_0$ as at any other point in $\man_0$.

Let us utilize this perspective in working out the transformation of the self-forces in Eqs.~\eqref{F1_generic} and \eqref{F2_generic}. Since the equations of motion~\eqref{z1_generic} and \eqref{z2_generic} are valid in any gauge, we can easily find the transformations of $\GW{F}^\mu_n$ by writing those equations in terms of quantities $\GW{z}'_{1\perp}=\GW{z}_{1\perp}+\Delta \GW{z}_{1\perp}$, $\GW{z}'_{2\ddagger}=\GW{z}_{2\ddagger}+\Delta \GW{z}_{2\ddagger}$, and $\GW{F}'^\mu_n=\GW{F}^\mu_n+\Delta\GW{F}^\mu_n$. Substituting Eqs.~\eqref{z1perp_transformation} and \eqref{z2ddagger_transformation} and then solving for $\Delta\GW{F}^\alpha_n$ yields 
\begin{align}
\Delta\GW{F}^\alpha_1 &=  -\frac{D^2\xi_{1\perp}^\alpha}{d\tau_0^2}-R^\alpha{}_{\mu\beta\nu}u_0^\mu \xi^\beta_{1\perp}u_0^\nu,\label{DF1}
\end{align}
and
\begin{align}
\Delta\GW{F}^\alpha_2 &= \frac{D^2\Delta\GW{z}_{2\ddagger}^\alpha}{d\tau_0^2}+R^\alpha{}_{\mu\beta\nu}u^\mu_0\Delta\GW{z}_{2\ddagger}^\beta u^\nu_0\nonumber\\
							&\quad	-2P^\alpha_{0\rho}R^\rho{}_{\mu\beta\nu}u_0^\nu\left(\dot\xi^\mu_{1\perp}\GW{z}^\beta_{1\perp}+u^\mu_{1\perp}\xi^\beta_{1\perp}
							-\dot\xi^\mu_{1\perp}\xi^\beta_{1\perp}\right)\nonumber\\
							&\quad +2P^\alpha_{0\rho}R^\rho{}_{\mu\beta\nu;\gamma}\left(\xi^{(\mu}_{1\perp}u^{\beta)}_0z^{[\nu}_{1\perp}u^{\gamma]}_0
							+z^{(\mu}_{1\perp}u^{\beta)}_0\xi^{[\nu}_{1\perp}u^{\gamma]}_0\right.\nonumber\\
							&\quad\left.-\xi^{(\mu}_{1\perp}u^{\beta)}_0\xi^{[\nu}_{1\perp}u^{\gamma]}_0\right),\label{DF2}
\end{align}
where $\Delta\GW{z}_{2\ddagger}^\alpha$ is given by Eq.~\eqref{z2ddagger_transformation}.

One could equally well derive these results by  directly substituting $h'^{\R n}_{\mu\nu}=h^{\R n}_{\mu\nu}+\Delta h^{\R n}_{\mu\nu}$ into Eqs.~\eqref{F1_generic} and \eqref{F2_generic} and then using the expressions~\eqref{DhRGW} for $\Delta h^{\R n}_{\mu\nu}$.

Now to the subtle issue, which is clearly seen from a geometrical standpoint. Consider a tensor $T$ defined only on the family of worldlines $\gamma_\e$. Figure~\ref{Fig:worldlines}  shows that the relationship between $\gamma_\e\subset\widetilde\man_\e$ and $\gamma_0\subset\man_0$ is unaltered by a gauge transformation: the shaded surface connecting them, call it $\mathcal{S}^*$, is gauge independent. It follows that the expansion of a tensor $T$ that lives only on the worldline is gauge invariant. Its expansion always reads $e^{\e\Lie_Z}T\big|_{\gamma_0}$. 

To write the quantities in this expansion in any particular gauge $X$, we can relate the situation back to the expansion of the worldline in Sec.~\ref{geodesic_expansion_in_h_and_dz}. The surface $\mathcal{S}$ described there is a gauge-dependent quantity, equal in the gauge $X$ to $(\phi_\e^X)^{-1}(\mathcal{S}^*)$, a surface connecting $\gamma_0$ to $\gamma_\e^X$. It is generated by the vector field $v$, which in analogy with Eq.~\eqref{left-xi} is now given by $v=\e(Z-X)+\tfrac{1}{2}\e^2 [X,Z]+\O(\e^3)$. So in gauge $X$ the expansion reads $e^{\e\Lie_Z}T = T_0 + \e (T_1+\Lie_v T_0)+\O(\e^2)$, where $T_1=\Lie_X T$. In a gauge $Y$ the expansion would read $e^{\e\Lie_Z}T = T_0 + \e (T'_1+\Lie_{v'} T_0)+\O(\e^2)$, where $T'_1=\Lie_Y T$ and  $v'=\e(Z-Y)+\tfrac{1}{2}\e^2 [Y,Z]+\O(\e^3)$. Clearly, the results in the two gauges must agree, since they both began from the gauge-invariant field $Z$. Put another way, $Z$ provides a preferred reference gauge in this scenario, and when we examine the expansion of quantities defined only on the worldline, we must always transform to this reference gauge, with the vector $v$ acting as the gauge vector. 

However, in most circumstances, this subtle point need not be minded. In a typical situation, we are faced with some quantity written in terms of tensors on $\gamma_0$, and we are only interested in whether the value of that quantity is altered when we calculate the tensors in a different gauge. We can then go ahead and adopt the simple viewpoint described above.

\subsection{Gauge freedom of another kind}\label{third-freedom}
In general, three kinds of ``gauge freedom'' exist in perturbation theory. In this paper, I have focused exclusively on one of them: the identification of points in the perturbed spacetime with points in the background spacetime. This is the gauge freedom usually discussed in the context of perturbation theory~\cite{Geroch:69, Stewart-Walker:74, Bruni-etal:96}; it is equivalent to expanding out the effect of a small coordinate transformation, such that the background is left unchanged and order-$\e$ transformations of background fields are treated as alterations of the perturbations to those fields. 

A second type of gauge freedom is the freedom to perform a coordinate transformation (or diffeomorphism) on the background, inducing, via one's identification map, a coordinate transformation in the full spacetime. 

However, prior to these two types of freedoms, there is a third type: the choice of the spacetime family itself. Given a particular metric ${\sf g}$ involving a numerically small parameter $\e$, one can embed it into different families ${\sf g}_\e$. This corresponds to choosing ``what to hold fixed'' while taking the limit $\e\to0$. As a trivial example, consider adding a linear, global mass perturbation $h^1=\frac{\partial g}{\partial M}\delta M$ to a black hole metric $g$. Such a perturbation can be incorporated into the background $g$ via a redefinition of the background mass, $M\to M+\delta M$, thereby choosing to hold $M+\delta M$, rather than $M$, fixed while $\e$ varies. Since the mass of a spacetime cannot be altered by a coordinate transformation, this trivial example illustrates that this freedom is independent of the usual ``small coordinate transformations''. 

In the context of self-force theory, this third freedom has important consequences. Consider the Gralla-Wald approximation. Choosing a family of spacetimes requires choosing the family of worldlines $z_\e^\mu$. As an example, suppose we are given a spacetime $({\sf g}_{\mu\nu}(x),\man)$ containing a small object of mass $m$ moving about a nonspinning black hole of mass $M$ in a quasicircular orbit. On a short timescale the orbit has an approximately constant orbital frequency $\Omega^*$ measured by an inertial observer at infinity. We can imagine embedding this spacetime into many different families. We might choose a family ${\sf g}_{\mu\nu}(x,\e;\Omega)$, satisfying ${\sf g}_{\mu\nu}(x,\e^*;\Omega)={\sf g}_{\mu\nu}(x)$ for some $\e^*>0$, in which all members share the same orbital frequency $\Omega=\Omega(\e^*) :=\Omega^*$. But we might instead choose a family ${\sf g}_{\mu\nu}(x,\e;\Omega(\e))$, satisfying ${\sf g}_{\mu\nu}(x,\e^*;\Omega(\e^*))={\sf g}_{\mu\nu}(x)$, in which the orbital frequency varies between family members, such that $\Omega(\e^*)=\Omega^*$ but $\Omega(\e\neq\e^*)\neq\Omega^*$. 

With all else the same, in both cases we write ${\sf g}_{\mu\nu}(x)$ as an expansion around the same background metric $g_{\mu\nu}(x)={\sf g}_{\mu\nu}(x,0;\Omega)={\sf g}_{\mu\nu}(x,0;\Omega(0))$, which in the present example is a Schwarzschild metric of mass $M$. But the background geodesic around which we expand the worldline $z^\mu$ differs between the two families. In the first case, we expand $z^\mu$ around a circular background geodesic of the same frequency, $\Omega^*$, while in the second case we expand it around a circular geodesic of a different frequency, $\Omega(0)\neq\Omega^*$. Either way, we describe the same physical spacetime $({\sf g}_{\mu\nu}(x),\man)$, but we do so by writing it as an expansion along flows in two different families of spacetimes. Because the orbital frequency (as measured by an inertial observer at infinity) is a gauge-invariant quantity, the two families are mathematically distinguishable; the difference between them is \emph{not} a difference that can be removed by a small coordinate transformation. 

This example is relevant for many practical calculations in the literature. For a canonical instance, refer to Refs.~\cite{Detweiler:08,Sago-Barack-Detweiler:08}. There the authors first expanded the circular orbit about a circular geodesic of the same Schwarzschild-coordinate radius $r_0$. This specifies a perfectly acceptable family of spacetimes, the members of which contain orbits of differing frequencies $\Omega(\e)$. However, the authors then re-expanded their expressions such that the end result is equivalent to expanding the circular orbit about a circular geodesic of the same orbital  frequency $\Omega=\Omega^*$, now placing the metric in a new family of spacetimes, each member of which contains an orbit of frequency $\Omega^*$.\footnote{Of course, one can instead expand at fixed frequency directly, without first expanding at fixed orbital radius.} This second choice is more useful in most cases, because it enables comparison to other formalisms, such as post-Newtonian theory, which can obtain results at fixed frequencies but, because they do not expand around a Schwarzschild background, cannot obtain results at fixed Schwarzschild-coordinate radius.  However, from the perspective of perturbation theory on a given background, the ``fixed-$r_0$'' and ``fixed-$\Omega$'' expansions are equally admissible, and they equally well approximate the particular spacetime $({\sf g}_{\mu\nu}(x),\man)$. 

For more details of how this type of freedom can play out, I refer the reader to Ref.~\cite{Pound:14c}.

\section{Summary and discussion of practical issues}\label{conclusion}

In this paper, my central concern has been elucidating the mathematical structure and gauge freedom of approximation schemes involving perturbed motion in general relativity. The presentation has been, for the most part, quite abstract. However, it has also led to several concrete, applicable results, including
\begin{enumerate}
\item a covariant and reparametrization-invariant expansion of an accelerated worldline around a neighbouring geodesic. This led to the second-order equation of motion~\eqref{z2_generic} and the worldline-deviation terms in the stress-energy tensor~\eqref{dT_zperp} in the Gralla-Wald approximation. Moreover, the methods can be applied in any problem in which one is interested in expanding one worldline about another. Such expansions appear in osculating-geodesic and two-timescale expansions, amongst others.
\item gauge transformation laws for the metric perturbation of the self-consistent expansion, given by Eq.~\eqref{DhSC}, as well as a general transformation law for tensors that live only on the worldline, given by Eq.~\eqref{varDelta}. 
\item natural transformation rules for the singular and regular fields, given by Eqs.~\eqref{DhRSC}--\eqref{DhSSC} [or Eqs.~\eqref{DhRGW}--\eqref{DhSGW} in the Gralla-Wald approximation], which preserve the essential properties of those fields and leave invariant all the governing equations of self-force theory. 
\item a transformation law for the second-order self-force, given by Eq.~\eqref{DFSC} [Eq.~\eqref{DF2} in the Gralla-Wald approximation].
\end{enumerate} 

In the remainder of the discussion, I describe some practical issues related to these results.

\subsection{Working in gauges other than Lorenz}
To clarify point 3, I  note that to make use of the fact that the governing equations are the same in all smoothly related gauges, one must first find an effective metric satisfying them in a particular gauge. And given that multiple effective metrics can satisfy the same governing equations, one must realize that under a gauge transformation, one is referring to the transformation of one's particular choice of effective metric.

Currently, an effective metric satisfying the desired properties through second order has been found only in the Lorenz gauge~\cite{Pound:12a,Pound:12b,Pound-Miller:14}. Furthermore, although a local expansion of the second-order field has been derived in an alternative gauge~\cite{Gralla:12}, it is available in a practical form only in the Lorenz gauge~\cite{Pound-Miller:14}. 

However, with the results of this paper,  in principle one can calculate the field in any gauge smoothly related to Lorenz. One need only apply Eq.~\eqref{DhSSC} to find the correct singular field in the non-Lorenz gauge. Of course, this requires knowing the gauge generator $\xi^\mu_1$ that brings the first-order field from Lorenz to non-Lorenz; depending on how $h^1_{\mu\nu}$ is found, obtaining this gauge generator can be quite difficult.
 
A larger issue is that the restriction to smooth transformations is quite severe. The gauges that are most convenient for explicit computations in a black hole background, the Regge-Wheeler and radiation gauges, are not related to Lorenz by smooth transformations. One might be able to utilize the simplicity of these gauges by working in a mixed gauge: adapting the ideas of Ref.~\cite{Gralla:12}, one might implement a puncture scheme in which the puncture is constructed from the singular field in the Lorenz gauge, but the residual field (the numerical variable in the scheme) is calculated in any desired gauge. The transformation to such a mixed gauge is not perfectly smooth on the worldline, but it is of nonnegative differentiability, since it only alters the gauge of the residual field, not the puncture. 

If one wishes to work in gauges further removed from Lorenz, substantially more analysis is required. In particular, if the gauge generator $\xi_1^\mu$ is not $C^1$ at the worldline, or if $\xi_2^\mu$ is not $C^0$ there, then the transformation law for the worldline is not given by Eq.~\eqref{z_transformation}. This spoils the reasoning used to choose the relation~\eqref{DhRSC} between the regular fields in the two gauges. In that case, one must derive an alternative to Eq.~\eqref{z_transformation} from first principles, and from there write new transformation laws for the singular and regular fields. This analysis has been carried out at first order in $\e$ in recent formulations of self-force theory in singular gauges~\cite{Gralla:11,Pound-Merlin-Barack:14,Shah-Pound:15}, but it remains to be done at second order.
 
\subsection{Gauge and motion on large scales}
Forget, for a moment, the question of which gauge we are using, and simply suppose we have computed the motion and metric in some gauge. We must next consider how the results depend on that choice of gauge. Because the worldline is a gauge-dependent quantity, the answer is not obvious, and we must do some work to disentangle the motion from the gauge.

First consider the Gralla-Wald case. Here it may seem that motion and gauge are entirely indistinguishable, since we can freely adjust each deviation vector $\GW{z}^\mu_n$ with a gauge transformation. Nevertheless, we can distinguish the two notions. Begin in a gauge where the first-order field $\GW{h}^1_{\mu\nu}(x;z_0)$ does not grow large with time; any of the typical gauge choices, such as Lorenz or Regger-Wheeler, will achieve this (putting aside spurious, growing gauge modes~\cite{Dolan-Barack:13}). In this gauge, the deviation vector $\GW{z}_1^\mu$ grows quadratically with time, according to Eq.~\eqref{z1_generic}, which causes the second-order field $\GW{h}^2_{\mu\nu}(x;z_0)$ to likewise grow quadratically with time, as we see in Eq.~\eqref{hs2GW-local}. Now suppose we perform a gauge transformation generated by $\xi_1^\mu=\GW{z}_1^\mu$. This removes the first-order deviation, setting $\GW{z}'^\mu_1=0$ and thereby eliminating the growing term in $\GW{h}'^2_{\mu\nu}(x;z_0)$. But according to Eq.~\eqref{Deltah1GW}, the transformation generically introduces growing terms into $\GW{h}'^1_{\mu\nu}(x;z_0)$. That is, the object's motion away from $z^\mu_0$ induces secular growth in the second-order field; if we force the object to stay on $z^\mu_0$ via a gauge choice, then we instead induce secular growth in the \emph{first-order} field.

This distinction between gauge and motion carries over to the self-consistent case, but the situation becomes more subtle. Here all of the fields $h^n_{\mu\nu}$ are functionals of a gauge-dependent worldline. Even for manifestly physical quantities such as the gravitational waveform, we hence seem to lack invariance. If we calculate, say, the Weyl scalar $\psi_0$ from the linearized field in a particular gauge,the result is invariant under the transformation $h^1_{\mu\nu}\to h^1_{\mu\nu}+\Lie_\xi g_{\mu\nu}$, but it is not invariant under $h^1_{\mu\nu}[z]\to h^1_{\mu\nu}[z']+\Lie_\xi g_{\mu\nu}$.

To resolve this, we may first consider behavior of the approximation on short, order-unity timescales $\sim\e^0$. On these timescales, the difference between $h^1_{\mu\nu}[z]$ and $h^1_{\mu\nu}[z_0]=\GW{h}^1_{\mu\nu}$ is of second order. Hence, the first-order information in $\psi_0$, as constructed from $h^1_{\mu\nu}[z]$, is invariant; there is merely an irrelevantly small bit of gauge-dependence buried within it. Similarly, if we construct a curvature invariant from $h^1_{\mu\nu}[z]$ plus $h^2_{\mu\nu}[z]$, it will be invariant through second order and contain a negligible third-order gauge-dependence. 

However, this description ignores the principal aim of the self-consistent expansion, which is to model long-term evolution. Specifically, look at an EMRI. A single orbit occurs on the timescale $\sim\e^0$, but the inspiral of the orbit occurs on a much longer timescale. Gravitational waves carry away orbital energy at the rate $\dot E/E\sim \e$, implying that the inspiral occurs on the time scale $\sim E/\dot E\sim 1/\e$. Hence, when modeling an EMRI we do not want to work on a fixed, $\e$-independent interval of time; instead, we want to look at the limit $\e\to 0$ in a domain ${\cal D}$ of size $\sim1/\e$ that blows up in the limit. 


Working on an $\e$-dependent domain forces us to revise our thinking about gauge and motion. For our approximation to be sensible, we must require that on the domain we work in, all excluded terms are much smaller than the included ones, and all included ones are well ordered. For example, to claim first-order accuracy, we must have $h^1_{\mu\nu}(x;z)\sim\e^0$ on $\mathcal{D}$, and we must have that all excluded terms are $\ll \e^0$. These criteria strongly restrict the allowed choices of gauge. If, for example, we adopt a gauge in which there is no self-force, then $h^1_{\mu\nu}(x;z)$ grows large with time; this is not an allowed gauge. Although these issues have not been studied in any detail, if we wish to stay within a class of gauges satisfying the above criteria, a natural requirement is to insist that all gauge vectors are uniformly small (and uniformly well ordered, such that $\e^{n+1}\xi^\mu_{n+1}\ll\e^n\xi^\mu_n$) in ${\cal D}$.

This implies that in well-behaved approximations in ${\cal D}$, the effects of the self-force are \emph{not} pure gauge. The accelerated worldline $z^\mu$ eventually deviates from any given geodesic $z_0^\mu$ by a very large amount, while with an allowable gauge transformation we may shift it only by a very small amount. In other words, although the self-forced deviation from $z_0^\mu$ is pure gauge on a domain of size $\sim\e^0$, it is not pure gauge in a domain of size $\sim 1/\e$. Furthermore, since the object is of size $\e$ and the gauge transformation may only shift the object by an amount of order $\e$, within the allowed class of gauges, \emph{the gauge ambiguity in the object's position is of the same order as the size of the object itself}. For the purposes of modeling long-term inspirals, this should not concern us; what concerns us is only the large, long-term changes in the object's position.

Let me summarize the practical relevance of all this. Suppose that in some gauge, one computes a metric perturbation  over a long timescale in the self-consistent approximation, using the self-accelerated orbit as a source. Further suppose that one ends up with a well-behaved approximation, in the sense that $\e^{n+1}h^{n+1}_{\mu\nu}\ll\e^n h^n_{\mu\nu}\ll g_{\mu\nu}$ for all included terms, and that error estimates suggest that all excluded terms are smaller. Then one has included the correct invariant information about the orbit's long-term evolution. One has also included some gauge-dependent information about the orbit, but on all scales, this information is negligibly small compared to the order of accuracy one should expect from one's approximation.

\begin{acknowledgments}
I thank Soichiro Isoyama for a thought-provoking correspondence and Carsten Gundlach for a helpful discussion. This work was supported by the European Research Council under the European Union's Seventh Framework Programme (FP7/2007-2013)/ERC Grant No. 304978. I also acknowledge support from the Natural Sciences and Engineering Research Council of Canada.
\end{acknowledgments}

\appendix

\section{Reparametrization transformations}\label{reparametrization}
Section~\ref{motion_expansions} describes how the coefficients in the expansion 
\beq
z^\mu(s,\e)=\sum_{n\geq0} \e^n \GW{z}^\mu_n(s)
\eeq
depend on the choice of parameter $s$. This appendix works out explicitly how the terms $\GW{z}^\mu_n(s)$ transform under a reparametrization $s\to s'(s,\e)$. 

Let $z'^\mu(s',\e)=z^\mu(s(s',\e),\e)$ be the coordinates of the worldline in the new parametrization. Their expansion reads
\beq
z'^\mu(s',\e)=\sum_{n\geq0} \e^n \GW{z}'^\mu_n(s'),
\eeq
with coefficients $\GW{z}'^\mu_n(s'):=\frac{1}{n!}\frac{\partial^nz'^\mu}{\partial\e^n}(s',0)$. 

We find the relationship between $\GW{z}^\mu_1$ and $\GW{z}'^\mu_1$ by writing
\begin{subequations}
\begin{align}
\frac{\partial z'^\mu(s',\e)}{\partial \e} &= \frac{dz^\mu(s(s',\e),\e)}{d\e} \\
			&= \frac{\partial z^\mu(s,\e)}{\partial\e}+\frac{\partial s(s',\e)}{\partial\e}\frac{\partial z^\mu(s,\e)}{\partial s}.
\end{align}
\end{subequations}
Evaluating at $\e=0$, we have
\beq\label{reparam-1}
\GW{z}'^\mu_1 = \GW{z}^\mu_1 + \dot z^\mu_0\left.\frac{\partial s}{\partial\e}\right|_{\e=0},
\eeq
where $\dot z^\mu_0:=\frac{dz_0^\mu}{ds}$. From this we immediately find that the perpendicular piece of $\GW{z}^\mu_1$, unlike $\GW{z}^\mu_1$ itself, is reparametrization invariant:
\beq
\GW{z}'^\mu_{1\perp} = \GW{z}^\mu_{1\perp}.
 \eeq 

Similarly, we find the relationship between $\GW{z}^\mu_2$ and $\GW{z}'^\mu_2$ by writing
\begin{subequations}
\begin{align}
\frac{\partial^2 z'^\mu(s',\e)}{\partial \e^2} &= \frac{d^2 z^\mu(s(s',\e),\e)}{d\e^2} \\
&=\frac{\partial^2 z^\mu(s,\e)}{\partial\e^2}+\frac{\partial^2 s(s',\e)}{\partial\e^2}\frac{\partial z^\mu(s,\e)}{\partial s}\nonumber\\
&\quad +2\frac{\partial s(s',\e)}{\partial\e}\frac{\partial^2 z^\mu(s,\e)}{\partial\e\partial s}\nonumber\\
&\quad +\left(\!\frac{\partial s(s',\e)}{\partial\e}\!\right)^{\!2}\frac{\partial^2 z^\mu(s,\e)}{\partial s^2}.
\end{align}
\end{subequations}
Evaluating this at $\e=0$ and specializing to a coordinate system that is normal along $\gamma_0$, we find
\begin{align}
\GW{z}'^\mu_{2\rm N} = \GW{z}^\mu_{2\rm N} +\frac{1}{2}\frac{\partial^2s}{\partial\e^2}\frac{dz_0^\mu}{ds}
+ \frac{\partial s}{\partial\e}\frac{D\GW{z}^\mu_1}{ds} 
								+\frac{1}{2}\!\left(\!\frac{\partial s}{\partial\e}\!\right)^{\!2}\!\frac{D^2z_0^\mu}{ds^2},
\end{align}
where the partial derivatives are evaluated at $\e=0$. By projecting orthogonally to $u_{0\mu}$, we find that unlike at first order, the perpendicular deviation $\GW{z}^\mu_{2\rm N\perp}$ is not parametrization independent:
\begin{align}
\GW{z}'^\mu_{2\rm N\perp} = \GW{z}^\mu_{2\rm N\perp} 
+ \frac{\partial s}{\partial\e}\frac{D\GW{z}^\mu_{1\perp}}{ds},
\end{align}
where I have used the geodesic equation $\frac{D^2z_0^\mu}{ds^2}=\kappa\frac{dz_0^\mu}{ds}$. However, after invoking Eq.~\eqref{z2cross-vs-z2perp} and Eq.~\eqref{reparam-1}, we find that the quantity $\GW{z}^\mu_{2\ddagger}$ \emph{is} parametrization independent: 
\beq
\GW{z}'^\mu_{2\ddagger}=\GW{z}^\mu_{2\ddagger}.
\eeq

\section{Identities on $\mathcal{S}$}\label{v-zdot-identities}

In Sec.~\ref{geodesic_expansion_in_h_and_dz}, we require various derivatives of quantities on a surface $\mathcal{S}=\{z^\mu(s,\e)|s\in\mathbb{R},\e\in[0,\infty)\}$. This appendix displays useful identities for that purpose.
 
The surface can be generated by the two tangent vector fields $v^\mu=\frac{\partial z^\mu}{\partial\e}$ and $\dot z^\mu=\frac{\partial z^\mu}{\partial s}$. An important relation between these fields follows from the equality $\mathcal{L}_{\dot z} v^\mu=\frac{\partial^2 x^\mu}{\partial s\partial\e}=\frac{\partial^2 x^\mu}{\partial\e\partial s}=\mathcal{L}_v \dot z^\mu$; since $\mathcal{L}_{\dot z}v^\mu=-\mathcal{L}_v\dot z^\mu$, this implies $\mathcal{L}_{\dot z} v^\mu=0=\mathcal{L}_v \dot z^\mu$ and hence
\begin{equation}\label{commutation}
v^\mu{}_{;\nu}\dot z^\nu = \dot z^\mu{}_{;\nu}v^\nu.
\end{equation}



Using the identity~\eqref{commutation}, we can derive a host of others. It immediately gives us
\beq\label{Liedsdtau}
\Lie_v \frac{ds}{d\tau}=\left(\!\frac{ds}{d\tau}\!\right)^{\!\!\!3} v_{\alpha}{}_{;\beta}\dot{z}^{\alpha} \dot{z}^{\beta},
\eeq
since $\frac{ds}{d\tau}=1/\sqrt{-g_{\mu\nu}\dot z^\mu\dot z^\nu}$. This in turn gives us
\begin{align}
\Lie_v \kappa &= -\left(\! \frac{ds}{d\tau}\!\right)^{\!\!\!2}\left( \ddot{z}^{\alpha} \dot{z}^{\beta} v_{\beta}{}_{;\alpha} +\ddot{z}^{\alpha} \dot v_{\alpha}
 - 2 \kappa \dot{z}^{\alpha} \dot v_{\alpha} \right.\nonumber\\&\quad\left.+ \dot{z}^{\alpha} \dot{z}^{\beta} \dot{z}^{\gamma} v_{\alpha}{}_{;\beta\gamma}\right),\label{Liekappa}
\end{align}
where $\kappa = -\dot z^\alpha \partial_\alpha\ln\frac{ds}{d\tau}$.

Similarly,
\begin{align}
\Lie_vu^\mu &= u^\mu u_\nu \frac{Dv^\nu}{d\tau},\label{Lievu} \\
v^\nu\nabla_\nu u^\mu &= P^\mu{}_\nu \frac{Dv^\nu}{d\tau},\label{vDu}
\end{align}
where $u^\mu=\frac{dz^\mu}{d\tau}=\frac{ds}{d\tau}\dot z^\mu$, and
\begin{subequations}\label{vDzddot}
\begin{align}
v^\nu \nabla_\nu \ddot z^\alpha &= \left(\dot z^\alpha{}_{;\beta}\dot z^\beta\right)_{;\gamma}v^\gamma\\
							&= \ddot v^\alpha+R^\alpha{}_{\mu\beta\nu}\dot z^\mu v^\beta \dot z^\nu,
\end{align}
\end{subequations}
where the second line follows from Eq.~\eqref{commutation} and the Ricci identity.

\section{Identities for gauge transformations of curvature tensors}\label{gauge-identities}
Let $A[g]$ be a tensor of any rank constructed from a metric $g$. (To streamline the presentation, I adopt index-free notation throughout most of this appendix.) Now define
\begin{align}
\delta^n& A[f_1,\ldots,f_n] := \nonumber\\
			&\frac{1}{n!}\frac{d^n}{d\lambda_1\cdots d\lambda_n}A[g+\lambda_1 f_1 +\cdots + \lambda_n f_n]\big|_{\lambda_i=0},
\end{align}
where ``$\lambda_i=0$'' stands for evaluation at $\lambda_i=\cdots=\lambda_n=0$. The tensor-valued functional $\delta^n A[f_1,\ldots,f_n]$ is linear in each of its arguments $f_1,\ldots,f_n$; it is also symmetric in them.  In the case that all the arguments are the same, we have $\delta^n A[h,\ldots,h] = \frac{1}{n!}\frac{d^n}{d\lambda^n}A[g+\lambda h]\big|_{\lambda=0}$, the piece of $A[g+h]$ containing precisely $n$ factors of $h$ and its derivatives. 

The following identities are easily proved (and will be proved below) by writing Lie derivatives as ordinary derivatives: 
\begin{align}
\Lie_\xi A[g] &=\delta A[\Lie_\xi g] , \label{Lie A}\\
\tfrac{1}{2}\Lie^2_\xi A[g] &=\tfrac{1}{2}\delta A[\Lie^2_\xi g] + \delta^2A[\Lie_\xi g,\Lie_\xi g] ,\label{Lie2 A}\\
\Lie_\xi\delta A[h] &= \delta A[\Lie_\xi h] + 2\delta^2 A[\Lie_\xi g, h],\label{Lie dA}
\end{align}
where we can also write $\delta^2 A[\Lie_\xi g,h]= \delta^2 A[h,\Lie_\xi g]=\frac{1}{2}\left(\delta^2 A[h,\Lie_\xi g]+\delta^2 A[\Lie_\xi g,h]\right)$. As an example, if $A$ is the Ricci tensor, then
\begin{align}
\Lie_\xi R_{\mu\nu}[g] &=\delta R_{\mu\nu}[\Lie_\xi g], \label{Lie R}\\
\tfrac{1}{2}\Lie^2_\xi R_{\mu\nu}[g] &=\tfrac{1}{2}\delta R_{\mu\nu}[\Lie^2_\xi g] + \delta^2R_{\mu\nu}[\Lie_\xi g,\Lie_\xi g] ,\label{Lie2 R}\\
\Lie_\xi\delta R_{\mu\nu}[h] &= \delta R_{\mu\nu}[\Lie_\xi h] + 2\delta^2 R_{\mu\nu}[h,\Lie_\xi g],\label{Lie dR}
\end{align}
where I have restored indices to avoid confusion with the Ricci scalar (though the same equation, with indices removed, would apply to the Ricci scalar). 

Before moving to the proof of these results, I note one of their consequences: when examining perturbations of a curvature tensor, we can derive transformation laws in two equally natural ways: directly from Eq.~\eqref{DeltaA} or from the transformations of the metric perturbations. For example, in a vacuum background, Eq.~\eqref{DeltaA} directly implies
\begin{align}
\Delta \delta R_{\mu\nu}[h^1] &= \Lie_{\xi_1}R_{\mu\nu}[g]=0,\label{DEFE1}\\
\Delta (\delta R_{\mu\nu}[h^2]+\delta^2 R_{\mu\nu}[h^1,h^1]) &= \Lie_\xi \delta R_{\mu\nu}[h^1];\label{DEFE2}
\end{align}
or the same equations can be found by instead applying  Eq.~\eqref{DeltaA} to the metric itself, writing
\begin{subequations}
\begin{align}\label{DEFE2-steps}
\Delta (\delta &R_{\mu\nu}[h^2]+\delta^2 R_{\mu\nu}[h^1,h^1])\nonumber\\
		&= \delta R_{\mu\nu}[h'^2]+\delta^2 R_{\mu\nu}[h'^1,h'^1] \nonumber\\
		&\quad- (\delta R_{\mu\nu}[h^2]+\delta^2 R_{\mu\nu}[h^1,h^1]) \\
		&= \delta R_{\mu\nu}[h^2\!+\!\Lie_{\xi_2}g\!+\!\tfrac{1}{2}\Lie^2_{\xi_1}g\!+\!\Lie_{\xi_1}h^1]\nonumber\\
		&\quad+\delta^2 R_{\mu\nu}[h^1\!+\!\Lie_{\xi_1}g,h^1\!+\!\Lie_{\xi_1}g]\\
		&=  \delta R_{\mu\nu}[\Lie_{\xi_2}g]+\tfrac{1}{2}\delta R_{\mu\nu}[\Lie^2_{\xi_1}g]+\delta R_{\mu\nu}[\Lie_{\xi_1}h^1]\nonumber\\
		&\quad	+2 \delta^2 R_{\mu\nu}[h^1,\Lie_{\xi_1}g]+\delta^2 R_{\mu\nu}[\Lie_{\xi_1}g,\Lie_{\xi_1}g]
\end{align}
\end{subequations}
and then applying Eqs.~\eqref{Lie R}--\eqref{Lie dR}.

Now, let us return to the proofs. To establish Eq.~\eqref{Lie A}, one can write the metric as a function of a parameter $\lambda$ along the flow generated by $\xi$ and then perform a Taylor expansion:
\begin{subequations}
\begin{align}
\Lie_\xi A[g] &= \frac{d}{d\lambda}A\!\left[g(0)+\lambda\frac{dg}{d\lambda}\Big|_{\lambda=0}\right]\!\!\bigg|_{\lambda=0}\\ 
&= \delta A\left[\frac{dg}{d\lambda}\big|_{\lambda=0}\right]\\
& = \delta A[\Lie_\xi g].
\end{align}
\end{subequations}
Similarly, to establish Eq.~\eqref{Lie2 A}, one can write
\begin{subequations}
\begin{align}
\tfrac{1}{2}\Lie^2_\xi A[g] &= \tfrac{1}{2}\frac{d^2}{d\lambda^2} A\!\left[g(0)+\lambda\frac{dg}{d\lambda}\Big|_{\lambda=0}+\tfrac{1}{2}\lambda^2\frac{dg}{d\lambda^2}\Big|_{\lambda=0}\right]\!\!\bigg|_{\lambda=0}\nonumber\\
&= \tfrac{1}{2}\delta A[\Lie^2_\xi g] + \delta^2A[\Lie_\xi g,\Lie_\xi g] ,
\end{align}
\end{subequations}
and to establish Eq.~\eqref{Lie dA}, one can write $g$ as a function of parameters $(\lambda,\e)$ along commuting flows, where $h:= \frac{dg}{d\e}\big|_{\e=0}$, and then write
\begin{subequations}
\begin{align}
\Lie_\xi\delta A[h] &= \frac{d^2}{d\lambda d\e}A\!\left[g(\lambda,0)+\e \frac{d g}{d\epsilon}(\lambda,0)\right]\!\!\bigg|_{\lambda=\e=0}\\
&=  \frac{d^2}{d\lambda d\e}A\!\left[g(0,0)+\lambda \frac{d g}{d\lambda}(0,0) + \e \frac{d g}{d\e}(0,0)\right.\nonumber\\
&\qquad\qquad\left.+\lambda \e \frac{d^2 g}{d\lambda d\e}(0,0)\right]\!\!\bigg|_{\lambda=\e=0}\\
&= \delta A[\Lie_\xi h] + 2\delta^2 A[h,\Lie_\xi g].
\end{align}
\end{subequations}







\section{Gauge transformation of the second-order stress-energy}\label{Delta T2}
According to Eq.~\eqref{DT}, the gauge transformation of the (trace-reversed) second-order stress-energy, $\Delta \bar T^2_{\mu\nu}$, is given by Lie derivatives of the first-order stress-energy. This appendix evaluates those derivatives.

The calculation can be performed following Sec.~\ref{T1_expansion}. For any vector $\xi_1^\mu$, the Lie derivative $\varLie_{\xi_1} T_1^{\mu\nu}(x;z)$ is given by Eq.~\eqref{dT_zperp} with the replacements $\GW{z}_1^\mu\to\xi_1^\mu$ and $u^\mu_0\to u^\mu$. The ordinary Lie derivative $\Lie_{\xi_1}T^{\mu\nu}_1$ can be evaluated by following very similar steps as were used to derive Eq.~\eqref{dT_zperp}, leading to
\begin{align}
\Lie_{\xi_1} T^{\alpha\beta}_1 &= -m\!\int_\gamma\! g^\alpha_{\alpha'}g^\beta_{\beta'}
					\bigg\{2u^{(\alpha'}\frac{D\xi_{1\perp}^{\beta')}}{d\tau'}\delta(x,z)\nonumber\\
					&\quad +u^{\alpha'}u^{\beta'}\left(\frac{d\xi_{1\parallel}}{d\tau}+\xi^{\rho'}_1{}_{;\rho'}\right)\delta(x,z)\nonumber\\
					&\quad-u^{\alpha'}u^{\beta'}\xi_{1\perp}^{\gamma'}g^\gamma_{\gamma'}\nabla_{\gamma}\delta(x,z)\bigg\}d\tau,\label{LieT}
\end{align}
where $\xi_{1\perp}^{\beta'}:= P^{\beta'}{}_{\alpha'}\xi_1^{\alpha'}$ and $\xi_{1\parallel}:= u_{\mu'}\xi_1^{\mu'}$. The total of the two Lie derivatives yields the simple result
\begin{align}\label{LieT+varLieT}
(\Lie_\xi+\varLie_\xi)T^{\mu\nu}_1 &= -m\int g^\mu_{\mu'}g^\nu_{\nu'}u^{\mu'}u^{\nu'}\nonumber\\
						&\quad \times\left(\frac{d\xi_{1\parallel}}{d\tau}+\xi^{\rho'}_1{}_{;\rho'}\right)\delta(x,z)d\tau.
\end{align}
The $\pm\xi_1^\mu\nabla_\mu\delta(x,z)$ terms in Eqs.~\eqref{LieT} and \eqref{dT_zperp} signal that the mass $m$ is displaced from $z^\mu$ by an amount $\pm\xi_1^\mu$; the lack of any such term in Eq.~\eqref{LieT+varLieT} signals that the displacements due to the two derivatives precisely cancel one another, leaving the mass $m$ moving on $z^\mu$.

Equation~\eqref{DT} involves derivatives of $\bar T^1_{\mu\nu}$, not of $T^{\mu\nu}_1$. So we have
\begin{align}
\Delta\bar T^2_{\mu\nu} &= (g_{\mu\alpha}g_{\nu\beta}\!-\!\tfrac{1}{2}g_{\mu\nu}g_{\alpha\beta})(\Lie_{\xi_1}\!+\!\varLie_{\xi_1})T_1^{\alpha\beta} 	\!\!
	+ 2\big(\xi^1_{(\mu;\alpha)}g_{\nu\beta}\nonumber\\
				&\quad+g_{\mu\alpha}\xi^1_{(\nu;\beta)} -\tfrac{1}{2}\xi^1_{(\mu;\nu)}g_{\alpha\beta} -\tfrac{1}{2}g_{\mu\nu}\xi^1_{\alpha;\beta}\big)T_1^{\alpha\beta},\!\!
\end{align}
with $(\Lie_{\xi_1}+\varLie_{\xi_1})T_1^{\alpha\beta}$ given by Eq.~\eqref{LieT+varLieT}.

These are the results for the self-consistent case. For the Gralla-Wald case, we simply drop the $\varLie_{\xi_1}$ term and set $z^\mu=z^\mu_0$:
\begin{align}
\Delta\GW{\bar T}^2_{\mu\nu} &= (g_{\mu\alpha}g_{\nu\beta}-\tfrac{1}{2}g_{\mu\nu}g_{\alpha\beta})\Lie_{\xi_1}\GW{T}_1^{\alpha\beta} 	
	+ 2\big(\xi^1_{(\mu;\alpha)}g_{\nu\beta}\nonumber\\
		&\quad+g_{\mu\alpha}\xi^1_{(\nu;\beta)}\! -\!\tfrac{1}{2}\xi^1_{(\mu;\nu)}g_{\alpha\beta}\! -\!\tfrac{1}{2}g_{\mu\nu}\xi^1_{\alpha;\beta}\big)\GW{T}_1^{\alpha\beta},
\end{align}
where $\Lie_{\xi_1}\GW{T}_1^{\alpha\beta}$ is given by Eq.~\eqref{LieT} with $z^\mu\to z_0^\mu$. Unlike in the self-consistent case, where the $\nabla_\mu\delta(x,z)$ terms cancelled in the final result, in the Gralla-Wald case there is a term $\propto\xi^\mu_{1\perp}\nabla_\mu\delta(x,z)$, corresponding to the center of mass having been displaced by an amount $\Delta z_1^\mu=-\xi^\mu_{1\perp}$ relative to $z_0^\mu$.

\bibliography{../bibfile}

\begin{thebibliography}{37}%
\makeatletter
\providecommand \@ifxundefined [1]{%
 \@ifx{#1\undefined}
}%
\providecommand \@ifnum [1]{%
 \ifnum #1\expandafter \@firstoftwo
 \else \expandafter \@secondoftwo
 \fi
}%
\providecommand \@ifx [1]{%
 \ifx #1\expandafter \@firstoftwo
 \else \expandafter \@secondoftwo
 \fi
}%
\providecommand \natexlab [1]{#1}%
\providecommand \enquote  [1]{``#1''}%
\providecommand \bibnamefont  [1]{#1}%
\providecommand \bibfnamefont [1]{#1}%
\providecommand \citenamefont [1]{#1}%
\providecommand \href@noop [0]{\@secondoftwo}%
\providecommand \href [0]{\begingroup \@sanitize@url \@href}%
\providecommand \@href[1]{\@@startlink{#1}\@@href}%
\providecommand \@@href[1]{\endgroup#1\@@endlink}%
\providecommand \@sanitize@url [0]{\catcode `\\12\catcode `\$12\catcode
  `\&12\catcode `\#12\catcode `\^12\catcode `\_12\catcode `\%12\relax}%
\providecommand \@@startlink[1]{}%
\providecommand \@@endlink[0]{}%
\providecommand \url  [0]{\begingroup\@sanitize@url \@url }%
\providecommand \@url [1]{\endgroup\@href {#1}{\urlprefix }}%
\providecommand \urlprefix  [0]{URL }%
\providecommand \Eprint [0]{\href }%
\providecommand \doibase [0]{http://dx.doi.org/}%
\providecommand \selectlanguage [0]{\@gobble}%
\providecommand \bibinfo  [0]{\@secondoftwo}%
\providecommand \bibfield  [0]{\@secondoftwo}%
\providecommand \translation [1]{[#1]}%
\providecommand \BibitemOpen [0]{}%
\providecommand \bibitemStop [0]{}%
\providecommand \bibitemNoStop [0]{.\EOS\space}%
\providecommand \EOS [0]{\spacefactor3000\relax}%
\providecommand \BibitemShut  [1]{\csname bibitem#1\endcsname}%
\let\auto@bib@innerbib\@empty
\bibitem [{\citenamefont {Pound}(2012{\natexlab{a}})}]{Pound:12a}%
  \BibitemOpen
  \bibfield  {author} {\bibinfo {author} {\bibfnamefont {A.}~\bibnamefont
  {Pound}},\ }\href {\doibase 10.1103/PhysRevLett.109.051101} {\bibfield
  {journal} {\bibinfo  {journal} {Phys. Rev. Lett.}\ }\textbf {\bibinfo
  {volume} {109}},\ \bibinfo {pages} {051101} (\bibinfo {year}
  {2012}{\natexlab{a}})},\ \Eprint {http://arxiv.org/abs/1201.5089}
  {arXiv:1201.5089 [gr-qc]} \BibitemShut {NoStop}%
\bibitem [{\citenamefont {Mino}\ \emph {et~al.}(1997)\citenamefont {Mino},
  \citenamefont {Sasaki},\ and\ \citenamefont
  {Tanaka}}]{Mino-Sasaki-Tanaka:97}%
  \BibitemOpen
  \bibfield  {author} {\bibinfo {author} {\bibfnamefont {Y.}~\bibnamefont
  {Mino}}, \bibinfo {author} {\bibfnamefont {M.}~\bibnamefont {Sasaki}}, \ and\
  \bibinfo {author} {\bibfnamefont {T.}~\bibnamefont {Tanaka}},\ }\href@noop {}
  {\bibfield  {journal} {\bibinfo  {journal} {Phys. Rev. D}\ }\textbf {\bibinfo
  {volume} {55}},\ \bibinfo {pages} {3457} (\bibinfo {year}
  {1997})}\BibitemShut {NoStop}%
\bibitem [{\citenamefont {Quinn}\ and\ \citenamefont
  {Wald}(1997)}]{Quinn-Wald:97}%
  \BibitemOpen
  \bibfield  {author} {\bibinfo {author} {\bibfnamefont {T.~C.}\ \bibnamefont
  {Quinn}}\ and\ \bibinfo {author} {\bibfnamefont {R.~M.}\ \bibnamefont
  {Wald}},\ }\href@noop {} {\bibfield  {journal} {\bibinfo  {journal} {Phys.
  Rev. D}\ }\textbf {\bibinfo {volume} {56}},\ \bibinfo {pages} {3381}
  (\bibinfo {year} {1997})}\BibitemShut {NoStop}%
\bibitem [{\citenamefont {Detweiler}\ and\ \citenamefont
  {Whiting}(2003)}]{Detweiler-Whiting:03}%
  \BibitemOpen
  \bibfield  {author} {\bibinfo {author} {\bibfnamefont {S.~L.}\ \bibnamefont
  {Detweiler}}\ and\ \bibinfo {author} {\bibfnamefont {B.~F.}\ \bibnamefont
  {Whiting}},\ }\href@noop {} {\bibfield  {journal} {\bibinfo  {journal} {Phys.
  Rev. D}\ }\textbf {\bibinfo {volume} {67}},\ \bibinfo {pages} {024025}
  (\bibinfo {year} {2003})}\BibitemShut {NoStop}%
\bibitem [{\citenamefont {Gralla}\ and\ \citenamefont
  {Wald}(2008)}]{Gralla-Wald:08}%
  \BibitemOpen
  \bibfield  {author} {\bibinfo {author} {\bibfnamefont {S.~E.}\ \bibnamefont
  {Gralla}}\ and\ \bibinfo {author} {\bibfnamefont {R.~M.}\ \bibnamefont
  {Wald}},\ }\href@noop {} {\bibfield  {journal} {\bibinfo  {journal} {Class.
  Quant. Grav.}\ }\textbf {\bibinfo {volume} {25}},\ \bibinfo {pages} {205009}
  (\bibinfo {year} {2008})}\BibitemShut {NoStop}%
\bibitem [{\citenamefont {Pound}(2010{\natexlab{a}})}]{Pound:10a}%
  \BibitemOpen
  \bibfield  {author} {\bibinfo {author} {\bibfnamefont {A.}~\bibnamefont
  {Pound}},\ }\href@noop {} {\bibfield  {journal} {\bibinfo  {journal} {Phys.
  Rev. D}\ }\textbf {\bibinfo {volume} {81}},\ \bibinfo {pages} {024023}
  (\bibinfo {year} {2010}{\natexlab{a}})}\BibitemShut {NoStop}%
\bibitem [{\citenamefont {Pound}(2012{\natexlab{b}})}]{Pound:12b}%
  \BibitemOpen
  \bibfield  {author} {\bibinfo {author} {\bibfnamefont {A.}~\bibnamefont
  {Pound}},\ }\href {\doibase 10.1103/PhysRevD.86.084019} {\bibfield  {journal}
  {\bibinfo  {journal} {Phys. Rev. D}\ }\textbf {\bibinfo {volume} {86}},\
  \bibinfo {pages} {084019} (\bibinfo {year} {2012}{\natexlab{b}})},\ \Eprint
  {http://arxiv.org/abs/1206.6538} {arXiv:1206.6538 [gr-qc]} \BibitemShut
  {NoStop}%
\bibitem [{\citenamefont {Gralla}(2012)}]{Gralla:12}%
  \BibitemOpen
  \bibfield  {author} {\bibinfo {author} {\bibfnamefont {S.~E.}\ \bibnamefont
  {Gralla}},\ }\href {\doibase 10.1103/PhysRevD.85.124011} {\bibfield
  {journal} {\bibinfo  {journal} {Phys. Rev. D}\ }\textbf {\bibinfo {volume}
  {85}},\ \bibinfo {pages} {124011} (\bibinfo {year} {2012})},\ \Eprint
  {http://arxiv.org/abs/1203.3189} {arXiv:1203.3189 [gr-qc]} \BibitemShut
  {NoStop}%
\bibitem [{\citenamefont {Harte}(2012)}]{Harte:12}%
  \BibitemOpen
  \bibfield  {author} {\bibinfo {author} {\bibfnamefont {A.~I.}\ \bibnamefont
  {Harte}},\ }\href {\doibase 10.1088/0264-9381/29/5/055012} {\bibfield
  {journal} {\bibinfo  {journal} {Class. Quant. Grav.}\ }\textbf {\bibinfo
  {volume} {29}},\ \bibinfo {pages} {055012} (\bibinfo {year} {2012})},\
  \Eprint {http://arxiv.org/abs/1103.0543} {arXiv:1103.0543 [gr-qc]}
  \BibitemShut {NoStop}%
\bibitem [{\citenamefont {Poisson}\ \emph {et~al.}(2011)\citenamefont
  {Poisson}, \citenamefont {Pound},\ and\ \citenamefont
  {Vega}}]{Poisson-Pound-Vega:11}%
  \BibitemOpen
  \bibfield  {author} {\bibinfo {author} {\bibfnamefont {E.}~\bibnamefont
  {Poisson}}, \bibinfo {author} {\bibfnamefont {A.}~\bibnamefont {Pound}}, \
  and\ \bibinfo {author} {\bibfnamefont {I.}~\bibnamefont {Vega}},\ }\href
  {http://www.livingreviews.org/lrr-2011-7} {\bibfield  {journal} {\bibinfo
  {journal} {Living Rev. Relativity}\ }\textbf {\bibinfo {volume} {14}},\
  \bibinfo {pages} {7} (\bibinfo {year} {2011})}\BibitemShut {NoStop}%
\bibitem [{\citenamefont {Pound}(2015)}]{Pound:15a}%
  \BibitemOpen
  \bibfield  {author} {\bibinfo {author} {\bibfnamefont {A.}~\bibnamefont
  {Pound}},\ }\enquote {\bibinfo {title} {Motion of small bodies in curved
  spacetimes: an introduction to gravitational self-force},}\ in\ \href@noop {}
  {\emph {\bibinfo {booktitle} {Equations of Motion in Relativistic
  Gravity}}},\ \bibinfo {series} {Fundamental Theories of Physics}, Vol.\
  \bibinfo {volume} {179},\ \bibinfo {editor} {edited by\ \bibinfo {editor}
  {\bibfnamefont {D.}~\bibnamefont {Puetzfeld}}, \bibinfo {editor}
  {\bibfnamefont {C.}~\bibnamefont {Lammerzahl}}, \ and\ \bibinfo {editor}
  {\bibfnamefont {B.}~\bibnamefont {Schutz}}}\ (\bibinfo  {publisher}
  {Springer},\ \bibinfo {year} {2015})\BibitemShut {NoStop}%
\bibitem [{\citenamefont {Mino}(2005)}]{Mino:05}%
  \BibitemOpen
  \bibfield  {author} {\bibinfo {author} {\bibfnamefont {Y.}~\bibnamefont
  {Mino}},\ }\href {\doibase 10.1143/PTP.113.733} {\bibfield  {journal}
  {\bibinfo  {journal} {Prog. Theor. Phys.}\ }\textbf {\bibinfo {volume}
  {113}},\ \bibinfo {pages} {733} (\bibinfo {year} {2005})},\ \Eprint
  {http://arxiv.org/abs/gr-qc/0506003} {arXiv:gr-qc/0506003 [gr-qc]}
  \BibitemShut {NoStop}%
\bibitem [{\citenamefont {Pound}\ and\ \citenamefont
  {Poisson}(2008)}]{Pound-Poisson:08a}%
  \BibitemOpen
  \bibfield  {author} {\bibinfo {author} {\bibfnamefont {A.}~\bibnamefont
  {Pound}}\ and\ \bibinfo {author} {\bibfnamefont {E.}~\bibnamefont
  {Poisson}},\ }\href {\doibase 10.1103/PhysRevD.77.044013} {\bibfield
  {journal} {\bibinfo  {journal} {Phys. Rev. D}\ }\textbf {\bibinfo {volume}
  {77}},\ \bibinfo {pages} {044013} (\bibinfo {year} {2008})},\ \Eprint
  {http://arxiv.org/abs/0708.3033} {arXiv:0708.3033 [gr-qc]} \BibitemShut
  {NoStop}%
\bibitem [{\citenamefont {Warburton}\ \emph {et~al.}(2012)\citenamefont
  {Warburton}, \citenamefont {Akcay}, \citenamefont {Barack}, \citenamefont
  {Gair},\ and\ \citenamefont {Sago}}]{Warburton-etal:12}%
  \BibitemOpen
  \bibfield  {author} {\bibinfo {author} {\bibfnamefont {N.}~\bibnamefont
  {Warburton}}, \bibinfo {author} {\bibfnamefont {S.}~\bibnamefont {Akcay}},
  \bibinfo {author} {\bibfnamefont {L.}~\bibnamefont {Barack}}, \bibinfo
  {author} {\bibfnamefont {J.~R.}\ \bibnamefont {Gair}}, \ and\ \bibinfo
  {author} {\bibfnamefont {N.}~\bibnamefont {Sago}},\ }\href {\doibase
  10.1103/PhysRevD.85.061501} {\bibfield  {journal} {\bibinfo  {journal} {Phys.
  Rev. D}\ }\textbf {\bibinfo {volume} {85}},\ \bibinfo {pages} {061501}
  (\bibinfo {year} {2012})},\ \Eprint {http://arxiv.org/abs/1111.6908}
  {arXiv:1111.6908 [gr-qc]} \BibitemShut {NoStop}%
\bibitem [{\citenamefont {Hinderer}\ and\ \citenamefont
  {Flanagan}(2008)}]{Hinderer-Flanagan:08}%
  \BibitemOpen
  \bibfield  {author} {\bibinfo {author} {\bibfnamefont {T.}~\bibnamefont
  {Hinderer}}\ and\ \bibinfo {author} {\bibfnamefont {E.~E.}\ \bibnamefont
  {Flanagan}},\ }\href@noop {} {\bibfield  {journal} {\bibinfo  {journal}
  {Phys. Rev. D}\ }\textbf {\bibinfo {volume} {78}},\ \bibinfo {pages} {064028}
  (\bibinfo {year} {2008})}\BibitemShut {NoStop}%
\bibitem [{\citenamefont {Pound}(2014)}]{Pound:14c}%
  \BibitemOpen
  \bibfield  {author} {\bibinfo {author} {\bibfnamefont {A.}~\bibnamefont
  {Pound}},\ }\href@noop {} {\bibfield  {journal} {\bibinfo  {journal} {Phys.
  Rev. D}\ }\textbf {\bibinfo {volume} {90}},\ \bibinfo {pages} {084039}
  (\bibinfo {year} {2014})},\ \Eprint {http://arxiv.org/abs/1404.1543}
  {arXiv:1404.1543 [gr-qc]} \BibitemShut {NoStop}%
\bibitem [{\citenamefont {Barack}\ and\ \citenamefont
  {Ori}(2001)}]{Barack-Ori:01}%
  \BibitemOpen
  \bibfield  {author} {\bibinfo {author} {\bibfnamefont {L.}~\bibnamefont
  {Barack}}\ and\ \bibinfo {author} {\bibfnamefont {A.}~\bibnamefont {Ori}},\
  }\href@noop {} {\bibfield  {journal} {\bibinfo  {journal} {Phys. Rev. D}\
  }\textbf {\bibinfo {volume} {64}},\ \bibinfo {pages} {124003} (\bibinfo
  {year} {2001})}\BibitemShut {NoStop}%
\bibitem [{\citenamefont {Pound}(2010{\natexlab{b}})}]{Pound:10b}%
  \BibitemOpen
  \bibfield  {author} {\bibinfo {author} {\bibfnamefont {A.}~\bibnamefont
  {Pound}},\ }\href@noop {} {\bibfield  {journal} {\bibinfo  {journal} {Phys.
  Rev. D}\ }\textbf {\bibinfo {volume} {81}},\ \bibinfo {pages} {124009}
  (\bibinfo {year} {2010}{\natexlab{b}})}\BibitemShut {NoStop}%
\bibitem [{\citenamefont {Gralla}(2011)}]{Gralla:11}%
  \BibitemOpen
  \bibfield  {author} {\bibinfo {author} {\bibfnamefont {S.~E.}\ \bibnamefont
  {Gralla}},\ }\href {\doibase 10.1103/PhysRevD.84.084050} {\bibfield
  {journal} {\bibinfo  {journal} {Phys. Rev. D}\ }\textbf {\bibinfo {volume}
  {84}},\ \bibinfo {pages} {084050} (\bibinfo {year} {2011})},\ \Eprint
  {http://arxiv.org/abs/1104.5635} {arXiv:1104.5635 [gr-qc]} \BibitemShut
  {NoStop}%
\bibitem [{\citenamefont {Pound}\ \emph {et~al.}(2014)\citenamefont {Pound},
  \citenamefont {Merlin},\ and\ \citenamefont
  {Barack}}]{Pound-Merlin-Barack:14}%
  \BibitemOpen
  \bibfield  {author} {\bibinfo {author} {\bibfnamefont {A.}~\bibnamefont
  {Pound}}, \bibinfo {author} {\bibfnamefont {C.}~\bibnamefont {Merlin}}, \
  and\ \bibinfo {author} {\bibfnamefont {L.}~\bibnamefont {Barack}},\ }\href
  {\doibase 10.1103/PhysRevD.89.024009} {\bibfield  {journal} {\bibinfo
  {journal} {Phys. Rev. D}\ }\textbf {\bibinfo {volume} {89}},\ \bibinfo
  {pages} {024009} (\bibinfo {year} {2014})},\ \Eprint
  {http://arxiv.org/abs/1310.1513} {arXiv:1310.1513 [gr-qc]} \BibitemShut
  {NoStop}%
\bibitem [{\citenamefont {Shah}\ and\ \citenamefont
  {Pound}(2015)}]{Shah-Pound:15}%
  \BibitemOpen
  \bibfield  {author} {\bibinfo {author} {\bibfnamefont {A.~G.}\ \bibnamefont
  {Shah}}\ and\ \bibinfo {author} {\bibfnamefont {A.}~\bibnamefont {Pound}},\
  }\href {\doibase 10.1103/PhysRevD.91.124022} {\bibfield  {journal} {\bibinfo
  {journal} {Phys. Rev. D}\ }\textbf {\bibinfo {volume} {91}},\ \bibinfo
  {pages} {124022} (\bibinfo {year} {2015})},\ \Eprint
  {http://arxiv.org/abs/1503.02414} {arXiv:1503.02414 [gr-qc]} \BibitemShut
  {NoStop}%
\bibitem [{\citenamefont {Barack}(2009)}]{Barack:09}%
  \BibitemOpen
  \bibfield  {author} {\bibinfo {author} {\bibfnamefont {L.}~\bibnamefont
  {Barack}},\ }\href {\doibase 10.1088/0264-9381/26/21/213001} {\bibfield
  {journal} {\bibinfo  {journal} {Class. Quant. Grav.}\ }\textbf {\bibinfo
  {volume} {26}},\ \bibinfo {pages} {213001} (\bibinfo {year} {2009})},\
  \Eprint {http://arxiv.org/abs/0908.1664} {arXiv:0908.1664 [gr-qc]}
  \BibitemShut {NoStop}%
\bibitem [{\citenamefont {Amaro-Seoane}\ \emph {et~al.}(2015)\citenamefont
  {Amaro-Seoane}, \citenamefont {Gair}, \citenamefont {Pound}, \citenamefont
  {Hughes},\ and\ \citenamefont {Sopuerta}}]{Amaro-Seoane-etal:14}%
  \BibitemOpen
  \bibfield  {author} {\bibinfo {author} {\bibfnamefont {P.}~\bibnamefont
  {Amaro-Seoane}}, \bibinfo {author} {\bibfnamefont {J.~R.}\ \bibnamefont
  {Gair}}, \bibinfo {author} {\bibfnamefont {A.}~\bibnamefont {Pound}},
  \bibinfo {author} {\bibfnamefont {S.~A.}\ \bibnamefont {Hughes}}, \ and\
  \bibinfo {author} {\bibfnamefont {C.~F.}\ \bibnamefont {Sopuerta}},\ }\href
  {\doibase 10.1088/1742-6596/610/1/012002} {\bibfield  {journal} {\bibinfo
  {journal} {J. Phys. Conf. Ser.}\ }\textbf {\bibinfo {volume} {610}},\
  \bibinfo {pages} {012002} (\bibinfo {year} {2015})},\ \Eprint
  {http://arxiv.org/abs/1410.0958} {arXiv:1410.0958 [astro-ph.CO]} \BibitemShut
  {NoStop}%
\bibitem [{\citenamefont {Pound}\ and\ \citenamefont
  {Miller}(2014)}]{Pound-Miller:14}%
  \BibitemOpen
  \bibfield  {author} {\bibinfo {author} {\bibfnamefont {A.}~\bibnamefont
  {Pound}}\ and\ \bibinfo {author} {\bibfnamefont {J.}~\bibnamefont {Miller}},\
  }\href {\doibase 10.1103/PhysRevD.89.104020} {\bibfield  {journal} {\bibinfo
  {journal} {Phys. Rev. D}\ }\textbf {\bibinfo {volume} {89}},\ \bibinfo
  {pages} {104020} (\bibinfo {year} {2014})},\ \Eprint
  {http://arxiv.org/abs/1403.1843} {arXiv:1403.1843 [gr-qc]} \BibitemShut
  {NoStop}%
\bibitem [{\citenamefont {Blanchet}(2014)}]{Blanchet:14}%
  \BibitemOpen
  \bibfield  {author} {\bibinfo {author} {\bibfnamefont {L.}~\bibnamefont
  {Blanchet}},\ }\href {http://www.livingreviews.org/lrr-2014-2} {\bibfield
  {journal} {\bibinfo  {journal} {Living Reviews in Relativity}\ }\textbf
  {\bibinfo {volume} {17}},\ \bibinfo {pages} {2} (\bibinfo {year}
  {2014})}\BibitemShut {NoStop}%
\bibitem [{\citenamefont {Poisson}\ and\ \citenamefont
  {Will}(2014)}]{Poisson-Will:14}%
  \BibitemOpen
  \bibfield  {author} {\bibinfo {author} {\bibfnamefont {E.}~\bibnamefont
  {Poisson}}\ and\ \bibinfo {author} {\bibfnamefont {C.~M.}\ \bibnamefont
  {Will}},\ }\href@noop {} {\emph {\bibinfo {title} {Gravity: {N}ewtonian,
  {P}ost-{N}ewtonian, and {R}elativistic}}}\ (\bibinfo  {publisher} {Cambridge
  University Press},\ \bibinfo {address} {Cambridge, United Kingdom},\ \bibinfo
  {year} {2014})\BibitemShut {NoStop}%
\bibitem [{\citenamefont {Barack}\ and\ \citenamefont
  {Golbourn}(2007)}]{Barack-Golbourn:07}%
  \BibitemOpen
  \bibfield  {author} {\bibinfo {author} {\bibfnamefont {L.}~\bibnamefont
  {Barack}}\ and\ \bibinfo {author} {\bibfnamefont {D.~A.}\ \bibnamefont
  {Golbourn}},\ }\href {\doibase 10.1103/PhysRevD.76.044020} {\bibfield
  {journal} {\bibinfo  {journal} {Phys. Rev. D}\ }\textbf {\bibinfo {volume}
  {76}},\ \bibinfo {pages} {044020} (\bibinfo {year} {2007})},\ \Eprint
  {http://arxiv.org/abs/0705.3620} {arXiv:0705.3620 [gr-qc]} \BibitemShut
  {NoStop}%
\bibitem [{\citenamefont {Vega}\ and\ \citenamefont
  {Detweiler}(2008)}]{Vega-Detweiler:07}%
  \BibitemOpen
  \bibfield  {author} {\bibinfo {author} {\bibfnamefont {I.}~\bibnamefont
  {Vega}}\ and\ \bibinfo {author} {\bibfnamefont {S.~L.}\ \bibnamefont
  {Detweiler}},\ }\href {\doibase 10.1103/PhysRevD.77.084008} {\bibfield
  {journal} {\bibinfo  {journal} {Phys. Rev. D}\ }\textbf {\bibinfo {volume}
  {77}},\ \bibinfo {pages} {084008} (\bibinfo {year} {2008})},\ \Eprint
  {http://arxiv.org/abs/0712.4405} {arXiv:0712.4405 [gr-qc]} \BibitemShut
  {NoStop}%
\bibitem [{\citenamefont {Vines}(2015)}]{Vines:14}%
  \BibitemOpen
  \bibfield  {author} {\bibinfo {author} {\bibfnamefont {J.}~\bibnamefont
  {Vines}},\ }\href {\doibase 10.1007/s10714-015-1901-9} {\bibfield  {journal}
  {\bibinfo  {journal} {Gen. Rel. Grav.}\ }\textbf {\bibinfo {volume} {47}},\
  \bibinfo {pages} {59} (\bibinfo {year} {2015})},\ \Eprint
  {http://arxiv.org/abs/1407.6992} {arXiv:1407.6992 [gr-qc]} \BibitemShut
  {NoStop}%
\bibitem [{\citenamefont {Geroch}(1969)}]{Geroch:69}%
  \BibitemOpen
  \bibfield  {author} {\bibinfo {author} {\bibfnamefont {R.}~\bibnamefont
  {Geroch}},\ }\href@noop {} {\bibfield  {journal} {\bibinfo  {journal}
  {Communications in mathematical Physics}\ }\textbf {\bibinfo {volume} {13}},\
  \bibinfo {pages} {180} (\bibinfo {year} {1969})}\BibitemShut {NoStop}%
\bibitem [{\citenamefont {Stewart}\ and\ \citenamefont
  {Walker}(1974)}]{Stewart-Walker:74}%
  \BibitemOpen
  \bibfield  {author} {\bibinfo {author} {\bibfnamefont {J.~M.}\ \bibnamefont
  {Stewart}}\ and\ \bibinfo {author} {\bibfnamefont {M.}~\bibnamefont
  {Walker}},\ }\href@noop {} {\bibfield  {journal} {\bibinfo  {journal} {Proc.
  R. Soc. Lond. A}\ }\textbf {\bibinfo {volume} {341}},\ \bibinfo {pages} {49}
  (\bibinfo {year} {1974})}\BibitemShut {NoStop}%
\bibitem [{\citenamefont {Bruni}\ \emph {et~al.}(1997)\citenamefont {Bruni},
  \citenamefont {Matarrese}, \citenamefont {Mollerach},\ and\ \citenamefont
  {Sonego}}]{Bruni-etal:96}%
  \BibitemOpen
  \bibfield  {author} {\bibinfo {author} {\bibfnamefont {M.}~\bibnamefont
  {Bruni}}, \bibinfo {author} {\bibfnamefont {S.}~\bibnamefont {Matarrese}},
  \bibinfo {author} {\bibfnamefont {S.}~\bibnamefont {Mollerach}}, \ and\
  \bibinfo {author} {\bibfnamefont {S.}~\bibnamefont {Sonego}},\ }\href
  {\doibase 10.1088/0264-9381/14/9/014} {\bibfield  {journal} {\bibinfo
  {journal} {Class.Quant.Grav.}\ }\textbf {\bibinfo {volume} {14}},\ \bibinfo
  {pages} {2585} (\bibinfo {year} {1997})},\ \Eprint
  {http://arxiv.org/abs/gr-qc/9609040} {arXiv:gr-qc/9609040 [gr-qc]}
  \BibitemShut {NoStop}%
\bibitem [{\citenamefont {Kevorkian}\ and\ \citenamefont
  {Cole}(1996)}]{Kevorkian-Cole:96}%
  \BibitemOpen
  \bibfield  {author} {\bibinfo {author} {\bibfnamefont {J.}~\bibnamefont
  {Kevorkian}}\ and\ \bibinfo {author} {\bibfnamefont {J.~D.}\ \bibnamefont
  {Cole}},\ }\href@noop {} {\emph {\bibinfo {title} {Multiple scale and
  singular perturbation methods}}}\ (\bibinfo  {publisher} {Springer},\
  \bibinfo {address} {New York},\ \bibinfo {year} {1996})\BibitemShut {NoStop}%
\bibitem [{\citenamefont {Kates}(1981)}]{Kates:81}%
  \BibitemOpen
  \bibfield  {author} {\bibinfo {author} {\bibfnamefont {R.}~\bibnamefont
  {Kates}},\ }\href@noop {} {\bibfield  {journal} {\bibinfo  {journal} {Ann.
  Phys. (N.Y.)}\ }\textbf {\bibinfo {volume} {132}},\ \bibinfo {pages} {1}
  (\bibinfo {year} {1981})}\BibitemShut {NoStop}%
\bibitem [{\citenamefont {Detweiler}(2008)}]{Detweiler:08}%
  \BibitemOpen
  \bibfield  {author} {\bibinfo {author} {\bibfnamefont {S.~L.}\ \bibnamefont
  {Detweiler}},\ }\href {\doibase 10.1103/PhysRevD.77.124026} {\bibfield
  {journal} {\bibinfo  {journal} {Phys. Rev. D}\ }\textbf {\bibinfo {volume}
  {77}},\ \bibinfo {pages} {124026} (\bibinfo {year} {2008})},\ \Eprint
  {http://arxiv.org/abs/0804.3529} {arXiv:0804.3529 [gr-qc]} \BibitemShut
  {NoStop}%
\bibitem [{\citenamefont {Sago}\ \emph {et~al.}(2008)\citenamefont {Sago},
  \citenamefont {Barack},\ and\ \citenamefont
  {Detweiler}}]{Sago-Barack-Detweiler:08}%
  \BibitemOpen
  \bibfield  {author} {\bibinfo {author} {\bibfnamefont {N.}~\bibnamefont
  {Sago}}, \bibinfo {author} {\bibfnamefont {L.}~\bibnamefont {Barack}}, \ and\
  \bibinfo {author} {\bibfnamefont {S.~L.}\ \bibnamefont {Detweiler}},\ }\href
  {\doibase 10.1103/PhysRevD.78.124024} {\bibfield  {journal} {\bibinfo
  {journal} {Phys. Rev. D}\ }\textbf {\bibinfo {volume} {78}},\ \bibinfo
  {pages} {124024} (\bibinfo {year} {2008})},\ \Eprint
  {http://arxiv.org/abs/0810.2530} {arXiv:0810.2530 [gr-qc]} \BibitemShut
  {NoStop}%
\bibitem [{\citenamefont {Dolan}\ and\ \citenamefont
  {Barack}(2013)}]{Dolan-Barack:13}%
  \BibitemOpen
  \bibfield  {author} {\bibinfo {author} {\bibfnamefont {S.~R.}\ \bibnamefont
  {Dolan}}\ and\ \bibinfo {author} {\bibfnamefont {L.}~\bibnamefont {Barack}},\
  }\href {\doibase 10.1103/PhysRevD.87.084066} {\bibfield  {journal} {\bibinfo
  {journal} {Phys. Rev. D}\ }\textbf {\bibinfo {volume} {87}},\ \bibinfo
  {pages} {084066} (\bibinfo {year} {2013})},\ \Eprint
  {http://arxiv.org/abs/1211.4586} {arXiv:1211.4586 [gr-qc]} \BibitemShut
  {NoStop}%
\end{thebibliography}%

\end{document}